\documentstyle[amsfonts,epsfig,12pt,cite]{article}

\newcommand{\vb}[1]{{\mathbf{#1}}}
\newcommand{\r}[1]{\ref{#1}}
\newcommand{\lb}[1]{\label{#1}}

\def\zed{{\mathbb{Z}}}
\def\real{{\mathbb{R}}}
\newcommand{\bc}{\begin{center}}
\newcommand{\ec}{\end{center}}
\newcommand{\be}{\begin{equation}}
\newcommand{\ee}{\end{equation}}
\newcommand{\bea}{\begin{eqnarray}}
\newcommand{\eea}{\end{eqnarray}}
\newcommand{\nn}{\nonumber}
\newcommand{\ba}[1]{\begin{array}{#1}}
\newcommand{\ea}{\end{array}}
\newcommand{\bz}{{\overline{z}}}

\newcommand{\bt}[1]{\begin{table}[ht]\centering\begin{tabular}{#1}}
\newcommand{\et}[1]{\end{tabular}\caption{\small#1}\end{table}}

\def\sphere{\vb{S}}
\def\disc{\vb{D}}
\def\torus{\vb{T}}
\def\cyl{\vb{C}}
\def\Bmatrix{{\sf B}}
\def\Fmatrix{{\sf F}}
\def\Cmatrix{{\sf C}}
\def\Smatrix{{\sf S}}
\def\nmatrix{{\sf n}}
\def\Nmatrix{{\sf N}}
\newcommand{\NO}{\,\mbox{$\circ\atop\circ$}\,} 

\newcommand{\fig}[3]{\begin{figure}[htb]\bigskip\epsfxsize=100mm
\centerline{\epsfbox{#1}}\caption{\baselineskip=12pt\small\it #2
  \label{#3}}\bigskip\end{figure}}

\newcommand{\Tr}{{\mathrm{Tr}}\,}

\newcommand{\id}{{1\!\!1}} 
\def\e{{\,\rm e}\,}

\begin{document}


\thispagestyle{empty}

\begin{flushright}

{\small
OUTP/03--21P\\
HWM--03--14 \\
EMPG--03--13\\
{\tt hep-th/0308101} \\ {\sl August 2003}}

\end{flushright}

\vspace{1 truecm}

\begin{center}

{\Large\bf{D-BRANES IN TOPOLOGICAL MEMBRANES}}\\

\vspace{1 truecm}

{\bf P. Castelo Ferreira}
\\[2mm]
{\small{\it Dep. de Matem\'atica -- Instituto
Superior T\'ecnico, Av. Rovisco Pais, 1049-001 Lisboa, Portugal}\\ and\\
{\it PACT -- University of Sussex, Falmer, Brighton BN1 9QJ, U.K.}}\\ {\tt
pcferr@math.ist.utl.pt}
\\[5mm]
{\bf I.I.\ Kogan}
\\[2mm]
{\small{\it Dept. of Physics, Theoretical Physics -- University of Oxford,
Oxford OX1 3NP, U.K.}}
\\[5mm]
{\bf R.J. Szabo}
\\[2mm]
{\small\it Dept. of Mathematics -- Heriot-Watt University, Riccarton, Edinburgh
EH14 4AS, U.K.}\\{\tt R.J.Szabo@ma.hw.ac.uk}
\\[10mm]

\vspace{1.5 truecm}

{\bf\sc Abstract}

\begin{minipage}{15cm}
\vspace{2mm}

It is shown how two-dimensional states corresponding to D-branes arise
in orbifolds of topologically massive gauge and gravity
theories. Brane vertex operators naturally appear in induced
worldsheet actions when the three-dimensional gauge theory is
minimally coupled to external charged matter and the orbifold
relations are carefully taken into account. Boundary states
corresponding to D-branes are given by vacuum wavefunctionals of the
gauge theory in the presence of the matter, and their various
constraints, such as the Cardy condition, are shown to arise through
compatibility relations between the orbifold charges and bulk gauge
invariance. We show that with a further conformally-invariant coupling
to a dynamical massless scalar field theory in three dimensions, the
brane tension is naturally set by the bulk mass scales and arises
through dynamical mechanisms. As an auxilliary result, we show how to
describe string descendent states in the three-dimensional theory
through the construction of gauge-invariant excited wavefunctionals.

\end{minipage}

\end{center}

\newpage

\addtolength{\baselineskip}{0.20\baselineskip}

\thispagestyle{empty}
{\baselineskip=12pt
  \tableofcontents}

\newpage

\setcounter{page}{1}

\setcounter{equation}{0}
\section{INTRODUCTION AND SUMMARY}

Holography has played an important role over the last few years in
many developments centered around string theory and quantum
gravity. In many of its incarnations it provides a powerful tool which
allows one to extract information about a strongly-coupled theory from
the perturbative sector of a holographic, or dual theory. For example,
the duality between string theory on anti-de~Sitter spacetimes and
supersymmetric Yang-Mills theory on the boundaries provides a means of
extracting information about the strong curvature limit of supergravity from
quantum gauge theory, and it is further hoped that string theory could
shed light in this context on the nature of the strong-coupling regime
of gauge theories.

In this paper we will study one of the simplest examples of
holography. It is based on the representations of spaces of conformal
blocks in two-dimensions as the quantum Hilbert spaces of Chern-Simons
gauge theories in three-dimensions~\cite{W_0,LR_1,O_1,BN_1,W_1}. It
extends to a duality between two-dimensional conformal field
theories and gravity and three-dimensional topologically massive gauge
and gravity theories~\cite{TMGT_01,TMGT_02,TMGT_03,TMGT_04}. This has
culminated into the topological membrane approach to string theory and
has provided various powerful, dynamical descriptions of processes in
perturbative string theory which are reachable in the language of
three-dimensional gauge and gravity theories. As has been extensively
studied in the
past~\cite{TM_00,TM_01,TM_02,TM_03,TM_04,TM_05,TM_06,TM_07,TM_08,TM_09,TM_10,TM_11,TM_12,TM_13,TM_14,TM_15,TM_16,TM_17,TM_18},
this formalism starts with a three-dimensional thickened worldsheet on
which there lives a topologically massive (or Maxwell-Chern-Simons) gauge
theory, together with topologically massive (or Einstein-Chern-Simons)
gravity and a propagating three-dimensional scalar field corresponding
to the string dilaton field. This entity is known as the ``topological
membrane''. The usual string worldsheet is split into the left and
right moving two-dimensional boundaries of the membrane. As is
well-known, even at the level of the topological quantum field theory,
pure Chern-Simons theory is not sufficient to describe all of the
freatures of a full conformal field theory (let alone the full string
theory). Including a Maxwell kinetic term into the theory lifts the
degeneracy of the phase space, leaving four canonical variables (in
contrast to the two of pure Chern-Simons theory). This lifting is
necessary to enable the proper incorporation of both left and right
moving degrees of freedom of the string theory.

The primary motivation behind this approach has been the unification
of the various string theories through one fundamental theory, the
topological membrane, using the wealth of analytic tools available to
field theories in three spacetime dimensions. This framework can
thereby be thought of as M-theory at the worldsheet level. In fact,
there are many indications already at this level that various
11-dimensional constructs, such as the Ho\v{r}ava-Witten embedding of the
$E_8\times E_8$ heterotic string into M-theory~\cite{HW_1,HW_2}, can
find fundamental realizations as a one dimension extension of the
string worldsheet. In other contexts this holographic correspondence
has been described using certain spin-foam models of quantum
gravity~\cite{FK_1}. In this paper we will concentrate on the
incorporation of D-branes and open string Wilson lines into the
topological membrane formalism.

\subsection{D-Branes}

To help place our analysis into context, we begin with a brief
overview of the nature of D-branes in string theory. Dirichlet
boundary conditions were originally introduced for all of the string
embedding coordinates in order to probe off-shell string
theory~\cite{Or_01,Or_02,Or_03,Or_04}. Coexisting Dirichlet and
Neumann boundary conditions were studied for the first time
in~\cite{Siegel} as a possible compactification scheme for string
theory. Dual string theories~\cite{Scherk} incorporate
point-like structures~\cite{Green_01,Green_02,Green_03}. This fact
strongly suggests that partonic behaviour and hence a temptative
realistic description of quantum chromodynamics in string theory is
introduced through point-like
sources~\cite{G-QCD_01,Z-QCD,Y-QCD_01,Y-QCD_02,W-QCD}.

It was noted in~\cite{Green_03} that duality interchanges Neumann and
Dirichlet boundary conditions. This fact strongly indicates that both
types of boundary conditions should coexist in dual
theories. Furthermore, it was argued in~\cite{POL_01} that the large
nonperturbative effects (of order $\e^{-1/g_s}$) in string theory have
their origin in the same sorts of mechanisms. In~\cite{POL_02} it was
finally established that D-branes are indeed intrinsic objects in
string theory and are required by consistency of the various string
dualities.

Recently, the study of D-branes at the level of worldsheet conformal
field theory and Wess-Zumino-Novikov-Witten (WZNW) $\sigma$-models have
received much
attention~\cite{RS1,AS1,WZW_1,WZW_2,WZW_3,PS_3}
(see~\cite{G} for a review). In these approaches, D-branes correspond
to vertex operators in boundary conformal field theory. The
fruitfulness of these models is that they provide an analytic and
tractable means of studying D-branes on curved spaces which are group
manifolds. The connection between the target space description of
D-branes and the worldsheet framework is best understood within a
worldsheet $\sigma$-model path integral (effective action)
formalism~\cite{Db_02,Db_03}. These descriptions were brought
together in~\cite{TD} where it was found that the tension of a D-brane
is proportional to the suitably regularized dimension of the physical
Hilbert space of the worldsheet boundary conformal field theory, and
agrees with the description in terms of an effective target space
action for the D-brane dynamics. It is these boundary conformal field
theory approaches that will naturally emerge from the topological
membrane formalism.

\subsection{Summary of Main Results}

In this paper we will expand on the analysis initiated in~\cite{TM_18}
which is based on constructing Schr\"odinger wavefunctionals of
topologically massive gauge theory that correspond to states of the
induced conformal field theory. Partition functions and correlators of
the two-dimensional $\sigma$-models are thereby obtained through inner
products between three-dimensional states. The construction of open
string states, in particular those describing D-branes, is based on
representing the open string sector as a worldsheet orbifold of an
associated closed string sector~\cite{SBH_03,SBH_04,SBH_05}. The
extension of these constructions to the three-dimensional setting was
first described in~\cite{H_1} and further elucidated
in~\cite{TM_16,TM_17,TM_18}. Recently, the three-dimensional
topological field theory description has been extended to incorporate
boundary states and amplitudes appropriate to D-branes
in~\cite{FS_1,FS_2,FS_3,FS_4}. This approach relies on a model-independent
axiomatic formalism which gives rise to a modular tensor category
describing the Moore-Seiberg data of a rational, chiral conformal
field theory. While some of the techniques we introduce in the
following are conceptually the same, our goal is to obtain {\it explicit}
representations of D-brane states directly from the physical
wavefunctions of the underlying three-dimensional gauge and gravity
theories.

In what follows we will only analyse the simplest instances
corresponding to D-branes in flat, compact backgrounds, or
equivalently WZNW models based on abelian Lie groups. The advantage of
these models over their non-abelian counterparts is that we can make
all constructions very explicit without too many technical
obstructions, and a lot of what we say can at least in principle be
applied also to arbitrary gauge groups. The extensions to non-abelian gauge
groups and D-branes in curved backgrounds represents one of the most
important generalizations of the ensuing analysis. The topological
membrane could thereby prove to be an indispensible tool for
discovering the properties of D-branes on group manifolds, which are
not yet understood in their entirety. Another important aspect which
is not addressed in the following is the emergence of D-brane charge
within the three-dimensional context. This requires the proper
construction of Ramond-Ramond states in the topological membrane and
hence an understanding of how target space supersymmetry arises, which
at present is not very well-understood. A better understanding of the
latter feature, along with the embedding of the $E_8\times E_8$
heterotic string in three-dimensions~\cite{TM_06,TM_15}, may help
elucidate the connections between the topological membrane and
M-theory.

The main accomplishments of the present paper can be summarized as
follows:
\begin{itemize}
\item For the first time, D-branes and open string Wilson lines are
  incorporated into the topological membrane approach. Their vertex
  operators naturally appear in vacuum membrane wavefunctionals when the
  topologically massive gauge theory is coupled to external sources,
  and the orbifold relations are carefully taken into account.
\item We find the membrane wavefunctionals corresponding to Ishibashi
  boundary states. The Cardy condition, which normally selects the
  appropriate D-brane states, in the three-dimensional case is found
  to be a compatibility requirement with bulk gauge invariance of the
  boundary state wavefunctionals.
\item In addition to the Cardy map, we find natural dynamical
  interpretations of the various constraints in boundary conformal
  field theory. For example, locality is a consequence of the cross
  channel duality in the scattering amplitudes of two external charged
  particles in the bulk. All of these properties stem from the
  representation of the Verlinde formula in terms of the Hopf linking
  of charged particle trajectories.
\item We show that the mass of a D-brane is naturally set by the bulk
  mass scales of the three-dimensional theory, and that it also arises as a
  dynamical property of the topological membrane.
\item We find the membrane wavefunctionals corresponding to string
  descendent states. They are gauge-invariant excited eigenstates of the
  three-dimensional field theoretic Hamiltonian. In particular, the
  wavefunctionals corresponding to the graviton vertex operator are
  those which are directly responsible for the appearence of the
  D-brane tension.
\item We acquire a new perspective on open/closed string duality.
\end{itemize}

\subsection{Outline}

The remainder of this paper is devoted to deriving the results
described above, along with many other auxilliary descriptions. It
naturally falls into three parts which describe, respectively, in
detail how the appropriate brane vertex operators emerge, what the
three-dimensional analogs of brane boundary states are, and how the
brane tension manifests itself in the topological membrane
framework. In the rest of this section we shall outline the structure
of each of these three parts.

In section~\r{Verts} we minimally couple external sources to the
quantum field theory living in the three-dimensional membrane. We
study their compatibility with the orbifolds of the membrane under its
allowed discrete symmetries $\sf PT$ and $\sf PCT$. The continuity
equations for the source currents are found to impose very stringent
conditions. Depending on the type of orbifold taken, the currents
$J_i$ can be of only two types. For the three-dimensional orbifolds taken with
respect to the $\sf PT$ symmetry, we find $J_i\propto\partial_iY_{\rm D}$
corresponding to Dirichlet boundary conditions on the induced open
string theory, while for the $\sf PCT$ orbifolds we find
$J_i\propto\epsilon_{ij}\,\partial^jY_{\rm N}$ corresponding to Neumann
boundary conditions. It is shown that the quantity $Y_{\rm D}$
corresponds to a D-brane collective coordinate, while $Y_{\rm N}$
parametrizes the photon field of the open string Wilson line. The
D-brane vertex naturally emerges in this framework as a boundary term
in the induced open string worldsheet action. The analysis of this
section motivates the conjecture that one way to include curved
backgrounds in this approach, for an {\it abelian} gauge group of the
three-dimensional theory, is through the proper incorporation of
auxilliary gauge fields as external currents.

In section~\r{bstates} we study, from the point of view of the
topological membrane, various aspects of boundary conformal field
theory and its description in terms of pure three-dimensional topological field
theory, i.e. the gauge theory with just a Chern-Simons term in the
action. Following~\cite{BT_1,GSST_1}, we study the correlation
functions of Polyakov loop operators in the corresponding finite
temperature gauge theory. The wavefunctionals describing the states of
conformal field theory are explicitly constructed, and in particular the ones
corresponding to Ishibashi boundary states. The Verlinde
diagonalization formula is described in a new vein and the constraints
defining the fundamental boundary states (corresponding to the D-brane
and open string Wilson line vertex operators) are chosen by
compatibility with bulk gauge invariance. The Cardy map is obtained by
the constraints on the allowed charges of the bulk matter fields.

Finally, in section~\r{Daction} we derive the D-brane tension directly
from the scales of the topological membrane. For this, it is also
necessary to introduce on the membrane topologically massive gravity
along with a coupled scalar field $D$ which induces the string
dilaton field in two-dimensions. The gravitational sector of the
membrane induces two-dimensional quantum gravity, i.e. a (deformed)
Liouville field theory on the string worldsheet, and the string coupling
constant is determined by the vacuum expectation value of the scalar
field $D$ as $g_s=\left<D^4\right>$. We derive the effective
worldsheet $\sigma$-model action including the D-brane collective
coordinates and the open string Wilson lines. We obtain explicitly the
Born-Infeld action for a wrapped D-brane and relate it to the
regularized dimension of the full Hilbert space of physical states of
the three-dimensional quantum field theory (including excited
states). The latter quantity is also explicitly computed by exploiting
a careful bulk renormalization of the topological graviton mass. By
doing so, we also solve the long-standing problem of describing
excited string states in terms of the topological membrane, and we
obtain a novel impetus on the worldsheet modular duality between the
open and closed string sectors of the theory.

\subsection{Note Added}

The work described in this paper and most of the writing of the
manuscript were completed by June 4 2003, the tragic day on which
Ian~Kogan suddenly and unexpectedly passed away. The other two named
authors have decided to proceed with submission of the paper as a
tribute to his memory, as this was a topic very close and dear to
him. He would have taken great pleasure in seeing it finally
completed.

\setcounter{equation}{0}
\section{BRANE VERTICES\label{Verts}}

In~\cite{TM_18} it was remarked for the first time that one could
construct collective coordinates for D-branes by minimally coupling
external currents to three-dimensional topologically massive gauge
fields. The purpose of this section is to make this observation
explicit and construct D-brane
vertex operators in detail from the vacuum Schr\"odinger wavefunctionals of
topologically massive gauge theory. The basic idea is that the
position of the D-brane can be changed by the addition to the
two-dimensional $\sigma$-model action of the boundary
operator $\partial_\perp^{~}x^I$ given by the normal derivative to the
boundary of the worldsheet of the open string embedding field. The
operator $a_I\,\partial_\perp^{~}x^I$ translates the D-brane by the
position vector $a^I$. Moreover, in this framework the string Wilson line
is given by the tangential derivative to the boundary and corresponds to
the operator $A_I\,\partial_\parallel^{~}x^I$.
The important observation of~\cite{TM_18} was that
these vertex operators can be induced in the three-dimensional setting
by the addition of bulk matter fields. This agrees with the general
ideology of topological membrane theory that a change in conformal
background, which is described in two-dimensional terms by a deformed
conformal field theory, is described in three-dimensions by adding
charged matter in the bulk~\cite{TM_09}. Thus the D-brane collective
coordinate, which controls the background, is now itself controlled by the bulk
distribution of charged matter. This unifies the topological membrane
pictures for all possible backgrounds. As we will see in the
following, the D-brane in this picture simply corresponds to charged
matter on an orbifold line in three dimensions.

\subsection{\lb{sec.basics}Hamiltonian Formulation}

We will begin by reviewing and expanding on some of the basic aspects of $U(1)$
topologically massive gauge theories~\cite{TMGT_01,TMGT_02,TMGT_03,TMGT_04}
that will be required in the following.
The gauge theory is defined on a three-dimensional manifold of the form
$M=\Sigma\times[0,1]$, where the finite interval $[0,1]$ parametrizes the time
$t$, and $\Sigma$ is a compact orientable Riemann surface whose local
complex coordinates will be denoted $\vb{z}=(z,\bz\,)$ with integration
measure $d^2z=|dz\wedge d\bz\,|$. We will adopt Gaussian normal
coordinates for the Minkowski three-geometry, in which the metric
takes the form
\be
ds_{(3)}^2=g_{\mu\nu}~dx^\mu~dx^\nu=-dt^2+h_{ij}~dx^i~dx^j \ .
\label{ds3}\ee
The three-manifold has two boundaries $\Sigma_0=\Sigma\times\{0\}$
and $\Sigma_1=\Sigma\times\{1\}$ at times $t=0$ and $t=1$, respectively,
both of which are copies of $\Sigma$ with opposite orientation.

The action is a sum of Maxwell and Chern-Simons terms for a
$U(1)$ gauge field $A$ with curvature $F$,
\be
S_{\rm TMGT}[A]=\int\limits_0^1dt~\int\limits_\Sigma d^2z~
\left[-\frac{\sqrt{-g}}{4\gamma}\,F_{\mu\nu}F^{\mu\nu}+\frac{k}{8\pi}\,
\epsilon^{\mu\nu\lambda}\,A_\mu\,\partial_\nu A_\lambda+
\sqrt{-g}\,A_\mu J^\mu\right] \ ,
\lb{STMGT}
\ee
where we have included the minimal coupling of the gauge fields to a conserved
current $J^\mu$,
\be
\partial_\mu J^\mu=0 \ .
\label{Jmuconserved}
\ee
It will be convenient to rescale the spatial components
of the current by the coupling constant $\gamma$ such that
the continuity equation reads
\be
\partial_t\rho=-\gamma\,\partial_zj^\bz-\gamma\,\partial_\bz j^z
\lb{J0}
\ee
in terms of the charge density $J^0=\rho$ and the current densities
\be
J^i=2\gamma\,j^i \ .
\label{Jigammaji}\ee
The bulk Levi-Civita antisymmetric tensor density
$\epsilon^{\mu\nu\lambda}$ induces the tensor density
$\epsilon^{ij}=\epsilon^{0ij}$ on the boundaries. As has been
extensively studied in the
past~\cite{TM_00,TM_01,TM_02,TM_03,TM_04,TM_05,TM_06,TM_07,TM_08,TM_09,TM_10,TM_11,TM_12,TM_13,TM_14,TM_15,TM_16,TM_17,TM_18,W_0,W_1,BN_1,LR_1,O_1},
the quantum field theory defined by (\ref{STMGT}) induces new degrees
of freedom on the boundaries which constitute chiral gauged WZNW models. They
are fields belonging to two-dimensional chiral conformal field
theories living on $\Sigma_0$ and $\Sigma_1$.

The action (\ref{STMGT}) can be written in a canonical splitting
$A_\mu=(A_0,A_i)$ as
\be
\ba{lll}
S_{\rm TMGT}[A]&=&\displaystyle\int\limits_0^1dt~\int\limits_\Sigma d^2z~
\left[-\frac{\sqrt{-g}}{2\gamma}\,F_{0i}F^{0i}-\frac{\sqrt{-g}}{4\gamma}\,
F_{ij}F^{ij}+\frac{k}{16\pi}\,\epsilon^{ij}\,A_0F_{ij}+\frac{k}{8\pi}\,
\epsilon^{ij}\,A_iF_{j0}\right.\\[2mm]&&\displaystyle+\biggl.\sqrt{-g}\,A_0J^0+
\sqrt{-g}\,A_iJ^i\biggr] \ .
\ea
\ee
The canonical momentum conjugate to $A_i$ is
\be
\Pi^i=-\frac{\sqrt{-g}}{\gamma}\,F^{0i}+\frac{k}{8\pi}\,\epsilon^{ij}\,A_j \ ,
\ee
while, as usual, the canonical momentum conjugate to $A_0$ is identically
zero. The field $A_0$ is therefore non-dynamical and serves as a Lagrange
multiplier which imposes the Gauss law constraint
\be
0=\int\limits_{\Sigma}
d^2z~\left(-\frac{\sqrt{h}}{\gamma}\,\partial_iF^{0i}+\frac{k}{8\pi}\,
\epsilon^{ij}\,F_{ij}+\sqrt h\,\rho\right)-\oint\limits_{\partial\Sigma}
\left(-\frac{\sqrt{h}}{\gamma}\,F^{0i}+\frac{k}{8\pi}\,\epsilon^{ij}\,
A_{j}\right)n_i \ ,
\lb{gauss}
\ee
where $n_i$ is a vector normal to the boundary of $\Sigma$. Note that the
boundary term in (\ref{gauss}) is only present when the two-dimensional
boundary $\Sigma$ of the underlying three-manifold itself has a
boundary. Of course for a smooth space this term doesn't appear
because the boundary of a boundary is empty. However, once we quotient
the theory by its discrete symmetries new boundaries can emerge at orbifold
singularities~\cite{H_1,TM_16,TM_17,TM_18}. This extra boundary term
is vital for the construction that we will present in the following,
because it allows for the imposition of the correct boundary conditions on
the induced conformal field theory. Moreover, conformal vertex
operators inserted on the boundary are thereby included in the full
three-dimensional theory as \textit{external} fluxes coupled to the
gauge fields through the conserved current $J^\mu$, in accordance with
the fact that closed string vertex operators correspond to Wilson
lines of the three-dimensional gauge theory. We will see later on that
the external charges actually allow one to introduce collective
coordinates of D-branes. It is also this mechanism that constrains the
open string gauge group and therefore the Chan-Paton degrees of
freedom~\cite{SBH_07,SBH_07a,SBH_07b}.

The Hamiltonian of the field theory is given by
\be
\ba{rcl}
H&=&\displaystyle\int\limits_\Sigma d^2z~\left\{
-A_0\left[\partial_i\left(\Pi^i-\frac{k}{8\pi}\,
\epsilon^{ij}\,A_j\right)+\frac{k}{4\pi}\,\epsilon^{ij}\,\partial_iA_j
+\sqrt h\,\rho\right]\right.\\[2mm]&&\displaystyle+\,
\partial_i\left[A_0\Pi^i
\right]+\frac{1}{8\gamma\,\sqrt{h}}\,\left(\epsilon^{ij}\,
F_{ij}\right)^2\\[2mm]&&\displaystyle+\left.
\frac{\gamma}{2\,\sqrt{h}}\,h_{ij}\left(\Pi^i-\frac{k}{8\pi}\,
\epsilon^{ik}\,A_k\right)\left(\Pi^j-\frac{k}{8\pi}\,\epsilon^{jl}\,A_l\right)
-2\,\sqrt h\, \gamma\,A_ij^i\right\} \ .
\label{Ham}
\ea
\ee
By defining the electric and magnetic fields as
\be
\ba{rcl}
E^i&=&-\displaystyle\frac{1}{\gamma}\,F^{0i} \ , \\[4mm]
B  &=&\partial_z A_\bz-\partial_\bz A_z \ ,
\ea
\ee
the bulk and boundary Gauss law constraints in (\ref{gauss}) read
\bea
\partial_iE^i+\frac{k}{4\pi}\,B&=&-\rho~~~~{\rm in}~~\Sigma \ , \nn\\
E^\perp&=&-i\,\frac k{8\pi}\,A^\perp~~~~{\rm on}~~\partial\Sigma \ .
\label{GaussEB}\eea
Here we denote components of fields normal and tangential to the
worldsheet boundary by the scripts $\perp$ and $\parallel$,
respectively, and we use the metric conventions that raised and
lowered indices at $\partial\Sigma$ correspond to the interchanges
$\perp\leftrightarrow\parallel$. In the quantum field theory, the
canonical commutation relations can be written
as
\be
\ba{rcl}
\Bigl[E^i(\vb{z})\,,\,E^j(\vb{z}')\Bigr]&=&\displaystyle-i\,{k\over{4\pi}}
\,\epsilon^{ij}\,\delta^{(2)}(\vb{z}-\vb{z}') \ , \\[2mm]
\Bigl[E^i(\vb{z})\,,\,B(\vb{z'})\Bigr]&=&-i\,\epsilon^{ij}\,\partial_j
\delta^{(2)}(\vb{z}-\vb{z}') \ ,
\ea
\lb{com_EB}
\ee
and the constraints (\ref{GaussEB}) lead to an equation that needs to be
satisfied by the physical (gauge invariant) states. We will use these equations
when we construct the wavefunctions of the quantum
field theory. Note that the $A_0$ dependent terms in (\ref{Ham}) vanish when
the Hamiltonian operator acts on such states.

The generators of time-independent local gauge transformations can be easily
defined, for smooth real-valued gauge parameter functions $\Lambda$,
as~\cite{TM_08}
\be
U_\Lambda=\exp\left\{i \int\limits_\Sigma d^2z~\sqrt h\,\Lambda(\vb{z})
\left(\partial_iE^i+{k\over{4\pi}}\,B+\rho\right)\right\} \ .
\lb{U}
\ee
For consistency with the boundary Gauss law in (\ref{GaussEB}), the
normal derivative of the gauge parameter function in (\ref{U}) must
obey Neumann boundary conditions at the boundary $\partial\Sigma$, i.e.
$\partial_\perp^{~}\Lambda=0$. The physical Hilbert space consists of
those quantum states which are invariant
under the actions of the operators (\ref{U}). In addition, when there are
topologically non-trivial gauge field configurations, we must take into account
the large gauge transformations of the theory. They are generated by the
operators (\ref{U}) obtained by taking $\Lambda=\theta$ to be the
multi-valued angle function of the Riemann surface $\Sigma$. Then
integrating by parts in (\ref{U}) yields the extra local
operator~\cite{V_01,V_02,V_03,V_04}
\be
V(\vb{z}_0)=\exp\left\{-i\int\limits_\Sigma d^2z~\left[\left(E^i+
\frac{k\,\sqrt h}{4\pi}\,\epsilon^{ij}\,A_j\right)\epsilon_{ik}\,\partial^k
\ln E(\vb{z},\vb{z}_0)-\sqrt h\,
\theta(\vb{z},\vb{z}_0)\,\rho\right]\right\} \ ,
\lb{V}
\ee
where we have dropped the boundary term using the Gauss law
(\ref{GaussEB}) on $\partial\Sigma$. Here $E(\vb{z},\vb{z}_0)$ is the
prime form of $\Sigma$, $\vb{z}_0$ is a fixed point on $\Sigma$, and
\be
\theta(\vb{z},\vb{z}_0)={\rm Im}\,\ln\frac{E(\vb{z},\vb{z}_0)}
{E(\vb{z},\vb{z}')E(\vb{z}',\vb{z}_0)}
\label{thetaImdef}\ee
with $\vb{z}'$ an arbitrary fixed reference point. Demanding invariance under
these operators, i.e. under large gauge transformations, further truncates the
physical Hilbert space of the quantum field theory.

{}From the commutation relations~(\ref{com_EB}) we can compute the
commutator
\be
\Bigl[B(\vb{z})\,,\,V^n(\vb{z}_0)\Bigr]=2\pi n\,\sqrt h\,V^n(\vb{z}_0)\,
\delta^{(2)}(\vb{z}-\vb{z}_0)
\ee
for any integer $n$. This means that the operator $V^n(\vb{z}_0)$
creates a pointlike magnetic vortex at $\vb{z}_0$ with magnetic flux
$\int_\Sigma d^2z~\sqrt h\,B=2\pi n$. These objects thereby generate
nonperturbative processes which constitute monopoles of the gauge
theory. Moreover, from Gauss' law (\ref{GaussEB}) we see that they
also carry a bulk electric charge
\be
\Delta Q=-\frac{nk}2 \ .
\label{DeltaQ}\ee

The electric charge spectrum of the quantum field theory is~\cite{TM_08}
\be
Q_{m,n}=m+\frac{k}{4}\,n \ ,
\lb{charge}
\ee
where $m$ and $n$ are integers representing, respectively, the contributions
from the usual Dirac charge quantization and the monopole flux. Due to the
existence of monopole induced processes and linkings between Wilson lines
(charge trajectories) it can be shown~\cite{TM_15} that, with the correct
relative boundary conditions, the insertion of the charge $Q_{m,n}$ at one
boundary $\Sigma_0$ (corresponding to a vertex operator insertion in the
boundary conformal field theory) necessitates an insertion of the charge
\be
\bar{Q}_{m,n}=m-\frac{k}{4}\,n
\lb{ccharge}
\ee
at the other boundary $\Sigma_1$. This fact will be assumed throughout the rest
of this paper.

Let us now turn to the quantization of the topologically massive gauge
theory in the functional Schr\"odinger picture, whereby the physical
states are the wavefunctionals $\Psi^{\rm phys}[A;j]$. By using the
canonical quantum commutators (\ref{com_EB}) and the representation
\be
\Pi^i=-i\,\sqrt h\,\frac\delta{\delta A_i} \ ,
\ee
we impose the Gauss laws (\ref{GaussEB}) as constraint equations on the
wavefunctionals. By assuming a well-defined decomposition of all fields
into independent bulk and boundary degrees of freedom in what follows,
we may then factorize the physical states as
\be
\Psi^{\rm phys}[A;j]=\chi\left[A^\parallel\,\right]~
\Psi\left[A_z,A_\bz\,;j^z,j^\bz\,\right] \ .
\label{chiphysdec}
\ee
For closed surfaces where $\partial\Sigma=\emptyset$, we take
$\chi=1$. But for a generic surface the functional $\chi$
solves the functional boundary Gauss law constraint in (\ref{GaussEB}),
\be
-i\,\frac{\delta}{\delta A^\perp}\chi\left[A^\parallel\,\right]=0 \ .
\label{bdryconstrPi}
\ee
The bulk gauge constraint in (\ref{GaussEB}) reads
\bea
&&\left[\partial_\bz\left(-i\,\frac{\delta}{\delta A_\bz}+\frac{k}{8\pi}\,
\tilde\epsilon^{\,z\bz}\,A_z\right)+\partial_z\left(-i\,\frac{\delta}
{\delta A_z}-\frac{k}{8\pi}\,\tilde\epsilon^{\,z\bz}\,A_\bz\right)\right.
\nn\\&&\left.~~~~~~~~~~+\,\frac{k}{4\pi}\,\tilde\epsilon^{\,z\bz}
\,F_{z\bz}+\rho\right]\Psi\left[A_z,A_\bz\,;j^z,j^\bz\,\right]~=~0 \ ,
\lb{gaussjj}\eea
where the two-dimensional antisymmetric tensor $\tilde\epsilon$ is induced from
the bulk by $\tilde\epsilon^{\,ij}=\epsilon^{0ij}/\sqrt{-g}$. By applying the
Hamiltonian (\ref{Ham}) to the physical wavefunctions
(\ref{chiphysdec}) in this polarization, we
find that the stationary states satisfy the functional Schr\"odinger equation
\be
\ba{ll}
&\displaystyle\int\limits_\Sigma d^2z~\sqrt h\,\left\{-\frac\gamma2\,h_{z\bz}\,
\left(-i\,\frac\delta{\delta A_\bz}-\frac k{8\pi}\,\tilde\epsilon^{\,z\bz}
\,A_z\right)\left(-i\,\frac\delta{\delta A_z}+\frac k{8\pi}\,\tilde
\epsilon^{\,z\bz}\,A_\bz\right)\right.\\[2mm]&~~~~~~~~~~\displaystyle
+\left.\frac1{8\gamma}\,\left(\tilde
\epsilon^{\,z\bz}\,F_{z\bz}\right)^2-\gamma\,A_zj^\bz-\gamma\,A_\bz\,j^z
\right\}\Psi\left[A_z,A_\bz\,;j^z,j^\bz\,\right]~=~{\cal E}\,\Psi
\left[A_z,A_\bz\,;j^z,j^\bz\,\right]
\ea
\label{Schreq}
\ee
where $\cal E$ is the energy of the state and we have appropriately normal
ordered the Hamiltonian density (This latter point is elucidated in
section~\ref{Daction}). Since the gauge constraint commutes with the
Hamiltonian, these last two equations can be consistently solved.

\subsection{Neutral Wavefunctionals\label{WS-FC}}

To establish the pertinent properties of the general solutions to
(\ref{gaussjj}) and (\ref{Schreq}), we will first briefly recall the
solution in the absence of external currents~\cite{TM_18} and derive
its extension to surfaces with boundary. Because of the appearence of
the magnetic field in (\ref{U}), in topologically massive gauge theory
the physical states are not gauge invariant. Instead, by integrating the
infinitesimal gauge constraint (\ref{gaussjj}) one finds that the
gauge symmetry is represented projectively on wavefunctionals as
\be
U_\Lambda\Psi[A_i;0]=\e^{i\alpha[A_i,\Lambda]}\,
\Psi[A_i+\partial_i\Lambda;0] \ ,
\label{projrep}
\ee
where the projective phase is given by the Polyakov-Wiegmann
one-cocycle of the $U(1)$ Lie algebra as
\be
\alpha[A_i,\Lambda]=\frac k{8\pi}\,\int\limits_\Sigma
d^2z~\epsilon^{ij}\,A_i\,\partial_j\Lambda \ .
\label{cocycle}
\ee
To separate out the gauge invariant part, one integrates the corresponding
cocycle condition
\be
\alpha[A_i,\Lambda+\Lambda'\,]=\alpha[A_i+\partial_i\Lambda,
\Lambda'\,]+\alpha[A_i,\Lambda]
\label{cocyclecond}\ee
to get
\be
\Psi[A_z,A_\bz\,;0]=\exp\left\{-\frac{ik}{8\pi}\,\int\limits_\Sigma d^2z~
\sqrt{h}\,\tilde{\epsilon}^{\,z\bz}\,A_zA_\bz\right\}~\psi[A_z]~\Phi[B] \ ,
\lb{Psi}
\ee
where $B$ is the magnetic field. The exponential prefactor in
(\ref{Psi}) is the K\"ahler potential measure which is required in the
holomorphic representation and which compensates the appropriate
boundary conditions.

The factor $\Phi[B]$ is the gauge-invariant solution of Gauss' law for the pure
Maxwell theory ($k=0$),
\be
\left[\partial_\bz\,\frac{\delta}{\delta A_\bz}+\partial_z\,
\frac{\delta}{\delta A_z}\right]\Phi=0 \ .
\ee
If the fields have non-trivial magnetic charge $\int_\Sigma d^2z~\sqrt h\,B
\neq 0$, then the wavefunction $\Phi[B]$ vanishes~\cite{AFC_1}. When
$\partial\Sigma=\emptyset$, this result is
simply the statement that there is overall charge conservation on the closed
surface $\Sigma$. However, locally non-zero magnetic field distributions are
still possible. The component $\psi[A_z]$ solves the gauge constraint of
pure Chern-Simons theory~($\gamma\to\infty$),
\be
\left[\partial_z\,\frac{\delta}{\delta A_z}-\frac{i k}{4\pi}\,
\tilde{\epsilon}^{\,z\bz}\,\partial_\bz A_z\right]\psi[A_z]=0 \ .
\label{CSsols}
\ee

In this section we will only be interested in the ground state of the
field theory (Excited states will be described later on in
section~\ref{Daction}). This corresponds to a projection onto the
lowest Landau level of the quantum spectrum, which is attained in the
topological limit $\gamma\to\infty$ (The mass gap between Landau levels is
proportional to $\gamma$). In this case, $\Phi=1$ and we recover the
wavefunctions of pure Chern-Simons gauge theory. These solutions
correspond to configurations with weak magnetic field,
$\tilde\epsilon^{\,z\bz}\,F_{z\bz}\simeq 0$. The stationary states are
then the eigenfunctions of the first term on the left-hand
side of (\ref{Schreq}). In particular, the vacuum state (${\cal E}=0$) is
determined by the zero mode equation
\be
\left[\frac\delta{\delta A_z}-\frac k{8\pi}\,A_\bz\right]
\Psi[A_z,A_\bz\,;0]=0 \ ,
\label{0mode}
\ee
where we have fixed a complex structure on $\Sigma$ determined by the
worldsheet metric whereby $\tilde\epsilon^{\,z\bz}=i$ (In complex
Euclidean space the antisymmetric tensor is purely
imaginary). Furthermore, by using (\ref{CSsols}) and (\ref{Psi}) with
$\Phi=1$, one finds that it obeys the gauge constraint
\be
\left[\partial_\bz\,\frac{\delta}{\delta A_\bz}+\frac{k}{8 \pi}\,
\partial_\bz A_z-\frac{k}{4\pi}\,F_{\bz z}\right]\Psi[A_z,A_\bz\,;0]=0 \ .
\label{WW}
\ee

A solution of (\ref{0mode}) and (\ref{WW}), which is compatible with the gauged
WZNW construction for an abelian gauge group, is given by a path integral over
an auxilliary, dimensionless worldsheet scalar field $\varphi$ as
\be
\Psi[A_z,A_\bz\,;0]=\int[D\varphi]~
\exp\left\{\,\frac{k}{8\pi}\,\int\limits_{\Sigma}d^2z~
\left[A_\bz\,A_z-2A_\bz\,\partial_z\varphi+
\partial_\bz\varphi\,\partial_z\varphi\right]\right\} \ .
\lb{WWS}
\ee
When the worldsheet $\Sigma$ has a boundary, there is a correction to
the projective phase (\ref{cocycle}) given by
\be
\alpha^{~}_{\rm b}[A_i,\Lambda]=-\frac k{8\pi}\,\int\limits_\Sigma
d^2z~\epsilon^{ij}\,\partial_i\left(\Lambda\,A_j\,\right)=
-\frac k{8\pi}\,\oint\limits_{\partial\Sigma}\Lambda\,A^\parallel \ .
\label{cocycleb}
\ee
Then the functional
$\chi\left[A^\parallel\,\right]$ is also given in terms of the
auxilliary scalar field $\varphi^{~}_{\rm
  b}=\varphi|_{\partial\Sigma}$, the restriction
of $\varphi$ to the boundary of the worldsheet $\Sigma$, by
\be
\chi\left[A^\parallel\,\right]=\int
\left[D\varphi^{~}_{\rm b}\right]~
\exp\left\{\frac{k}{8\pi}\,\oint\limits_{\partial\Sigma}
\varphi^{~}_{\rm b}\,A^\parallel\right\} \ .
\label{chi}
\ee
The full wavefunction (\ref{chiphysdec}) is thereby given as
\be
\Psi^{\rm phys}[A_z,A_\bz\,;0]=\int[D\varphi]~
\exp\left\{\,\frac{k}{8\pi}\,\left[\,\int\limits_{\Sigma}d^2z~
\left(A_\bz\,A_z-2A_\bz\,\partial_z\varphi+\partial_\bz\varphi\,
  \partial_z\varphi\right)+\oint\limits_{\partial\Sigma}
\varphi^{~}_{\rm b}\,A^\parallel\right]\right\} \ .
\lb{WWSb}
\ee
Under a gauge transformation $A_\mu\to A_\mu+\partial_\mu\Lambda$,
this wavefunctional twisted by the $U(1)$ one-cocycle is indeed
invariant after the field redefinition
$\varphi\to\varphi+\Lambda$. The wavefuntional~(\r{WWS}) together with
the cocycle~(\r{cocycle}) is invariant up to a boundary term
$\oint_{\partial\Sigma}\varphi^{~}_{\rm
  b}\,\partial^\parallel\Lambda$~\cite{TM_17,TM_18}. By adding this
factor together with the correction~(\r{cocycleb}) to the cocycle
phase we find that the argument of the wavefuntional~(\r{chi}) transforms to
$(\varphi^{~}_{\rm b}-\Lambda)\,A^\parallel$, and the change can be
removed by the shift $\varphi\to\varphi+\Lambda$.

In these expressions the functional integration measure is assumed to
factorize over the bulk and boundary degrees of freedom of $\varphi$
on $\Sigma$, and is defined by
\be
[D\varphi]= \sqrt{{\cal A}_\Sigma}~\prod_{\vb{z}\in\Sigma}
d\varphi(\vb{z})~\delta\left(\,\int\limits_\Sigma d^2z~\varphi(\vb{z})
\,\phi^{~}_0\right)~\prod_{x\in\partial\Sigma}
d\varphi^{~}_{\rm b}(x)
\label{varphimeas}
\ee
with ${\cal A}_\Sigma=\int_\Sigma d^2z~\sqrt h$ the area of $\Sigma$.
The factors in the integration measure
(\ref{varphimeas}) remove the zero modes of the field $\varphi$ on the Riemann
surface $\Sigma$, which is required for a well-defined functional integral
because, by charge conservation on a compact space, the exponential in
(\ref{WWS}) is independent of them. Here $\phi^{~}_0=1/\sqrt{{\cal A}_\Sigma}$
is the normalized zero mode eigenfunction of the scalar Laplace operator
$\nabla^2$ on $\Sigma$, and the delta-function in (\ref{varphimeas}),
whose argument is the coefficient of $\phi^{~}_0$ in an arbitrary field
configuration $\varphi(\vb{z})$, projects out the zero mode
integration from the bulk measure
$\prod_{\vb{z}\in\Sigma}d\varphi(\vb{z})$. The worldsheet area factor
is included to make the overall combination dimensionless. A more
convenient way to use this measure is to change variables from
$\varphi$ to its worldsheet derivatives and compute the Jacobian to get
\be
[D\varphi]= \sqrt{\frac{{\cal A}_\Sigma}{{\det}'\,\nabla^2}}~
\prod_{\vb{z}\in\Sigma}d\Bigl(\partial_z
\varphi(\vb{z})\Bigr)~d\Bigl(\partial_\bz\varphi(\vb{z})\Bigr)~
\prod_{x\in\partial\Sigma}d\varphi^{~}_{\rm b}(x) \ ,
\label{measchange}
\ee
where the prime on the determinant means that zero-modes are excluded
in its evaluation.

\subsection{Boundary Conditions}

In~\cite{TM_18} it was demonstrated that the wavefunctions (\ref{WWS})
are the \textit{building blocks} of the boundary theories, in that by
inserting such states on the boundaries they \textit{act} as boundary
conditions and effectively select the boundary world. The fields
$\varphi$ introduce new propagating degrees of freedom on the
boundaries which are absolutely necessary for the consistency of the
full three-dimensional theory (on a manifold with boundary) as a
well-defined gauge theory. In other words, the requirement of gauge
invariance induces new scalar fields $\varphi$ on the two-dimensional
boundaries whose dynamics are governed by chiral, gauged $U(1)$ WZNW
conformal quantum field theories which have partition function
given by the path integral in (\ref{WWS}).

However, once one introduces external
sources into the quantum field theory it is not generally possible to
obtain a WZNW model on the boundaries. In terms of the boundary
wavefunctions this means that we no longer have purely holomorphic or
antiholomorphic functionals, and the nice holomorphic factorization
property of the quantum field theory is lost. In the case at hand,
this problem may be avoided by bearing in mind that we will eventually
orbifold the field theory and identify holomorphic and
anti-holomorphic degrees of freedom in boundary operators which are
described through insertions of external currents. To this end,
we assume that in the chiral sector of the worldsheet field theory
only the anti-holomorphic current component $j^\bz$ is non-vanishing,
while in the anti-chiral sector only holomorphic components $j^z$
remain, i.e. we take the boundary conditions
\be
j^z\Bigm|_{\Sigma_0}=0 \ , ~~ j^\bz\Bigm|_{\Sigma_1}=0 \ .
\lb{jbc}
\ee
These conditions are absolutely essential, because the presence of a
holomorphic source in the chiral boundary sector would couple extra
anti-holomorphic components of the gauge field and hence
drastically alter the nature of the degrees of freedom which live there.
We will also assume that the electric charge distribution decomposes
into holomorphic and anti-holomorphic components as the curl of a
vector field $\tilde Y_i$,
\be
\rho=\frac12\,\epsilon^{ij}\,\partial_i\tilde Y_j=i\,\partial_z
\tilde Y^z-i\,\partial_\bz\tilde Y^\bz \ .
\lb{rho0decomp}
\ee
As for the spatial components of the current, one imposes the conditions
\be
\tilde Y^\bz\Bigm|_{\Sigma_0}=0 \ , ~~ \tilde Y^z\Bigm|_{\Sigma_1}=0 \ .
\lb{Ybc}
\ee

Let us stress at this stage that in order to orbifold the theory later
on we shall have to extend the wavefunctions to the full, three-dimensional
bulk manifold $\Sigma\times[0,1]$, as in~\cite{TM_18}. In the bulk
theory the holomorphic and anti-holomorphic degrees of freedom will
mix at each time slice. Moreover, in the absence of spatial currents
the imposition of the Gauss law is enough to extract the complete ground state
wavefunctional. In the present situation one has to be more careful
because the presence of spatial currents $j^i$ does not alter the gauge
constraint in (\ref{gaussjj}). In order to properly implement our
program we must construct a family of distinct Hamiltonian operators
$H_t$ and infinitesimal gauge transformation generators ${\cal G}_t$ which vary
along the time slices $t\in[0,1]$ of the quantum field theory. The
corresponding vacuum states $\Psi_t$ at different times obey different
Schr\"odinger equations and Gauss laws given by
\be
H_t\Psi_t={\cal G}_t\Psi_t=0 \ .
\label{HtGt0}\ee
This will be carried out explicitly later on in this section.

In particular, for the boundary theories with wavefunctions $\Psi_0$
and $\Psi_1$ we have the operators
\bea
H_0&=&\int\limits_{\Sigma_0} d^2z~\left\{-\frac\gamma2\,
\left(\,\frac{\delta}{\delta A_\bz}+\frac k{8\pi}\,A_z\right)
\left(-\,\frac\delta{\delta A_z}+\frac k{8\pi}\,A_\bz\right)
-\gamma\,A_zj^\bz\right\} \ , \lb{H0}\\
{\cal G}_0&=&-i\left.\left[\partial_z
\left(\frac{\delta}{\delta A_z}+\frac{k}{8\pi}\,A_\bz\right)+
\partial_\bz\left(\frac{\delta}{\delta A_\bz}-\frac{k}{8\pi}\,
A_z\right)-\frac{k}{4\pi}\,F_{z\bz}-\partial_z\tilde Y^z\right]
\right|_{\Sigma_0} \ , \lb{G0}\\
H_1&=&\int\limits_{\Sigma_1} d^2z~\left\{\frac\gamma2\,
\left(\,\frac{\delta}{\delta A_z}+\frac k{8\pi}\,A_\bz\right)
\left(-\,\frac\delta{\delta A_\bz}+\frac k{8\pi}\,A_z\right)
+\gamma\,A_\bz\,j^z\right\} \ , \lb{H1}\\
{\cal G}_1&=&i\left.\left[\partial_z
\left(\frac{\delta}{\delta A_z}+\frac{k}{8\pi}\,A_\bz\right)+
\partial_\bz\left(\frac{\delta}{\delta A_\bz}-\frac{k}{8\pi}\,
A_z\right)+\frac{k}{4\pi}\,F_{z\bz}+\partial_\bz \tilde Y^\bz\right]
\right|_{\Sigma_1} \ ,
\lb{G1}
\eea
where, as in the previous subsection, we have fixed an appropriate
complex structure on the worldsheet $\Sigma$ and taken the topological
limit. We have also substituted in~(\ref{rho0decomp}) and imposed the
boundary conditions~(\ref{jbc}) and~(\ref{Ybc}). The overall changes
in sign between the time $t=0$ and $t=1$ operators are due to the
reversal of the orientations between the initial and final worldsheets
$\Sigma_0$ and $\Sigma_1$. The solutions of the four corresponding
equations (\ref{HtGt0}) will then work as functional boundary
conditions for the bulk wavefunctionals $\Psi_t$ and thereby constrain
the dynamics of the bulk quantum field theory.

\subsection{Charged Wavefunctionals\label{WCM}}

We will now carry out in detail the computation of the new wavefunctionals
for the boundaries $\Sigma_0$ and $\Sigma_1$ in the
presence of the external sources. By using (\ref{H0}) the
Schr\"odinger equation in (\ref{HtGt0}) at $t=0$ reads explicitly
\be
\int\limits_{\Sigma_0} d^2z~\left\{-\frac\gamma2\,
\left(\,\frac\delta{\delta A_\bz}+\frac k{8\pi}\,A_z\right)
\left(-\,\frac\delta{\delta A_z}+\frac k{8\pi}\,A_\bz\right)
-\gamma\,A_zj^\bz\right\}\Psi_0\left[A_z,A_\bz\,;j^\bz\,\right]=0 \
{}.
\lb{Schreq0}
\ee
Following the analysis of section~\ref{WS-FC}, we seek a functional
solution of the equation (\ref{Schreq0}) similar to~(\ref{WWS})
through the ansatz
\be
\Psi_0\left[A_z,A_\bz\,;j^\bz\,\right]=\int [D\varphi]~
\e^{iI_0[\varphi,A_z,A_\bz\,;j^\bz\,]} \ ,
\label{S0def}\ee
where
\bea
I_0\left[\varphi,A_z,A_\bz\,;j^\bz\,\right]&=&-i\int\limits_{\Sigma_0}d^2z~
\left[\frac{k}{8\pi}\,A_\bz\,A_z+\eta_1\,A_zj^\bz-\left(
\frac{k}{4\pi}\,A_\bz+\eta_2\,j^\bz-\tilde Y^z\right)\,\partial_z\varphi
\right.\nn\\&&+\left.\frac{k}{8\pi}\,\partial_\bz\varphi\,
\partial_z\varphi\right] \ .
\label{WWS0a}\eea
The unknown parameters $\eta_1$ and $\eta_2$ of this ansatz will be found by
imposing (\ref{Schreq0}) along with gauge invariance.

The terms in (\ref{WWS0a}) involving the vector field $\tilde Y^i$ arise as
follows. In the bulk field theory, the minimal coupling term of the
action (\ref{STMGT}) transforms, after an integration by parts over
time, under bulk gauge transformations as
\be
\int\limits_0^1dt~\int\limits_\Sigma d^2z~(A_\mu+\partial_\mu
\Lambda)J^\mu=\int\limits_0^1dt~\int\limits_{\Sigma_0}d^2z~A_\mu
J^\mu+\left\{\,\int\limits_{\Sigma_1}-\int\limits_{\Sigma_0}
\,\right\}d^2z~\Lambda\,\rho \ .
\label{mintransf}\ee
If we add the field $-i\varphi\,\rho$ to the Lagrangian of (\ref{WWS}),
then the inhomogeneous term in (\ref{mintransf}) can be absorbed by
shifting the scalar field $\varphi\to\varphi+\Lambda$, and the full
partition function of the topologically massive gauge theory is thereby
gauge invariant~\cite{TM_18}. Indeed, this
modification of the wavefunctional (\ref{WWS}) solves the Gauss law
(\ref{gaussjj}) with sources. Upon substituting in the
parameterization (\ref{rho0decomp}) for the charge density and
integrating by parts over the worldsheet $\Sigma_0$, we arrive at the
$\tilde Y$-field dependent terms in (\ref{WWS0a}).

Let us now compute the action of the Hamiltonian operator (\ref{H0})
on the wavefunction (\ref{S0def}). An easy calculation gives
\bea
H_0\Psi_0\left[A_z,A_\bz\,;j^\bz\,\right]&=&\int
[D\varphi]~\int\limits_{\Sigma_0}d^2z~\left[\frac{\gamma}{2}
\,\left(\frac{\delta}{\delta A_\bz}+\frac{k}{8\pi}\,
A_z\right)\,\eta_1\,j^\bz-\gamma\,A_zj^\bz\right]~
\e^{iI_0[\varphi,A_z,A_\bz\,;j^\bz\,]}\nn\\
         &=&\int [D\varphi]~
\e^{iI_0[\varphi,A_z,A_\bz\,;j^\bz\,]}\nn\\&&\times\,\int
\limits_{\Sigma_0}d^2z~\left[\gamma\left(\frac{k\,\eta_1}{8\pi}-1
\right)A_zj^\bz -\frac{k\,\gamma\,\eta_1}{8\pi}\,j^\bz\,\partial_z
\varphi\right] \ .
\lb{contasH0}\eea
The second term of the second equality in (\ref{contasH0}) vanishes
because we can write the functional integration measure as in
(\ref{measchange}) and shift the
new integration variable
\be
\partial_\bz\varphi~\longrightarrow~\partial_\bz\varphi+2A_\bz+\frac{8\pi}k\,
\Bigl(\eta_2\,j^\bz-\tilde Y^z\Bigr)
\label{partialphishift}\ee
to write (\ref{WWS0a}) as a quadratic form in
$\partial_\bz\varphi\,\partial_z\varphi$, while
leaving both the integration measure and pre-exponential factors in
(\ref{contasH0}) unchanged. The resulting functional Gaussian
integration then clearly vanishes. If we now set
\be
\eta_1=\frac{8\pi}{k} \ ,
\lb{alpha1}
\ee
then the first term also vanishes and we indeed
obtain a ground state solution to (\ref{Schreq0}).

To fix the remaining constant $\eta_2$ in (\ref{WWS0a}), we will
compute the Gauss law constraint using (\ref{G0}). Another simple
calculation gives
\bea
{\cal G}_0(\vb{z})\Psi_0\left[A_z,A_\bz\,;j^\bz\,\right]&=&
-i\int [D\varphi]~\left[\partial_z\left(\frac{k}{4\pi}\,
A_\bz(\vb{z})+\eta_1\,j^\bz(\vb{z})\right)-\frac{k}{4\pi}\,
\partial_\bz\partial_z\varphi(\vb{z})\right.\nn\\&&-\left.
\frac{k}{4\pi}\,F_{z\bz}(\vb{z})-\partial_z\tilde Y^z(\vb{z})\right]~
\e^{iI_0[\varphi,A_z,A_\bz\,;j^\bz\,]}\nn\\&=&-i\int [D\varphi]~
\left[\frac{k}{8\pi}\,\partial_\bz\partial_z\varphi(\vb{z})-
\frac{k}{4\pi}\,F_{z\bz}(\vb{z})+i\,\frac{\delta I_0}{\delta
\varphi(\vb{z})}\right.\nn\\&&+\Biggl.(\eta_1-\eta_2)\,j^\bz(\vb{z})
\Biggr]~\e^{iI_0[\varphi,A_z,A_\bz\,;j^\bz\,]} \ .
\lb{contasG0}\eea
The third term in the second equality vanishes via an integration by
parts in $\varphi$ space. The last term vanishes if we take
\be
\eta_1=\eta_2 \ .
\label{eta12eq}\ee
The remaining terms may be eliminated for smooth field configurations
by writing the derivative terms in $\varphi$ as
$\partial_\bz\partial_z\varphi+\partial_z\partial_\bz\varphi$ and then
shifting $\partial_z\varphi$ and $\partial_\bz\varphi$ by $-A_z$ and
$A_\bz$, respectively. This eliminates the local magnetic flux term in
(\ref{contasG0}). Then, as was done in (\ref{contasH0}), the remaining
functional integrals vanish from shifting derivatives of $\varphi$
analogously to (\ref{partialphishift}) to
obtain quadratic forms in (\ref{WWS0a}). This is consistent with the
integrated form of (\ref{contasG0}), which reads
\bea
\int\limits_{\Sigma_0}d^2z~{\cal G}_0\Psi_0
\left[A_z,A_\bz\,;j^\bz\,\right]&=&
-i\int [D\varphi]~\e^{iI_0[\varphi,A_z,A_\bz\,;j^\bz\,]}\nn\\
&&\times\,\int\limits_{\Sigma_0}d^2z~\left[
\partial_z\left(\frac k{4\pi}\,A_\bz+\eta_1\,j^\bz-\tilde Y^z-
\frac k{4\pi}\,\partial_\bz\varphi\right)-\frac k{4\pi}\,F_{z\bz}
\right] \ . \nn\\&&
\label{G0int}\eea
As discussed in section~\ref{WS-FC}, the total magnetic flux on
$\Sigma_0$ must vanish, and hence so does the last term in
(\ref{G0int}). The remaining terms may then be absorbed into an
appropriate shift of $\partial_\bz\varphi$ analogous to
(\ref{partialphishift}), which leaves a
quadratic form in (\ref{WWS0a}) proportional to
$\partial_\bz\varphi\,\partial_z\varphi$ {\it only} if (\ref{eta12eq})
holds. As in (\ref{contasH0}), the corresponding functional Gaussian
integral in the measure (\ref{measchange}) then vanishes. Therefore,
for the parametric values (\ref{alpha1}) and (\ref{eta12eq}), the wavefunction
(\ref{S0def}) is a physical ground state of the topologically massive
gauge theory.

The equality (\ref{eta12eq}) is not at all surprising, since
it is required to absorb the gauge transform $A_i\to
A_i+\partial_i\Lambda$ in (\ref{WWS0a}) through the shift
$\varphi\to\varphi+\Lambda$~\cite{TM_18}. Let us also remark that the
assumption that there are no singular $\varphi$-field configurations
in the bulk is natural because singularities or discontinuities would
account for external sources which are responsible for creating
additional flux, and would therefore be present in the Gauss law {\it a
  priori}. Such fluxes will be present only at the orbifold point on
the boundary of the string worldsheet. In other words, they will
correspond to insertions of vertex operators in the induced boundary
conformal field theory. This aspect will be
developed in detail in the next section.

To summarize, the full physical wavefunction (\ref{chiphysdec})
corresponding to the initial boundary surface $\Sigma_0$ is given by
\bea
\Psi_0^{\rm phys}[A;j]&=&\exp\left\{\,\int\limits_{\Sigma_0}d^2z~
\left(\frac{k}{8\pi}\,A_\bz+\frac{8\pi}{k}\,j^\bz\right)A_z\right\}\nn\\&&
\times\,\int [D\varphi]~\exp\left\{\frac{k}{8\pi}\,
\int\limits_{\Sigma_0}d^2z~\left(\partial_\bz\varphi-2A_\bz-
\frac{64\pi^2}{k^2}\,j^\bz+\frac{8\pi}{k}\,\tilde Y^z\right)\,
\partial_z\varphi\right\} \ . \nn\\&&
\lb{WWS0c}\eea
A completely analogous calculation for the final boundary $\Sigma_1$
using~(\ref{H1}) and~(\ref{G1}) gives
\bea
\Psi_1^{\rm phys}[A;j]^\dag&=&\exp\left\{\,\int\limits_{\Sigma_1}d^2z~
\left(-\frac{k}{8\pi}\,A_z-\frac{8\pi}{k}\,j^z\right)A_\bz\right\}\nn\\&&
\times\,\int [D\varphi]~\exp\left\{\frac{k}{8\pi}\,\int
\limits_{\Sigma_1}d^2z~\left(-\partial_z\varphi+2A_z+
\frac{64\pi^2}{k^2}\,j^z-\frac{8\pi}{k}\,\tilde Y^\bz\right)\,
\partial_\bz\varphi\right\} \ . \nn\\&&
\lb{WWS1c}\eea
This is actually not the end of the story, because $j^i$ and $\tilde Y^i$ are
in fact related through the continuity equation (\ref{J0}) if one
demands $\sf PT$ or ${\sf PCT}$ invariance of the three-dimensional quantum
field theory. These symmetries are related to the two possible types
of orbifolds of the topological membrane, as has been extensively studied
in~\cite{TM_16,TM_17,TM_18} and which will be our focal point for the
remainder of this section.

In the presence of a non-empty boundary $\partial\Sigma$
the wavefunctions are given by~(\r{chiphysdec}) with
\be
\Psi^{\rm phys}\left[A;j\right]=\int [D\varphi]~
\exp\left\{i\oint\limits_{\partial\Sigma}
\varphi^{~}_{\rm b}\,\left(\,\tilde Y^\parallel-A^\parallel\right)\right\}
{}~\e^{iI[\varphi,A_z,A_\bz\,;j^z,j^\bz\,]} \ .
\label{WWSbdry}\ee
In the topological membrane picture, the surfaces $\Sigma_0$ and $\Sigma_1$
are already the two-dimensional boundaries of the three-dimensional
manifold $M=\Sigma\times[0,1]$, and thus they cannot have a
boundary. Only after an appropriate orbifold operation such as those
in~\cite{TM_16,TM_17,TM_18} can we obtain a surface with boundary at
the orbifold fixed point $t=1/2$ in time. Therefore, the extra
boundary factor in (\ref{WWSbdry}) will only be present at that
specific point.

\subsection{Orbifold Relations\label{OR}}

We will now describe the orbifolds of the topological membrane
obtained by gauging the discrete symmetries of topologically massive
gauge theory. In this paper we will only consider worldsheet orbifolds
that give rise to {\it oriented} surfaces. We are interested in
constructing open (oriented) Riemann surfaces
$\Sigma^{\rm o}$ as the ${\mathbb{Z}}_2$-quotients $\Sigma^{\rm
  o}=\Sigma/{\mathbb{Z}}_2$ of a closed worldsheet $\Sigma$ (the
Schottky double of $\Sigma^{\rm o}$) by an
anti-conformal involution $\sigma:\Sigma\to\Sigma$ whose fixed points
correspond to the boundary points of $\Sigma^{\rm o}$. Given
$\Sigma^{\rm o}$, the Schottky double is constructed as the quotient
space $\Sigma=(\Sigma^{\rm o}\times\{0,1\})/\sim$ with respect to the
equivalence relation $\sim$ defined by $(x,0)\sim(x,1)$ if and only if
$x\in\partial\Sigma^{\rm o}$.

We shall combine the involution $\sigma$ with time-reversal in the
bulk, so that the open worldsheet $\Sigma^{\rm o}$ may be regarded
from the topological membrane perspective as the connecting three-manifold
$M/\zed_2=(\Sigma\times[0,1])/{\mathbb{Z}}_2$~(fig.~\ref{fig.orbi}). Due to the
presence of both Maxwell and Chern-Simons terms in the action
(\ref{STMGT}), the only combinations of discrete spacetime symmetries
in this context which are compatible with the three-dimenisonal action
are the $\sf PT$ and $\sf PCT$ automorphisms of the gauge
theory~\cite{TM_16}. The roles of the time reversal operation ${\sf
  T}:t\mapsto1-t$ on the geometry $\Sigma\times[0,1]$ are to create a
single new boundary surface $\Sigma_{1/2}$ at the $\sf T$ orbifold
fixed point $t=1/2$, and also to identify the initial and final
surfaces $\Sigma_0\equiv\Sigma_1$. The open worldsheet $\Sigma^{\rm
  o}$ is then obtained by quotienting $\Sigma_{1/2}$ by the discrete
$\zed_2$ symmetries generated by $\sf P$ and $\sf PC$ in two
dimensions, and its boundary is situated at the branch point locus
$x^\perp=0$.

\fig{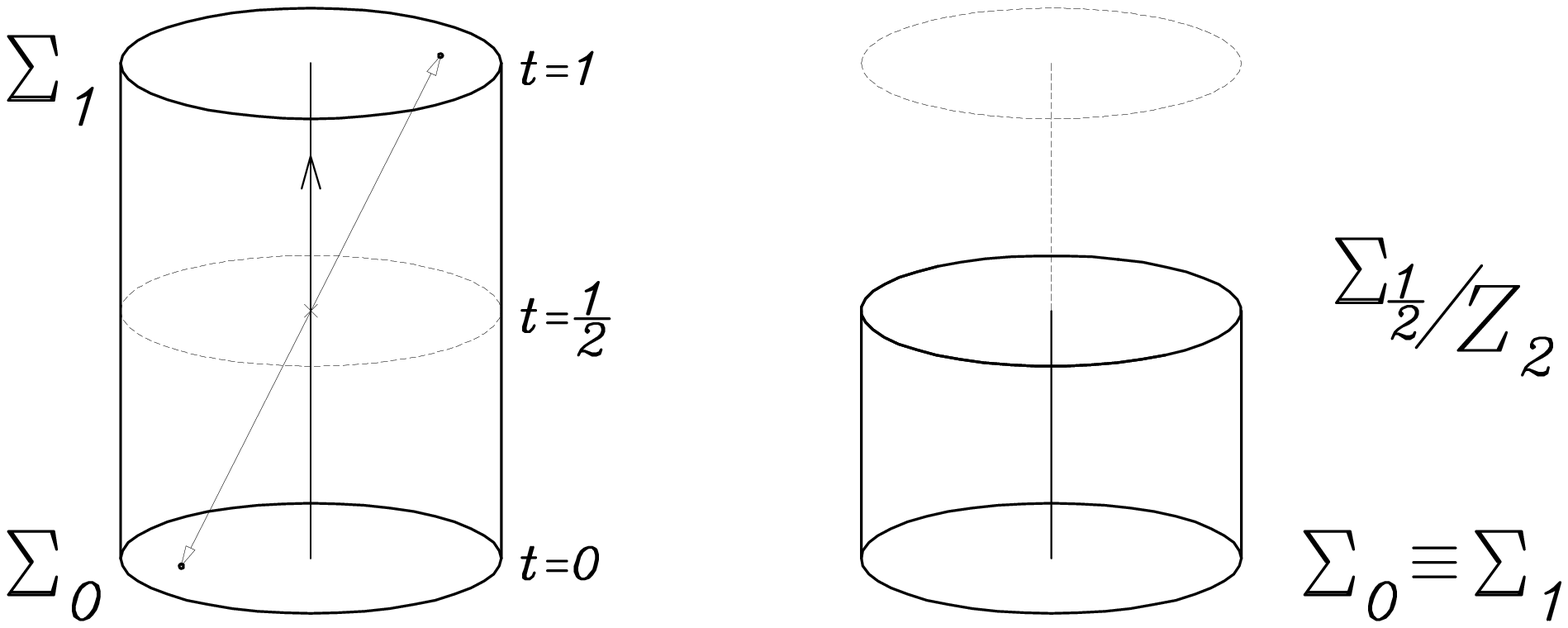}{Orbifold of the topological membrane. The worldsheet
$\Sigma_{1/2}$ only feels the discrete symmetries $\sf PC$ and
$\sf P$ which each generate the cyclic group
${\mathbb{Z}}_2$, whose action yields the open string worldsheet
$\Sigma^{\rm o}=\Sigma_{1/2}/\zed_2$.}{fig.orbi}

The topological degrees of freedom of the membrane are reduced by the
orbifold operations in the following way. The real homology group
$H_1(\Sigma,\real)\cong\real^{2g}$ of the closed genus $g$ Riemann
surface $\Sigma$ is generated as a vector space by canonical
homology cycles $\alpha_\ell,\beta^\ell$, $\ell=1,\dots,g$ which have
intersection pairings
$\alpha_\ell\cap\alpha_{\ell'}=\beta^\ell\cap\beta^{\ell'}=0$ and
$\alpha_\ell\cap\beta^{\ell'}=-\beta^{\ell'}\cap\alpha_\ell=
\delta^{\ell'}_\ell$. The intersection form makes $H_1(\Sigma,\real)$
into a real symplectic vector space. The orientation-reversing
homeomorphism $\sigma:\Sigma\to\Sigma$ induces an involutive
isomorphism $\sigma_*:H_1(\Sigma,\real)\to H_1(\Sigma,\real)$ which
gives the homology group a $\zed_2$-grading as
\be
H_1(\Sigma,\real)={\cal L}_+(\Sigma)\oplus{\cal L}_-(\Sigma) \ ,
\label{Lagrangiansplit}\ee
where ${\cal L}_\pm(\Sigma)$ are Lagrangian subspaces of
$H_1(\Sigma,\real)$, i.e. subspaces of maximal dimension on which the
intersection form vanishes, defined as the $\pm\,1$ eigenspaces of
$\sigma_*$, $\sigma_*{\cal L}_\pm(\Sigma)=\pm\,{\cal
  L}_\pm(\Sigma)$. We will identify $H_1(\Sigma^{\rm o},\real)={\cal
  L}_-(\Sigma)$. This is the canonical choice for the real homology
group of the open worldsheet $\Sigma^{\rm o}$ in the following
sense~\cite{FS_2}. Let $M_{\Sigma^{\rm o}}$ be the ``solid donut'' obtained by
filling in the Shottky double of $\Sigma^{\rm o}$,
i.e. $M_{\Sigma^{\rm o}}=(\Sigma^{\rm o}\times[0,1])/\sim$, where
$(x,t)\sim(x,1-t)~~\forall t\in[0,1]$ if and only if
$x\in\partial\Sigma^{\rm o}$. Then $\partial M_{\Sigma^{\rm
    o}}=\Sigma$, and the canonical inclusion $\imath:\partial M_{\Sigma^{\rm
    o}}\hookrightarrow M_{\Sigma^{\rm o}}$ induces a homomorphism
$\imath_*:H_1(\Sigma,\real)\to H_1(M_{\Sigma^{\rm o}},\real)$ with
kernel
\be
\ker(\imath_*)={\cal L}_-(\Sigma) \ .
\label{kericalL}\ee
In other words, the Lagrangian subspace ${\cal L}_-(\Sigma)$ consists
of those homology cycles of $\Sigma$ which are contractible in
$M_{\Sigma^{\rm o}}$. From the point of view of the membrane geometry
$M=\Sigma\times[0,1]$, the canonical inclusion $\jmath$ of
$\partial(M/\zed_2)\cong\Sigma\amalg\Sigma^{\rm o}$ into $M/\zed_2$
induces a homomorphism into the $\zed_2$-equivariant homology of the
topological membrane as
\be
H_1(\Sigma,\real)\oplus{\cal L}_-(\Sigma)~\stackrel{\jmath_*}
{\longrightarrow}~H_1^{\zed_2}\Bigl(\Sigma\times[0,1]\,,\,\real
\Bigr) \ .
\label{membranehomo}\ee

Here and in the following we will choose local complex coordinates at
the orbifold line $x^\perp=0$ which are defined such that
$z=x^\parallel+i\,x^\perp$ and $\bz=x^\parallel-i\,x^\perp$ (and
similarly for the vector fields). This choice is a matter of convention
in the definition of the parity operator $\sf P$. For example, in the
case of the sphere $\Sigma=\sphere^2$ represented stereographically as
the complex plane, with these conventions the boundary of the disk
$\Sigma^{\rm o}=\disc^2$ obtained by
orbifolding the sphere is the real axis~\cite{TM_16}. In the remainder
of this subsection we will summarize the orbifold transformation rules
derived in~\cite{TM_16,TM_18} that we will require in the following.

For the $\sf PT$ symmetry the field transformation rules are given by
\be
\ba{crcl}
{\sf PT}\,:\,&\Lambda&\longmapsto&-\Lambda\vspace{.1 cm}\\
         &\varphi&\longmapsto&-\varphi\vspace{.1 cm}\\
         &A_0    &\longmapsto&A_0\vspace{.1 cm}\\
         &A_\perp&\longmapsto&A_\perp\vspace{.1 cm}\\
         &A_\parallel&\longmapsto&-A_\parallel\vspace{.1 cm}\\
         &A_z&\longmapsto&A_\bz\vspace{.1 cm}\\
         &\partial_iE^i &\longmapsto&\partial_iE^i\vspace{.1 cm}\\
          &B      &\longmapsto&B\vspace{.1 cm}\\
                  &Q_{m,n}   &\longmapsto&-Q_{m,n} \ .
\ea
\lb{T2}
\ee
The orbifold obtained from the
quotient under this symmetry corresponds to
\textit{Dirichlet} boundary conditions on the string fields which restricts the
charge spectrum to only winding modes
\be
Q_{0,n}=\frac{kn}4 \ .
\label{Qm0}\ee
Under the $\sf PCT$ symmetry the fields transform as
\be
\ba{crcl}
{\sf PCT}\,:\,&\Lambda&\longmapsto&\Lambda\vspace{.1 cm}\\
          &\varphi&\longmapsto&\varphi\vspace{.1 cm}\\
          &A_0    &\longmapsto&-A_0\vspace{.1 cm}\\
          &A_\perp &\longmapsto&-A_\perp\vspace{.1 cm}\\
          &A_\parallel&\longmapsto&A_\parallel\vspace{.1 cm}\\
          &A_z&\longmapsto&-A_\bz\vspace{.1 cm}\\
          &\partial_iE^i &\longmapsto&-\partial_iE^i\vspace{.1 cm}\\
          &B      &\longmapsto&-B\vspace{.1 cm}\\
          &Q_{m,n}    &\longmapsto&Q_{m,n} \ .
\ea
\lb{PCT}
\ee
The orbifold obtained from the quotient under this symmetry corresponds to
\textit{Neumann} boundary conditions on the string fields which restricts the
charge spectrum to only Kaluza-Klein modes
\be
Q_{m,0}=m \ .
\label{Q0n}\ee
Note that T-duality in this setting is just the interchange of the
$\sf PT$ and $\sf PCT$ orbifolds in three-dimensions.

{}From what we have just described it follows that the orbifold construction
is compatible with the constraints on the external current as long as
$\rho|_{\Sigma_0}=-\rho|_{\Sigma_1}$, which is consistent
with~(\ref{rho0decomp}) and~(\ref{Ybc}). The transformations for the
current components are given by
\be
\ba{crcl}
{\sf PT}\,:\,&j^z&\longmapsto&j^\bz\\[1mm]
&\tilde Y^z&\longmapsto&\tilde Y^\bz
\ea
\ee
and
\be
\ba{crcl}
{\sf PCT}\,:\,&j^z&\longmapsto&-j^\bz\\[1mm]
&\tilde Y^z&\longmapsto&-\tilde Y^\bz \ .
\ea
\ee
Of course, these transformations are expected since the current
components should transform as vectors under the discrete symmetry
operations. We can now use these orbifold relations to systematically
construct the Schr\"odinger wavefunctionals corresponding to the open
surface $\Sigma^{\rm o}$ from those derived in section~\ref{WCM}.

Note that at the boundary $\partial\Sigma$ the scalar field
$\varphi^{~}_{\rm b}$ is not \textit{a priori} constrained by the orbifold
relations, and it has free boundary conditions. It is only after
integrating out the gauge field degrees of freedom that we obtain
boundary conditions for it~\cite{TM_18}. We will derive this fact
in the presence of external currents, along with the appropriate
extensions of the wavefunctionals to $\partial\Sigma$, in
section~\r{OrbPart}.

\subsection{The Continuity Equation}

Following~\cite{TM_17,TM_18} we will now derive the extension of
the wavefunctionals computed above to the entire bulk three-manifold
$\Sigma\times[0,1]$. For this, we consider two time-dependent
functions $f_0(t)$ and $f_1(t)$ which obey the boundary conditions
\be
\ba{rrl}
f_0(0)~=~-f_1(1)&=&-1 \ , \\[1mm]f_0(1)~=~f_1(0)&=&0 \ .
\ea
\lb{fs}
\ee
In addition, to ensure compatibility between gauge invariance and the
orbifold constructions described above, we must also
require that these functions be related by~\cite{TM_16,TM_18}
\bea
f_0(1-t)&=&-f_1(t) \ , \nn\\f_0(1/2)&=&-f_1(1/2)~=~\frac12 \ .
\label{f0f1rels}\eea
Then any integral over the boundaries $\Sigma_0$ and $\Sigma_1$ of the
three-manifold can be extended to the bulk as
\be
\int\limits_{\Sigma_1}d^2z~X_1+\int\limits_{\Sigma_0}d^2z~X_0
=\int\limits_0^1dt~\int\limits_\Sigma d^2z~
\partial_t\left(f_1\,X_1-f_0\,X_0\right) \ ,
\ee
where $X_0$ and $X_1$ stand for any combinations of the fields
at the worldsheet boundaries $\Sigma_0$ and $\Sigma_1$, respectively.

Within this simple framework one can now evaluate the worldsheet integral
at any fixed time slice $\Sigma_\tau\equiv\Sigma\times\{\tau\}$,
$\tau\in[0,1]$ by splitting the covering cylinder $\Sigma\times[0,1]$
over $\Sigma$ into two pieces, corresponding to the time intervals
$t\in[0,\tau]$ and $t\in(\tau,1]$, such that one has
\be
\int\limits_{\Sigma_\tau}d^2z~X_\tau=\int\limits_\Sigma
d^2z~\Bigl(f_1(\tau)\,X_1-f_0(\tau)\,X_0\Bigr) \ ,
\ee
where $X_\tau$, $\tau\in[0,1]$ is a one-parameter family of fields
defined on the time slices $\Sigma_\tau$. Of course, for this we need to
know the precise dependences of $f_0(t)$ and $f_1(t)$ as functions of
the time coordinate $t$. In~\cite{TM_18} this problem was not addressed, since
for the considerations there it sufficed to know their values solely at the
orbifold point $t=1/2$. In this subsection we will use the continuity
equation (\ref{J0}) for the external currents to derive differential equations
for these functions.

For the stationary states of the topologically massive gauge theory,
we assume that the components $j^z$, $j^\bz$
and $\rho$ are all time-independent fields. Then, according to the framework
just described, the original currents which appear in the
action (\ref{STMGT}) are given by the expressions
\bea
J^0&=&\partial_t\left(F_0\,\rho_0+F_1\,\rho_1\right) \ , \nn\\J^z&=&2\gamma\,
\partial_t\left(F_1\,j^z\right) \ , \nn\\
J^\bz&=&2\gamma\,\partial_t\left(F_0\,j^\bz\,\right) \ ,
\eea
where the functions $F_\tau$, $\tau=0,1$ are defined by
\be
F_\tau(t)=\int\limits_{1-\tau}^tdt'~f_\tau(t'\,) \ ,
\ee
and we have allowed two different but fixed charge distributions
$\rho_\tau$ to live on the two boundaries at $\tau=0,1$. Then the
sources indeed do reduce to the desired ones on the boundary surfaces,
which we summarize as
\be
\ba{lllllll}
&\underline{\Sigma_0}&&~~~~&&\underline{\Sigma_1}&\\[4mm]
J^0&=&-\rho_0&~~~~&J^0&=&\rho_1\\[1mm]
J^z&=&0&~~~~&J^z&=&2\gamma\,j^z\\[1mm]
J^\bz&=&-2\gamma\,j^\bz&~~~~&J^\bz&=&0 \ .
\ea
\label{sourceids}\ee

The continuity equation (\ref{J0}) thereby becomes
\be
\frac{1}{\gamma}\,\partial_t\left(f_0\,\rho_0+f_1\,\rho_1\right)
=f_0\,\partial_zj^\bz-f_1\,\partial_\bz j^z \ .
\ee
We can split this equation into two independent differential equations
\bea
\partial_tf_0&=&-(f_0+c_0)\lambda_0 \ , ~~
\lambda_0~=~-\frac{\gamma\,\partial_zj^\bz}{\rho_0} \ , \nn\\
\partial_tf_1&=&-(f_1+c_1)\lambda_1 \ , ~~
\lambda_1~=~\frac{\gamma\,\partial_\bz j^z}{\rho_1} \ ,
\eea
along with a single constraint which couples them through
\be
c_0\lambda_0+c_1\lambda_1=0 \ .
\label{couplingconstr}\ee
The source parameters $c_\tau$ and $\lambda_\tau$ ($\tau=0,1$) are constant
in time. In this way we obtain the explicit solutions
\be
f_\tau(t)=-c_\tau+c_\tau'~\e^{-\lambda_\tau\,t} \ .
\label{ftaut}\ee
The parameters $\lambda_\tau$, $c_\tau$ and $c_\tau'$ ($\tau=0,1$) of
this solution can be unambiguously fixed once we place appropriate
orbifold conditions on the field theory.

For this, we note that compatibility with the orbifold identifications
requires that the two sets of charge densities be related through
$\rho_0=-\rho_1$. We further need to consider
two sets of functions $f_\tau$ and $\tilde f_\tau$ defined
respectively on the two branches $t\in[0,1/2]$ and $t\in(1/2,1]$. By
setting $t=0$, $1$ and $1/2$ in (\ref{ftaut}) and using (\ref{fs})
and (\ref{f0f1rels}), we arrive at the boundary conditions
\be
\ba{rcl}
\displaystyle-c_\tau+c_\tau'&=&\displaystyle1-\tau \ , \\[3mm]
\displaystyle-c_\tau+c_\tau'~\e^{-\lambda_\tau/2}&=&\displaystyle\frac12-
\tau \ , \\[3mm]\displaystyle-\tilde{c}_\tau+\tilde{c}_\tau'~
\e^{-\tilde{\lambda}_\tau/2}&=&\displaystyle\frac12-\tau \ , \\[3mm]
\displaystyle-\tilde{c}_\tau+\tilde{c}_\tau'~\e^{-\tilde{\lambda}_
\tau}&=&\displaystyle-\tau \ .
\ea
\label{bcs}\ee
These equations admit a unique solution in terms of the
$\lambda$'s. The constraint (\ref{couplingconstr}) which couples the
two differential equations can be solved by noting the relations
(\ref{sourceids}) between the sources on the various boundary
surfaces, which imply
\be
\left.\frac{\partial_zj^\bz}{\rho_0}\,\right|_{\Sigma_t}=-\left.
\frac{\partial_\bz j^z}{\rho_1}\,\right|_{\Sigma_{1-t}} \ .
\label{sourcetrels}\ee
This is in fact a global constraint which holds on the entire covering cylinder
$\Sigma\times[0,1]$, since the variation along the three-manifold is
accounted for by the temporal functions which solve the continuity
equation above. It follows that $\lambda_0=\lambda_1$ and the required
constraint (\ref{couplingconstr}) is satisfied by the solution of
(\ref{bcs}). The quantities (\ref{sourcetrels}) are dimensionless, and
as such determine an integration constant. We will fix this
last arbitrary parameter to the Chern-Simons coupling $k/8\pi$, as it
is the natural expansion coefficient of the model. This choice will
conveniently simplify things in the following.

By repeatedly applying the functional relation (\ref{f0f1rels}) we thereby
finally arrive at the full solution
\be
\ba{rclcl}
f_0(t)&=&\displaystyle-c_0+(c_0+1)~\e^{-k\,t/8\pi}&
\mathrm{~,~~}&t\in[0,1/2] \ , \\[3mm]\tilde{f}_0(t)&=&\displaystyle-
\frac{1}{2}+f_0(t-1/2)&\mathrm{~,~~}&t\in(1/2,1] \ , \\[3mm]
f_1(t)&=&\displaystyle\frac12-f_0(3/2-t)&\mathrm{~,~~}&t\in[0,1/2] \ , \\[2mm]
\tilde{f}_1(t)&=&\displaystyle-f_0(2-t)&\mathrm{~,~~}&t\in(1/2,1] \ ,
\ea
\ee
where
\be
c_0=-\,\frac{1/2}{1-\e^{-k/16\pi}} \ .
\ee
In addition, from the constraints (\ref{sourcetrels}) we can deduce an
important relation. Namely, under the orbifold involution, the
currents $j^i$ and the charge distribution vector fields $\tilde Y^i$
are related in a very simple way through
\be
j^z=\frac{k}{8\pi}\,\tilde Y^\bz \ , ~~ j^\bz=\frac{k}{8\pi}\,\tilde Y^z \ .
\label{jYrel}\ee
This relationship will drastically simplify the form of the
Schr\"odinger wavefunctionals for the orbifolded gauge theory, whose
computation we come to next.

\subsection{The Orbifold Partition Function\label{OrbPart}}

We can now finally demonstrate the explicit appearence of the D-brane
vertex operator in the induced two-dimensional conformal field
theory. We will assume that the initial (and final) Riemann
surface $\Sigma$ is closed, $\partial\Sigma=\emptyset$, and denote its
genus by $g$. The partition function for the topological membrane,
i.e. for topologically massive gauge theory on the three-geometry
$M=\Sigma\times[0,1]$, is given by the overlap between the initial and
final states (\ref{WWS0c}) and (\ref{WWS1c}) as~\cite{TM_18}
\be
Z^{\rm c}=\left.\left\langle\Psi_1\right|\Psi_0\right\rangle\equiv
\int\left[DA_z~DA_\bz\,\right]~\e^{iS_{\rm TMGT}[A]}~
\Psi_1^\dag\left[A_z,A_\bz\,;j^z\right]\,\Psi_0\left[A_z,A_\bz
\,;j^\bz\,\right] \ ,
\label{Zcdef}\ee
where the functional integral is taken with an appropriately defined
gauge-fixed integration measure. To explicitly evaluate (\ref{Zcdef}),
one decomposes the gauge field $A_i$ according to a representative of
its gauge orbit as
\be
A_i=\bar A_i+\partial_i\Lambda \ ,
\label{Aigaugeorbit}\ee
where $\bar A_i$ is the gauge-fixed field and $\Lambda$ is an
arbitrary real-valued gauge parameter function. The corresponding
change of measure is determined by the relation
(\ref{measchange}). For $\Sigma$ closed and $J^\mu=0$, by construction
the integrand of (\ref{Zcdef}) is independent of the gauge parameter
$\Lambda$, and one finds that the path integral factorizes as
\be
Z^{\rm c}=Z_{\rm bulk}~Z_\Sigma \ ,
\label{Zcfact}\ee
where $Z_{\rm bulk}$ comes from integrating out the bulk parts of the
gauge fields $\bar A_i$ and involves only the topologically massive
gauge theory action $S_{\rm TMGT}[\bar A\,]$, while $Z_\Sigma$ contains
only boundary degrees of freedoms of the fields and coincides with the
partition function of the usual full (non-chiral) gauged $c=1$ WZNW
model defined on the closed string worldsheet $\Sigma$. It is in
this way that the closed string sector is reproduced from the bulk
dynamics of the topological membrane. In what follows we will
normalize the full partition function by setting $Z_{\rm bulk}=1$.

We will now transform the vacuum amplitude (\ref{Zcdef}) into the appropriate
partition function for the orbifold $\sigma$-model on $\Sigma^{\rm
  o}$. For this, we extend the wavefunctionals (\ref{WWS0c}) and
(\ref{WWS1c}) to the full covering cylinder over $\Sigma$ by using the
prescriptions of the previous subsection to express it in the form
\bea
Z^{\rm c}&=&\int\left[DA_z~DA_\bz\,\right]~\int[D\varphi]~\e^{iS_{\rm TMGT}[A]}
\nn\\&&\times\,\exp\left\{-\frac k{8\pi}\,\int\limits_0^1dt~
\int\limits_\Sigma d^2z~\partial_t\left(\left[A_\bz\,A_z+\partial_\bz\varphi\,
\partial_z\varphi+\frac{8\pi}k\,\epsilon^{ij}
\,\partial_j\left(\varphi\,\tilde Y_i\right)\right]\left(f_0+f_1\right)
\right.\right.
\nn\\&&-\,\left[\left(A_\bz-\frac{8\pi}k\,\tilde Y^z+
\frac{64\pi^2}{k^2}\,j^\bz\,\right)\,\partial_z\varphi-
\frac{64\pi^2}{k^2}\,A_z\,j^\bz\,\right]f_0\nn\\&&-\left.\left.
\left[\left(A_z-\frac{8\pi}k\,\tilde Y^\bz+\frac{64\pi^2}
{k^2}\,j^z\right)\,\partial_\bz\varphi-\frac{64\pi^2}{k^2}\,
A_\bz\,j^z\right]f_1\right)\right\} \ .
\label{Zcext}\eea
By using the orbifold identifications of the fields described in
section~\ref{OR}, the explicit time dependences of the continuity
equation solutions of the previous subsection, and the relationship
(\ref{jYrel}), we find that the currents $j^i$ completely drop out of
the wavefunctionals giving the orbifold amplitude
\bea
Z^{\rm o}&=&\left.\left\langle\Psi_{1/2}\right|\Psi_0\right\rangle_{\rm
  orb}\nn\\&\equiv&\int[DA_z~DA_\bz\,]~
\e^{2iS_{\rm TMGT}^{\rm orb}[A]}~
\Psi_{1/2}^{\rm orb}\left[A_z,A_\bz\,;\tilde Y^z,\tilde Y^\bz,\tilde
Y^\parallel\,\right]^\dag\,
\Psi_0^{\rm orb}\left[A_z,A_\bz\,;\tilde Y^z\right] \ , \nn\\&&
\label{Zodef}\eea
where the orbifold wavefunction on the initial closed surface
$\Sigma_0\equiv\Sigma_1$ is given by
\bea
\Psi_0^{\rm orb}\left[A_z,A_\bz\,;\tilde Y^z\right]&=&\exp\left\{\,
\int\limits_\Sigma d^2z~\left(\frac k{4\pi}\,A_\bz+\tilde Y^z\right)A_z
\right\}\nn\\&&\times\,\int[D\varphi]~\exp\left\{\,\frac k{4\pi}\,
\int\limits_\Sigma d^2z~\left(\partial_\bz\varphi-2A_\bz\right)\,
\partial_z\varphi\right\} \ ,
\label{Psi0orb}\eea
while the wavefunctional at the orbifold fixed point $t=1/2$
corresponding to the open surface $\Sigma_{1/2}/\zed_2=\Sigma^{\rm o}$ is
\bea
&&\Psi_{1/2}^{\rm orb}\left[A_z,A_\bz\,;\tilde Y^z,\tilde Y^\bz,\tilde
Y^\parallel\,
\right]^\dag~=~\int[D\varphi]~\exp\left\{~\oint
\limits_{\partial\Sigma^{\rm o}}\varphi^{~}_{\rm b}\,
\left(\tilde Y^\parallel-\frac{k}{8\pi}\,A^\parallel\right)
\right\}\nn\\&&~~~~~\times\,\exp\left\{-\frac k{8\pi}\,\int
\limits_{\Sigma^{\rm o}}d^2z~\left[A_z\left(\partial_\bz\varphi-
\frac{8\pi}k\,\tilde Y^z\right)-A_\bz\,\left(\partial_z\varphi-\frac{8\pi}k
\,\tilde Y^\bz\,\right)\right]\right\} \ .
\label{Psi12orb}\eea

Here $S_{\rm TMGT}^{\rm orb}[A]$ is the topologically massive gauge
theory action (\ref{STMGT}) for the orbifold of the three-dimensional
theory, with the identification $2S_{\rm TMGT}^{\rm orb}[A^{\rm
  orb}\,]=S_{\rm TMGT}[A]$ and $A$ the extension of the gauge field
$A^{\rm orb}$ from $(\Sigma\times[0,1])/\zed_2$ to the covering
cylinder. Analogous relations hold for the other actions, and for
brevity we omit the orbifold superscript on all fields.
The structure of the wavefunctionals (\ref{Psi0orb}) and
(\ref{Psi12orb}) shows the physical significance of the specific
solutions to the continuity equations that we obtained in the previous
subsection. Namely, we can remove the external currents from the bulk
by the shifts $\partial_z\varphi\to\partial_z\varphi+8\pi\,\tilde Y^\bz/k$ and
$\partial_\bz\varphi\to\partial_\bz\varphi+8\pi\,\tilde Y^z/k$. In the
language of the induced conformal field theory, this means that the
change of conformal background is determined solely by boundary
deformations, as we hope to find in the end.

We can now use the decomposition (\ref{Aigaugeorbit}) to compute the
conformal field theory partition function $Z_{\Sigma^{\rm o}}$
for the induced open string WZNW model on the worldsheet $\Sigma^{\rm
  o}$. After gauge-fixing, the Hodge decomposition on the cover
$\Sigma$ of the gauge field one-form (\ref{Aigaugeorbit}) is given
by~\cite{TM_18}
\be
\bar A^i=a^i+\epsilon^{ij}\,\partial_j\xi\ ,
\label{barAfixed}\ee
where $a_i$ are the harmonic degrees of freedom, with quantized
periods in $H^1(\Sigma,\zed)\cong\zed^{2g}$, and $\xi$ is a real,
compact scalar field on $\Sigma$. The exact term which usually appears
in a Hodge decomposition has been set to $0$ here by an appropriate
gauge choice~\cite{TM_18}. We will also use a Hodge decomposition for
the external currents of the form
\be
\tilde{Y}^i=\frac{k}{4\pi}\,\Bigl(
\partial^iY_{\rm D}+\epsilon^{ij}\,\partial_jY_{\rm N}\Bigr) \ ,
\lb{HodgeY}
\ee
where again $Y_{\rm D}$ and $Y_{\rm N}$ are real compact fields. As
the currents are local operators, there are no harmonic degrees of
freedom present in this decomposition. While they do change the
topology of the worldsheet since a local insertion adds a new boundary
to the surface, it is a local deformation only and not a global
correction to the topology~\cite{POL_book}.

For convenience, let us now list the orbifold relations for the
various fields appearing in these decompositions. They are given by
\be
\ba{crclccrcl}
{\sf PT}\,:\,&a_z&\longmapsto&a_\bz&\hspace{12mm}\ &{\sf
  PCT}\,:\,&a_z&\longmapsto&-a_\bz\\[1mm]
&Y_{\rm D}&\longmapsto&Y_{\rm D}& & &Y_{\rm D}&\longmapsto&-Y_{\rm D}\\[3mm]
&\xi&\longmapsto&-\xi& & &\xi&\longmapsto&\xi\\[1mm]
&Y_{\rm N}&\longmapsto&-Y_{\rm N}& & &Y_{\rm N}&\longmapsto&Y_{\rm N} \ .
\ea
\ee
Defining complex fields in terms of their components tangent and
transverse to the worldsheet boundary as $X^z=X^\parallel+i\,X^\perp$,
$X^\bz=X^\parallel-i\,X^\perp$, we obtain for the $\sf PT$ orbifold of the
topologically massive gauge theory the boundary conditions
$a^\perp=\xi=Y_{\rm N}=0$ on $\partial\Sigma^{\rm o}$, while
$a^\parallel$ and $Y_{\rm D}$ are arbitrary. For the
$\sf PCT$ orbifold we have $a^\perp=a^\parallel=Y_{\rm D}=0$ with
arbitrary $\xi$ and $Y_{\rm N}$ on $\partial\Sigma^{\rm o}$.
The reason why $a^\perp=0$ at the boundary for both orbifolds is due
to the fact that $a^\perp$ encodes the harmonic degrees of freedom,
which depend only on the homology of the Riemann surface
$\Sigma$. When approaching a boundary, $a^\perp$ cannot have a
component normal to it since this would constitute a local
singularity, i.e. the failure of a canonical homology cycle of
$\Sigma$ to be closed. This is basically the same
argument as to why there is no harmonic component in the Hodge
decomposition of the current in~(\r{HodgeY}). Note also that for the
$\sf PCT$ orbifold $\partial^\parallel\xi=0$ at the boundary, because
$A^\perp=a^\perp-i\,\partial^\parallel\xi$ from the
decomposition~(\r{barAfixed}) and for this orbifold
$A^\perp|_{\partial\Sigma^{\rm o}}=0$.

Because of the remarks just made, the calculation of the orbifold
partition function (\ref{Zodef}) proceeds similarly to that described
in~\cite{TM_18}, except that now the wavefunction~(\ref{Psi12orb}) is
completely gauge-invariant, because here we have solved the gauge
constraint on $\partial\Sigma^{\rm o}$ in (\ref{gauss}) and so the
gauge parameter function $\Lambda$ gets absorbed into the boundary
scalar field degree of freedom $\varphi^{~}_{\rm b}$. We note again
that the boundary field $\varphi^{~}_{\rm b}$ has for now
free boundary conditions, and is arbitrary for both kinds of
orbifolds. Only after integration over the gauge degrees of freedom
will we encode its boundary conditions in the form of functional Dirac
delta-functions in the path integral.

Given the above decompositions, the orbifold
wavefunction~(\ref{Psi12orb}) is now rewritten as
\bea
&&\Psi_{1/2}^{\rm orb}\left[a,\xi\,;Y_{\rm D},Y_{\rm N}\,
\right]^\dag\nn\\&&~~~~~~=~\int[D\varphi]~\exp\left\{\,\frac{k}{8\pi}\,\oint
\limits_{\partial\Sigma^{\rm o}}\Bigl[2\,\varphi^{~}_{\rm b}\,
\left(\partial^\perp Y_{\rm D}+i\,\partial^\parallel Y_{\rm N}
\right)-i\,\varphi\,a^\parallel-i\,\xi\,\partial^\perp\varphi^{~}_{\rm
  b}\Bigr]\right\}\nn\\
&&~~~~~~~~~~\times\,\exp\left\{-\frac {i\,k}{8\pi}\,\int
\limits_{\Sigma^{\rm o}}d^2z~\Bigl[\epsilon^{ij}\,a_i\,\left(\partial_j\varphi-
2\,\partial_jY_{\rm D}-2h_{jl}\,\epsilon^{lk}\,\partial_kY_{\rm N}
\right)\Bigr.\right.\nn\\&&~~~~~~~~~~+\Biggl.\Bigl.h^{ij}\,
\partial_i\xi\,\left(\partial_j\varphi-
2\,\partial_jY_{\rm D}-2h_{jl}\,\epsilon^{lk}\,\partial_kY_{\rm N}\,
\right)\Bigr]\Biggr\} \ .
\label{Psi12orb2}\eea
By integrating by parts in the bulk of $\Sigma^{\rm o}$ it follows
that this wavefunction factorizes into bulk and boundary components as
\be
{\Psi_{1/2}^{\rm
    orb}}^\dag~=~\int[D\varphi]~\exp\left\{\,-\frac{k}{8\pi}\,
\int\limits_{\Sigma^{\rm o}}d^2z~\xi\,\nabla^2(\varphi-2\,Y_{\rm D})
\right\}~\Psi_{\partial\Sigma^{\rm o}}^\dag \ ,
\label{Psi12orbfact}\ee
where the boundary wavefunctional for both orbifold types is given by
\bea
{\sf PT}\,:\,&&\Psi_{\partial\Sigma,{\rm D}}^\dag~=~
\int[D\varphi^{~}_{\rm b}]~\exp\left\{\,\frac{k}{4\pi}\,\oint
\limits_{\partial\Sigma^{\rm o}}\Bigl[\varphi^{~}_{\rm b}\,
\partial^\perp Y_{\rm D}
-i\,a^\parallel\,\left(\varphi-Y_{\rm D}\right)\Bigr]\right\} \ ,
\label{Psi12orb2D} \\
{\sf PCT}:&&\Psi_{\partial\Sigma,{\rm N}}^\dag~=~
\int[D\varphi^{~}_{\rm b}]~\exp\left\{\,\frac{i\,k}{4\pi}\,\oint
\limits_{\partial\Sigma^{\rm o}}\left[\varphi^{~}_{\rm b}\,
\partial^\parallel Y_{\rm N}
+\xi\,\partial^\perp\varphi^{~}_{\rm b}\right]\right\} \ .
\label{Psi12orb2N}\eea
We thereby find that the
integrations over the gauge field degrees of freedom $a^\parallel$ and
$\xi$ produce, for the two types of orbifold involutions, Dirichlet
and Neumann boundary conditions on the scalar field $\varphi$ as
\be
\ba{rcl}
{\mathrm{Dirichlet}}~({\sf PT})&:&\displaystyle~\delta_{\Sigma^{\rm
    o}}\Bigl(\nabla^2(\varphi-2\,Y_{\rm D})\Bigr)~
\delta_{\partial\Sigma^{\rm o}}\left(\varphi^{~}_{\rm b}-Y_{\rm D}
\right) \ , \\[3mm]{\mathrm{Neumann}}~({\sf PCT})&:&
\displaystyle~\delta_{\Sigma^{\rm o}}\Bigl(\nabla^2(\varphi-2\,Y_{\rm
  D})\Bigr)~\delta_{\partial\Sigma^{\rm o}}\left(\partial^\perp
\varphi^{~}_{\rm b}\right) \ .
\ea
\label{DNbcs}\ee

It is the $\Lambda$ independence of (\ref{Psi12orb2}) that leads to
boundary conditions in (\ref{DNbcs}) which differ from
those found in~\cite{TM_18}. But just as in~\cite{TM_18}, the roles played by
the wavefunctional (\ref{Psi12orb2}) at the orbifold branch point are
to enforce the bulk equation of motion for the free scalar field
$\varphi$ on $\Sigma^{\rm o}$, and also to select the boundary
conditions of the open string theory. Otherwise it simply corresponds
to inserting the identity character state $|1\rangle$ into the inner
product~(\ref{Zodef}).

The main new result here is the appearence of the first exponential
factors in the boundary wavefunctionals (\ref{Psi12orb2D}) and
(\ref{Psi12orb2N}) which have emerged from a careful and specific
incorporation of the external sources and were absent in the analysis
of~\cite{TM_18}. After an integration by parts in (\ref{Psi12orb2D}),
for Dirichlet boundary conditions (${\sf PT}$) it becomes
\be
{\mathcal{V}}_{\rm D}=\exp\left\{-\frac{k}{4\pi}\,\oint
\limits_{\partial\Sigma^{\rm o}}Y_{\rm D}\,
\partial^\perp\varphi^{~}_{\rm b}\right\} \ .
\lb{Dvertexop}
\ee
The object (\ref{Dvertexop}) is just the vertex operator for a D-brane
wrapped around a single compact dimension of radius $R$ and described by
the collective coordinate $Y_{\rm D}$. This holds with the usual
identification of the Chern-Simons coupling constant as
\be
k=\frac{2R^2}{\alpha'} \ ,
\label{CScouplingid}\ee
where $\alpha'$ is the string slope. On the other hand, for Neumann
boundary conditions (${\sf PCT}$), the boundary wavefunction
(\ref{Psi12orb2N}) yields the vertex operator
\be
{\mathcal{V}}_{\rm N}=\exp\left\{-\frac{i\,k}{4\pi}\,\oint
\limits_{\partial\Sigma^{\rm o}}Y_{\rm N}\,
\partial^\parallel\varphi^{~}_{\rm b}\right\} \ ,
\label{Nvertexop}\ee
which is just the Wilson line for the target space gauge field
component $R\,Y_{\rm N}/2\alpha'$ around the compact direction. As
expected by T-duality, the vertex operators (\ref{Dvertexop}) and
(\ref{Nvertexop}) map into each other under
$\partial^\perp\varphi^{~}_{\rm b}\leftrightarrow
i\,\partial^\parallel\varphi^{~}_{\rm b}$.

Thus, as we anticipated at the beginning of this section, the coupling
of topologically massive gauge theory to an external current encodes
both the Wilson lines and the D-brane collective coordinates of string
theory. In the case of a more general topologically massive gauge
theory with abelian structure group $U(1)^d\times U(1)^D$, the orbifold of the
fields of the $U(1)^d$ sector by the ${\sf PT}$ involution of the bulk
quantum field theory would produce vertex operators of the type
$\partial_\perp^{~}\varphi_{\rm b}^I$, with $I=1,\dots,d$ corresponding to
the transverse directions to the brane, while the orbifold of the
$U(1)^D$ sector by the ${\sf PCT}$ involution of the three-dimensional
gauge fields $A^I$ would produce target space gauge fields $Y_{\rm
  N}^I$, with $I=d+1,\ldots,d+D$ running along the brane worldvolume
directions. Geometrically, both involutions produce the same open
surface $\Sigma_{1/2}/\zed_2$, but the charge conjugation operation
$\sf C$ acts on the charges and the gauge fields $A^I$, and not on the
membrane coordinates. In this way we have explicitly demonstrated the
emergence of D-branes in the topological membrane approach to string
theory through a derivation of the brane vertex operators in the
corresponding induced conformal field theory action on an open
surface. In section~\r{Daction} we will analyse D-branes at the level
of worldsheet effective actions which are derived from the topological
membrane by using the procedure outlined above.

\setcounter{equation}{0}
\section{\lb{bstates}BRANE STATES}

The purpose of this section is to construct the appropriate
generalizations of the Schr\"odinger wavefunctionals of the previous
section which correspond to boundary states of D-branes in the
induced boundary conformal field theory. We will first describe how
these functionals reproduce the well-known three-dimensional
descriptions of the fusion algebra and the corresponding Verlinde
formula in the closed string sector. Then we shall show how to modify
these calculations in the presence of an orbifold of the topological
membrane. We shall see that D-branes and string Wilson lines are
naturally selected by gauge invariance of the bulk three-dimensional
theory as fundamental boundary states. Moreover, in this setting the
Cardy map between the set of admissible boundary states and conformal
blocks is determined entirely by the spectrum of allowed charges of
dynamical matter inside the topological membrane, i.e. by the number
of possible ground states in Chern-Simons theory, which after
orbifolding give distinctive boundary conditions.

\subsection{The Fusion Ring}

The $U(1)$ topologically massive gauge theory has a multiply-connected
gauge group, and so it corresponds to a $c=1$ conformal field theory
with an extended chiral algebra, the chiral algebra of the
rational circle. For this, we take the Chern-Simons coefficient
(\ref{CScouplingid}) to be the rational number
\be
k=\frac{2p}q \ ,
\label{k2pq}\ee
where $p$ and $q$ are positive integers. For simplicity we assume
that $p$ is even with $p/2$ and $q$ coprime. The spectrum of charges on
a closed Riemann surface $\Sigma$ of genus $g$ is given by
\bea
Q_{\vec\lambda}^\ell&\equiv&\frac{\lambda^\ell}q~=~m^\ell+
\frac{k\,n^\ell}4 \ , \nn\\\bar Q_{\vec{\bar\lambda}}^\ell&\equiv&
\frac{\bar\lambda^\ell}q~=~m^\ell-\frac{k\,n^\ell}4
\label{rationalcharges}\eea
with $m^\ell,n^\ell=0,1,\dots,\frac{pq}2-1$, $\ell=1,\dots,g$ the winding and
monopole numbers of a charge as it moves around canonical homology
cycles $\beta^\ell$ of $\Sigma$. We assume, for each $\ell$, that the
integers $(m^\ell,n^\ell)$ form a Bezout pair with respect to the
decompositions (\ref{rationalcharges}). The resulting conformal field theory
is rational as the $U(1)$ charges are now restricted to
$\vec\lambda,\vec{\bar\lambda}\in(\zed_{pq})^g$. For any integer $n$,
we shall denote by $[n]\in\zed_{pq}$ its integer part modulo $pq$.

On the punctured (Riemann) sphere $\Sigma=\sphere_0^2$ (equivalently
the infinite cylinder $\Sigma=\real\times\sphere^1$), there are $pq$
primary blocks $\phi_\lambda(z)=\e^{i\,\lambda\,\varphi(z)/q}$,
$\lambda=0,1,\dots,pq-1$ of this chiral algebra. Members of the
conformal family $[\phi_\lambda]$ have charges $Q_\lambda+p\,l$ with
$l\in\zed$. The corresponding fusion rules
\be
[\phi_\lambda]\times[\phi_\mu]=\sum_{\nu=0}^{pq-1}N_{\lambda\mu}^{~~\nu}
{}~[\phi_\nu]
\label{fusionalggen}\ee
may be determined through the representation theory of the group
algebra ${\mathbb{C}}[\zed_{pq}]$ of the cyclic subgroup $\zed_{pq}$
in the $U(1)$ gauge group. In this simple case the fusion coefficients
are given by
\be
N_{\lambda\mu}^{~~\nu}=\delta_{[\lambda+\mu-\nu]} \ ,
\label{fusioncoeffs}\ee
where $\delta_{[\lambda]}$ is the (periodic) delta-function on the
finite group $\zed_{pq}$ which is defined by
\be
\delta_{[\lambda]}=
\frac1{pq}\,\sum_{\mu=0}^{pq-1}\e^{2\pi i\,\mu\lambda/pq} \ .
\label{Zpqdeltafn}\ee
The natural metric on the fusion algebra is the charge conjugation
matrix $\Cmatrix:[\phi_\lambda]\to[\phi_{\lambda^+}]$ and is given here by
\be
\Cmatrix_{\lambda\mu}=\delta_{[\lambda+\mu]} \ .
\label{fusionmetric}\ee
Because of the triviality of (\ref{fusionmetric}), we shall not
distinguish between indices which are raised and lowered with
this metric.

\subsection{Punctured Wavefunctionals\label{WilsonCorr}}

The main goal of this section is to provide a three-dimensional description
of the fusion algebra of the previous subsection and its open string
counterpart, with an eye to describing those conformal field theoretic
states which correspond to D-branes. As we have shown in the previous
section, D-brane states arise when we introduce external bulk matter
into the topologically massive gauge theory. When these currents
correspond to the propagation of charged particles in
three-dimensions, they can be equivalently described in terms of the
source-free gauge theory in the presence of Wilson lines
\be
W_{Q_i}[A]=\exp\left\{i\,Q_i\,\int\limits_{C_i}A\right\}
\label{WQiA}\ee
corresponding to the bulk propagation of charges $Q_i=\lambda_i/q$
along the oriented worldlines $C_i\subset M$. In the following we will
describe some generic properties of the correlators
\be
\left\langle\,\prod_{i=1}^sW_{Q_i}\right\rangle_M=
\int[DA]~{\cal W}_M[A]~\prod_{i=1}^sW_{Q_i}[A]~
\e^{iS_{\rm TMGT}^M[A]}
\label{WQicorr}\ee
evaluated in the topologically massive gauge theory defined on
various three-manifold geometries $M$. As before, all such correlation
functions will be evaluated in the topological limit of the quantum
field theory. The weight ${\cal W}_M$ depends on the particular
three-manifold that we are working on.

In this subsection we will be interested in (\ref{WQicorr})
for the case of the vertical Wilson lines of the topological
membrane~(fig.~\ref{Wilsonlines}(a)), whereby the three-geometry is taken
to be $M=\Sigma\times[0,1]$. Then the operators (\ref{WQiA}) are not
invariant under gauge transformations $A_i\to A_i+\partial_i\Lambda$
but instead change by a phase factor which is given by the boundary
values of $\Lambda$ on $\Sigma_0$ and $\Sigma_1$. Since such gauge
non-invariant terms should be absorbed as usual by shifting the WZNW field as
$\varphi\to\varphi+\Lambda$, it follows that insertions of the Wilson
line operators (\ref{WQicorr}) correspond to insertions of the primary
fields $\e^{i\,Q_i\,\varphi}$ at the boundary points $z_i,\bz_i$ on
$\Sigma_0$ and $\Sigma_1$ corresponding to the two endpoints of the
worldlines $C_i$. For a collection of unlinked and unknotted particle
trajectories in $M$, these Wilson lines thereby correspond to
gauge-invariant states of the topological membrane whose
wavefunctionals can be built by inserting the tachyon fields
$\e^{i\,Q_i\,\varphi}$ at the points $z_i$ into the vacuum functional
(\ref{WWS}) to give
\bea
&&\Xi\left[A_z,A_\bz\,;\left\{{\matrix{z_1&\cdots&z_s\cr Q_1&\cdots&
Q_s\cr}}\right\};0\right]~=~\exp\left\{\frac k{8\pi}
\,\int\limits_\Sigma d^2z~A_\bz\,A_z\right\}\nn\\&&~~~~~\times\,
\int[D\varphi]~\prod_{i=1}^s\e^{i\,Q_i\,\bigl(\varphi(z_i)+h_\varphi(z_i)
\bigr)}\,\exp\left\{\frac k{8\pi}\,\int\limits_\Sigma d^2z~\left(
\partial_\bz\varphi-2A_\bz\right)\,\partial_z\varphi\right\} \ .
\label{Xiexcited}\eea
The function $h_\varphi(z)$ is given by an integral over the harmonic
parts $a_i$ of the gauge field one-forms and it takes into account the large
gauge transformations $\Sigma\to U(1)$ which wind around the canonical
homology cycles of $\Sigma$. The functionals
(\ref{Xiexcited}) simply coincide with the gauge-invariant charged vacuum
states constructed in the previous section in the case that the
non-dynamical bulk matter fields correspond to static point
charges, as they clearly solve the vacuum wave equation
(\ref{Schreq0}) and gauge constraint (\ref{gaussjj}) with
\be
j^i=0 \ , ~~ \rho(\vb{z})=\sum_{i=1}^sQ_i~
\delta^{(2)}(\vb{z}-\vb{z}_i) \ .
\label{currentstatic}\ee
They consist of $s$ gauge-invariant combinations of external particles
in the bulk which pierce $\Sigma$ at the points $z_i$.

\fig{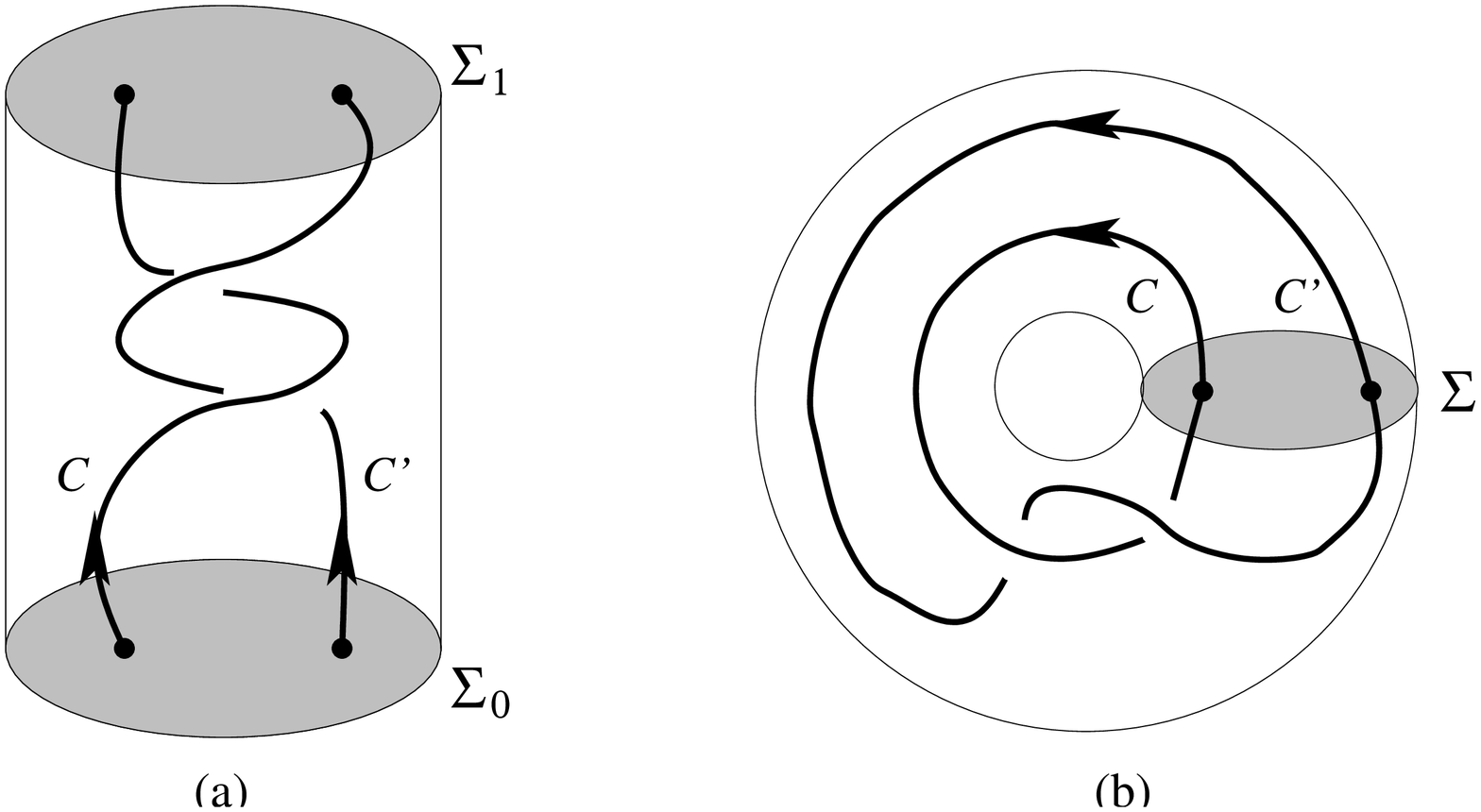}{{\rm (a)} The topological membrane with two
  linked Wilson lines $C$ and $C'$. {\rm (b)} Identifying the initial
  and final surfaces of (a) produces the
  three-geometry $\Sigma\times\sphere^1$ with two linked Polyakov
  loops.}{Wilsonlines}

In the present case, invariance under large gauge transformations in
the bulk is taken care of by inserting into (\ref{WQicorr}) the weight
functional~\cite{TM_18}
\be
{\cal W}_{\Sigma\times[0,1]}\left(\Delta\vec Q_{\vec\lambda}\right)
=\prod_aV^{n_a}(\vb{z}_a)=\prod_{\ell=1}^g
\exp\left\{i\,\Delta Q_{\vec\lambda}^\ell\,\oint\limits_{\beta^\ell}A
\right\} \ ,
\label{calWTM}\ee
with $\Delta Q_{\vec\lambda}^\ell=-n^\ell k/2$ corresponding to a set
of primary charges $\vec\lambda\in(\zed_{pq})^g$. This operator also
incorporates the appropriate non-perturbative linking and
monopole-instanton transitions in the bulk. With this definition, the
partition function (\ref{Zcdef}) of the topological membrane yields,
in the absence of sources, the inner product
\be
Z^{\rm c}\left(\Gamma,\overline{\Gamma}\,\right)
=\langle1\rangle_{\Sigma\times[0,1]}^{~}=k^{g/2}\,
\sum_{\vec\lambda\in(\zed_{pq})^g}\Psi^\dag_{\vec{\bar\lambda}}(
\,\overline{\Gamma}\,)~\Psi_{\vec\lambda}(\Gamma) \ ,
\label{partfnTM}\ee
where $\Psi_{\vec\lambda}(\Gamma)$ are the effective topological wavefunctions
of topologically massive gauge theory which depend only on the periods
$\Gamma$ of the compact Riemann surface $\Sigma$~\cite{TM_18}. They
are proportional to the characters of the extended Kac-Moody group in
this case. The key feature of (\ref{partfnTM}) is that it manifestly
admits a holomorphic factorization into components depending only on
the left and right moving worldsheets $\Sigma_0$ and $\Sigma_1$. In
exactly the same way, the inner product $\langle\Xi_1|\Xi_0\rangle$ of
the charged wavefunctionals (\ref{Xiexcited}), defined as in (\ref{Zcdef}),
reproduces the $c=1$ conformal field theory partition function on
$\Sigma$ with primary state insertions.

\subsection{Topological Correlation Functions\label{TopCorrs}}

To describe the chiral algebra of the rational circle, we can exploit
the holomorphic factorization of (\ref{Zcdef}) (or (\ref{partfnTM}))
to restrict it to a single chiral sector of the worldsheet
theory. This is achieved by introducing a fixed-point free gluing automorphism
along the time direction of the membrane $\Sigma\times[0,1]$ which identifies
$\Sigma_0\equiv\Sigma_1$, i.e. we replace $\Sigma\times[0,1]$ by the
quotient space $(\Sigma\times[0,1])/\sim$ with respect to the
equivalence relation $\sim$ defined by $(z,0)\sim(z,1)~~\forall
z\in\Sigma$. We then Wick rotate the time direction to
Euclidean signature. In this way, the
vertical Wilson lines of the topological membrane become Polyakov
loops of finite temperature gauge theory on $M=\Sigma\times
\sphere^1$~(fig.~\ref{Wilsonlines}(b))~\cite{BT_1}. Formally, the
chiral correlation functions on $\Sigma\times\sphere^1$ are related to
those of the topological membrane (on $\Sigma\times[0,1]$) as
follows. The crucial point is that in the finite-temperature gauge
theory it is no longer possible to fix the temporal gauge $A_0=0$, as
this is incompatible with invariance under large gauge transformations
which wind around the $\sphere^1$. Instead, a natural consistent gauge
choice is $\partial_tA_0=0$, in which the membrane amplitudes can be
evaluated as in the previous section with $A_0$ regarded as an
external static source.

For a collection of unlinked and unknotted loops $C_i$, the
  correlators (\ref{WQicorr}) with weight functionals ${\cal
  W}_{\Sigma\times\sphere^1}=1$ depend only on the insertion points
  $x_i\in\Sigma$. Then, by incorporating the standard Faddeev-Popov
  gauge-fixing determinant, the trace of the wavefunction correlator
  (\ref{Zcdef}) on $\Sigma\times[0,1]$ in the states (\ref{Xiexcited})
  can be used to generate the Polyakov loop correlation functions on
$\Sigma\times\sphere^1$ through the trace formula
\bea
&&\left\langle\,\prod_{i=1}^sW_{Q_i}(x_i)\right\rangle^{~}_{\Sigma
\times\sphere^1}~=~\Tr\,\langle\Xi_1|
\Xi_0\rangle\nn\\&&{~~~~}^{~~}_{~~}
\nn\\&&~~~~~=~\int[DA_0]~\int[DA_z~DA_\bz\,]~
{\det}'\,\partial_t\Bigm|_{\sphere^1}~\e^{iS_{\rm
    TMGT}^{\Sigma\times[0,1]}[A]}\nn\\&&~~~~~~~~~~\times\,
\Xi^\dag\left[A_z,A_\bz\,;\left\{{\matrix{x_1&\cdots&x_s\cr Q_1&\cdots&
Q_s\cr}}\right\};A_0\right]\,\Xi\left[A_z,A_\bz\,;\left\{{\matrix{x_1&\cdots&
x_s\cr Q_1&\cdots&Q_s\cr}}\right\};A_0\right] \ , \nn\\&&
\label{tempmembrane}\eea
where we have made the identification $\Xi=\Xi_0\equiv\Xi_1$ of
initial and final membrane states, corresponding to the gluing
$\Sigma_0\equiv\Sigma_1$, which gives a trace. We can now gauge
away the field component $A_0\to A_0-\partial_t\Lambda_t=0$ from
the wavefunctionals in (\ref{tempmembrane}) via the time-dependent
gauge transformation $\Lambda_t=tA_0$ with $\Lambda_0=0$ and
$\Lambda_1=A_0$. In the membrane picture, this gauge transform only
affects the anti-holomorphic spatial components $A_\bz$ in the
wavefunctions $\Xi_1^\dag$ on $\Sigma_1$. The Schr\"odinger amplitudes
(\ref{Zcdef}) are gauge-invariant, and whence up to an irrelevant
normalization the trace (\ref{tempmembrane}) becomes
\bea
&&\left\langle\,\prod_{i=1}^sW_{Q_i}(x_i)\right\rangle_{\Sigma
\times\sphere^1}~=~\int[DA_z~DA_\bz\,]~
\e^{iS_{\rm TMGT}^{\Sigma\times[0,1]}[A]}\nn\\&&~~~~~\times\,
\Xi^\dag\left[A_z,A_\bz\,;\left\{{\matrix{x_1&\cdots&x_s\cr Q_1&\cdots&
Q_s\cr}}\right\};0\right]\,\Xi\left[A_z,A_\bz\,;\left\{{\matrix{x_1&\cdots&
x_s\cr Q_1&\cdots&Q_s\cr}}\right\};0\right] \ .
\label{tempmembranefinal}\eea

Since the source-free Hamiltonian $H=H_0$ in (\ref{H0}) of this
  thermodynamical system vanishes in the vacuum sector of the quantum
  gauge theory, this definition turns the topological inner product
  into a trace of the identity operator over the finite-dimensional
  Hilbert space ${\cal H}_{g,s,k}(\lambda_1,\dots,\lambda_s)$ of
  physical ground states of the gauge theory in the presence of
  Polyakov loop insertions. In two-dimensional language, this space is
  the same as the vector space of holomorphic conformal blocks of the
  corresponding WZNW model on a closed surface $\Sigma$ of genus $g$
  with $s$ vertex operator insertions~\cite{W_0,FK_1,BT_1}, or
  alternatively as the fiber space of the appropriate Friedan-Shenker
  holomorphic vector bundle over the moduli space of $s$-punctured
  Riemann surfaces of genus $g$ (with $\Psi_{\vec\lambda}(\Gamma)$
  above particular sections of this bundle). It follows that the
  dimension of the vacuum sector of the matter-coupled topologically
  massive gauge theory is given by
\be
\dim{\cal H}_{g,s,k}(\lambda_1,\dots,\lambda_s)=
\Tr^{~}_{{\cal H}_{g,s,k}(\lambda_1,\dots,\lambda_s)}(\id)=
\left\langle\,\prod_{i=1}^s
W_{\lambda_i/q}(x_i)\right\rangle_{\Sigma\times\sphere^1} \ .
\label{dimcalHgsk}\ee

In this simple model it is easy to evaluate the topological numbers
(\ref{dimcalHgsk}) by using the trace formula
(\ref{tempmembranefinal}). The functional integrations arising after
gauge-fixing and insertion of the
explicit forms of the states (\ref{Xiexcited}) are all Gaussian, as
this is a free field theory. By using the decompositions
(\ref{Aigaugeorbit}) and (\ref{barAfixed}), the bulk gauge field
integrations are readily seen to eliminate the charge independent terms in the
scalar fields $\varphi$ after appropriate gauge transformations
parametrized by them. Integration over the
insertions $\varphi(x_i)$ then produces a delta-function constraint
enforcing bulk charge (or flux) conservation on a closed space,
$\sum_iQ_i=0$, and we arrive at
\be
\dim{\cal H}_{g,s,k}(\lambda_1,\dots,\lambda_s)=k^{g/2}~
\delta_{[\lambda_1+\dots+\lambda_s]}~\sum_{\vec\lambda\in
(\zed_{pq})^g}\,\Bigl(\Psi_{\vec\lambda}(\Gamma)\,,\,
\Psi_{\vec\lambda}(\Gamma)\Bigr)_a \ ,
\label{dimcalHint}\ee
where $(~,~)_a$ is the inner product on the space of harmonic
one-forms which is given by an integral over the Jacobian variety of
the Riemann surface $\Sigma$. The topological wavefunctions are
orthogonal with respect to this inner product~\cite{TM_18}
\bea
\Bigl(\Psi_{\vec\lambda}(\Gamma)\,,\,
\Psi_{\vec\lambda'}(\Gamma)\Bigr)_a&\equiv&\frac1{2^g\,\sqrt{\det
\Gamma_2}}~\int\limits_{{\rm Jac}(\Sigma)}\,\prod_{\ell=1}^g
da^\ell~d\overline{a}^{\,\ell}~\Psi_{\vec\lambda}^\dag\left.\left(\vec a\,,\,
\vec{\overline{a}}~\right|\,\Gamma\right)\,\Psi_{\vec\lambda'}
\left.\left(\vec a\,,\,\vec{\overline{a}}~\right|
\,\Gamma\right)\nn\\&=&\frac1{k^{g/2}}~
\delta_{\vec\lambda\,,\,\vec\lambda'} \ ,
\label{topinnerprod}\eea
where $\Gamma_2$ denotes the positive-definite imaginary part of the
period matrix $\Gamma$. The dimension (\ref{dimcalHint}) is therefore
given by
\be
\dim{\cal H}_{g,s,k}(\lambda_1,\dots,\lambda_s)=(pq)^g~
\delta_{[\lambda_1+\dots+\lambda_s]} \ ,
\label{dimcalHfinal}\ee
as expected since here there are $(pq)^g$ linearly independent topological
wavefunctions $\Psi_{\vec\lambda}(\Gamma)$.

\subsection{Surgical Wavefunctionals\label{Surgical}}

The final quantities that we shall need to know are the contributions from
the linkings of charged particle trajectories, as illustrated in
fig.~\ref{Wilsonlines}. This is a purely bulk effect which can be
computed by noting that in the neighbourhood of any such link, the
three-manifold may be regarded topologically as the three-sphere
$\sphere^3$ and the linking effect as generated by closed contours
$C_i$. With ${\cal W}_{\sphere^3}=1$ and the bulk normalization of
section~\ref{OrbPart} for the partition function, we may then invoke
the standard formula~\cite{W_0}
\be
\left\langle\,\prod_{i=1}^sW_{Q_i}[A]\right\rangle_{\sphere^3}=
\prod_{i,j=1}^s\e^{\frac{2\pi i}k\,Q_iQ_j\,\#(C_i,C_j)} \ ,
\label{WcorrS3}\ee
where
\be
\#(C_i,C_j)=\oint\limits_{C_i}dx_\mu~\oint\limits_{C_j}
dy_\nu~\epsilon^{\mu\nu\lambda}\,\frac{(x-y)_\lambda}{|x-y|^3}
\label{Gausslink}\ee
is the Gauss linking number of the closed particle trajectories $C_i$
and $C_j$ which counts the number of signed intersections of $C_i$
with the surface spanned by $C_j$. When $i=j$ the self-linking integral
(\ref{Gausslink}) must be suitably regularized by a framing of the contour
$C_i$~\cite{W_0}. More precisely, such effects are computed via surgery
prescriptions, by gluing three-manifolds with Wilson lines onto the
ones of interest to give invariants of links in $\sphere^3$ and by
using functoriality of the gauge theory amplitudes~\cite{W_0,FS_2}. In
what follows, however, we will only be concerned with deriving
relationships between the conformal blocks of open and closed string
theories, for which the correspondence described above will suffice.

For the membrane linkings of the sort illustrated in
fig.~\ref{Wilsonlines}(a), the linking effect can be incorporated
analytically by introducing the full, non-chiral wavefunctionals obtained
by gluing the left and right moving punctured states (\ref{Xiexcited})
together using the monopole operators (\ref{calWTM}) and the
multi-valued angle function (\ref{thetaImdef}) of $\Sigma$. The vacuum
functionals associated with the vertical (and possibly linked)
propagation of charged particles from the boundary $\Sigma_0$ to
the boundary $\Sigma_1$ are thereby given as
\bea
&&\Phi\left[A_z,A_\bz\,;\left\{\matrix{z_1&\cdots&z_s&\bz_1&\cdots&
\bz_s\cr Q_1&\cdots&Q_s&\bar Q_1&\cdots&\bar Q_s\cr}\right\}\right]
\nn\\&&~~~~~~=~\prod_{i,j=1}^s\e^{\frac ik\,Q_iQ_j\,\bigl(
\theta_{ij}(1)-\theta_{ij}(0)\bigr)}~{\cal W}_{\Sigma\times[0,1]}
\left(\Delta\vec Q_{\vec\lambda}\right)\,\Xi_0\left[A_z,A_\bz\,;
\left\{{\matrix{z_1&\cdots&z_s\cr Q_1&\cdots&Q_s\cr}}\right\};0\right]
\nn\\&&~~~~~~~~~~~\otimes\,
\Xi_1^\dag\left[A_z,A_\bz\,;\left\{{\matrix{\bz_1&\cdots&\bz_s\cr
\bar Q_1&\cdots&\bar Q_s\cr}}\right\};0\right] \ ,
\label{PsiWilsonnonchiral}\eea
where
\bea
\theta_{ij}(t)&=&\theta\Bigl(x_i(t)\,,\,x_j(t)\Bigr)\nn\\&&+\,2\,
\sum_{\ell,\ell'=1}^g\left(\Gamma_2^{-1}\right)^{\ell\ell'}\,
{\rm Im}\left[\,\int\limits_{x'}^{x_i(0)}\omega_\ell~\int
\limits_{x_j(0)}^{x_j(t)}\left(\omega_{\ell'}+\overline{\omega_{\ell'}}
\,\right)+\int\limits_{x'}^{x_j(0)}\omega_\ell~\int
\limits_{x_i(0)}^{x_i(t)}\left(\omega_{\ell'}+\overline{\omega_{\ell'}}
\,\right)\right]\nn\\&&
\label{angleij}\eea
is the angle function for particles $i$ and $j$~\cite{TM_10} with
$x_i(0)=z_i$ and $x_i(1)=\bz_i$, and $\omega_\ell=\omega_\ell(z)~dz$,
$\ell=1,\dots,g$ form a basis of holomorphic one-differentials on the
Riemann surface $\Sigma$. The function (\ref{angleij})
changes by $2\pi$ whenever the worldline of particle $i$ wraps exactly
once around that of particle $j$, as this induces an adiabatical rotation of
coordinates $(z_i-z_j)\to\e^{2\pi i}\,(z_i-z_j)$ on $\Sigma$. For other
purely bulk linkings such as those associated with closed Wilson loops
in the interior of the membrane, one needs to use the Gauss linking
formula (\ref{WcorrS3}) explicitly. Again we stress that by the usage of
(\ref{WcorrS3}) to compute linking effects we will always be implicitly
assuming an appropriate surgery prescription. With this, we have
thereby arrived at a complete description of all gauge-invariant
vacuum states in the matter-coupled topologically massive gauge
theory, and hence at an exact, three-dimensional version of the chiral
algebra of the $c=1$ conformal field theory underlying the rational circle.

\subsection{The Verlinde Formula\label{VerlindeForm}}

In this subsection we will exploit the simplicity of the present
abelian model to describe some physical properties of the Verlinde
diagonalization formula for (\ref{dimcalHfinal})~\cite{V_1}. While technically
this formula unveils no surprises for the case of the rational circle,
we will interpret it here as a
non-trivial statement about the requirement of charge conservation in
linking processes among charged particles in the bulk. From the
perspective of vacuum Schr\"odinger wavefunctionals, the topological
Verlinde numbers express which combinations of the insertions give
wavefunctionals (\ref{Xiexcited}) with non-vanishing inner products,
and hence which are the true dimensions of the Hilbert spaces spanned
by the states $\Xi$.

Our starting point is with the topological wavefunctions at genus~1 which
are given by~\cite{TM_18}
\be
\Psi_\lambda(\tau)=\frac1{(k\tau_2)^{1/4}\,\eta(\tau)}~
\sum_{r=-\infty}^\infty\exp\left\{\frac{2\pi i\tau}k\,\left(
rp+\frac\lambda q\right)^2\right\} \ , ~~ \lambda\in\zed_{pq} \ ,
\label{Psigenus1}\ee
where the modular parameter $\tau$ is a complex number of imaginary
part $\tau_2>0$ and
\be
\eta(\tau)=\e^{\pi i\tau/12}\,\prod_{r=1}^\infty\left(1-\e^{2\pi i
    r\tau}\right)
\label{Dedekind}\ee
is the Dedekind function. Using the completeness of these states, we
define the $\Smatrix$-matrix through the modular transform
\be
\Psi_\lambda\left(-\frac1\tau\right)=\sum_{\lambda'=0}^{pq-1}
\Smatrix_\lambda^{~\lambda'}\,\Psi_{\lambda'}(\tau) \ .
\label{Psimodular}\ee
It can be computed explicitly by using the definition (\ref{k2pq}),
the modular transformation properties
\be
\left(-\frac1\tau\right)_2=\frac{\tau_2}{|\tau|^2} \ , ~~
\eta\left(-\frac1\tau\right)=\sqrt{-i\tau}~\eta(\tau) \ ,
\label{modtransfprops}\ee
and the Poisson resummation formula
\be
\sum_{r'=-\infty}^\infty\e^{-\pi hr'^2-2\pi ibr'}=\frac1{\sqrt h}~
\sum_{r=-\infty}^\infty\e^{-\pi(r-b)^2/h}
\label{Poisson}\ee
to get
\be
\Psi_\lambda\left(-\frac1\tau\right)=\frac1{\sqrt{pq}~(k\tau_2)^{1/4}\,
\eta(\tau)}~\sum_{r'=-\infty}^\infty\exp\left\{\frac{2\pi ir'}{pq}\,
\left(\frac{\tau r'}2+\lambda\right)\right\} \ .
\label{PsiPoisson}\ee
By writing $r'=\lambda'+pqr$ and summing (\ref{PsiPoisson}) over
$\lambda'=0,1,\dots,pq-1$ and $r\in\zed$, we arrive at
(\ref{Psimodular}) with
\be
\Smatrix_\lambda^{~\lambda'}=\frac1{\sqrt{pq}}~\e^{2\pi
  i\,\lambda\lambda'/pq} \ .
\label{Smatrix}\ee
Thus the modular transformation (\ref{Psimodular}) is just a
finite Fourier transform on the cyclic group $\zed_{pq}$. The $\Smatrix$-matrix
(\ref{Smatrix}) is symmetric and unitary,
\be
\sum_{\lambda'=0}^{pq-1}\Smatrix_\lambda^{~\lambda'}~
\Smatrix_{\lambda'\mu}^\dag=\delta_{[\lambda-\mu]} \ .
\label{Sunitary}\ee

The key property of (\ref{Smatrix}) within the present context is that
it can be defined in terms of a completely non-perturbative,
three-dimensional bulk process. Namely, it corresponds to the
statistical exchange phase between two charged particles whose bulk
trajectories $C$ and $C'$ are linked according to $\#(C,C')=+1$ and
$\#(C,C)=\#(C',C')=0$, as illustrated in
fig.~\ref{Hopflink}. This equality follows from the Gauss linking
formula (\ref{WcorrS3}) along with (\ref{k2pq}),
(\ref{rationalcharges}) and (\ref{Smatrix}) which determine the loop
correlator in fig.~\ref{Hopflink} as $\e^{4\pi
  i\,QQ'/k}=\Smatrix_{\lambda\lambda'}/\Smatrix_{0\lambda}$.
This expression for the topological invariant of the
Hopf link in $\sphere^3$ can be derived more generally by using a
surgery prescription~\cite{W_0}, in which a solid torus surrounding
the link is cut out of $\sphere^3$ and then glued back after
performing a modular transformation on its boundary.

\fig{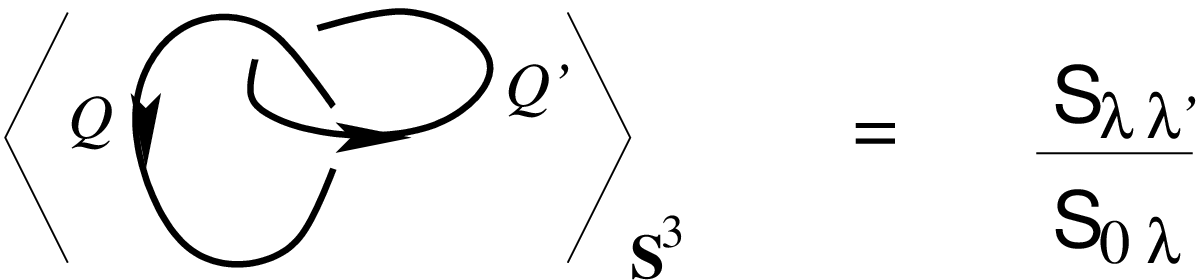}{The Hopf linking of two charges $Q=\lambda/q$ and
  $Q'=\lambda'/q$ in the bulk can be used to define the modular
  $\Smatrix$-matrix.}{Hopflink}

Let us now examine the fusion coefficients (\ref{fusioncoeffs}) for
the rational circle. By using (\ref{Zpqdeltafn}) they may be expressed
as a sum of statistical exchange phases from linked particle
trajectories in the bulk, which from (\ref{WcorrS3}) and
(\ref{Smatrix}) may each be written as
\be
\e^{4\pi i\,Q(Q_1+Q_2+Q_3)/k}=\frac{\Smatrix_{\lambda_1\lambda}\,
\Smatrix_{\lambda_2\lambda}\,\Smatrix_{\lambda_3\lambda}}
{(\Smatrix_{0\lambda})^3} \ .
\label{3ptphases}\ee
The fusion coefficients can thereby be written as
\be
N_{\lambda_1\lambda_2\lambda_3}=\sum_{\lambda=0}^{pq-1}
\frac{\Smatrix_{\lambda_1\lambda}\,\Smatrix_{\lambda_2\lambda}\,
\Smatrix_{\lambda_3\lambda}}{\Smatrix_{0\lambda}} \ ,
\label{Verlinde3pt}\ee
which is the celebrated Verlinde diagonalization for the three-punctured
Riemann sphere~\cite{V_1}. It implies, in particular, that the eigenvalues
$\nmatrix_{\lambda'}^{(\lambda)}$ of the fusion matrix
$(\Nmatrix_\lambda)_{\lambda'\mu}=N_{\lambda\lambda'\mu}$ are given by
\be
\nmatrix_{\lambda'}^{(\lambda)}=\frac{\Smatrix_{\lambda\lambda'}}
{\Smatrix_{0\lambda}} \ .
\label{fusioneigen}\ee
The statistical exchange phases (\ref{fusioneigen}) are discrete
characters of the $\zed_{pq}$ gauge subgroup of the
topologically massive gauge theory which themselves obey the fusion
algebra
\be
\nmatrix_{\lambda'}^{(\lambda)}\,\nmatrix_\mu^{(\lambda)}=
\sum_{\nu=0}^{pq-1}N_{\lambda'\mu}^{~~~\nu}\,\nmatrix_\nu^{(\lambda)}
\label{ellfusionalg}\ee
for all $\lambda\in\zed_{pq}$.

On the other hand, according to (\ref{dimcalHgsk}) and
(\ref{dimcalHfinal}), the fusion algebra of the rational circle is
determined by the three-point function of Polyakov loops on the
three-geometry $\sphere^2\times\sphere^1$~\cite{BT_1,GSST_1}. A
heuristic derivation of (\ref{Verlinde3pt}) is depicted schematically in
fig.~\ref{Verlinde}. Again it can be derived more generally via
surgery on $\sphere^2\times\sphere^1$~\cite{W_0}.

\fig{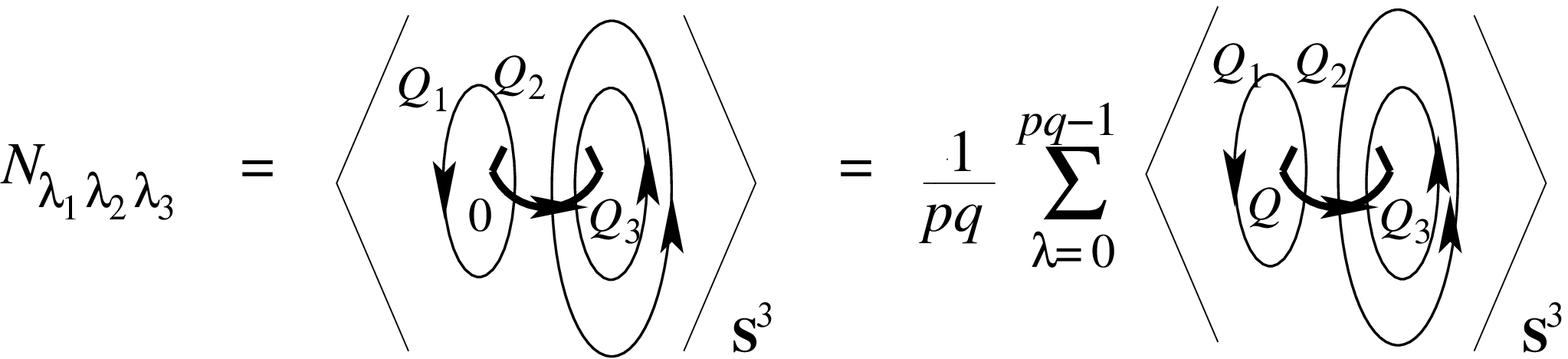}{Schematic derivation of the Verlinde formula for the
 three-punctured two-sphere. In the first equality we note that the puncturing
of $\Sigma=\sphere^2$ by Polyakov loops, as represented in
fig.~\ref{Wilsonlines}(b), can be interpreted as the single linking
of each of the three loops with a loop in the trivial $Q=0$
representation of the cyclic group $\zed_{pq}$. In the second equality
we use (\ref{Zpqdeltafn}) to replace this trivial loop by the sum over
a complete set of Polyakov loops. By (\ref{WcorrS3}), each term in
 this sum coincides with the linking phases (\ref{3ptphases}).}{Verlinde}

The general case (\ref{dimcalHfinal}) is treated similarly, giving the
general Verlinde formula
\be
\dim{\cal H}_{g,s,k}(\lambda_1,\dots,\lambda_s)=
\sum_{\lambda=0}^{pq-1}\left(\Smatrix_{0\lambda}\right)^{2-2g-s}\,
\prod_{i=1}^s\Smatrix_{\lambda_i\lambda}
\label{Verlindegen}\ee
for the dimension of the corresponding vacuum Hilbert space of
the three-dimensional gauge theory. Physically, we consider $s$ static
charges $Q_1,\dots,Q_s$ moving in the bulk of $\Sigma\times \sphere^1$ which
puncture the genus $g$ Riemann surface $\Sigma$ at fixed, distinct
points $x_1,\dots,x_s$. Such a punctured Riemann
surface can be represented by $2g+s-2$ ``pants'' (three-holed or
three-punctured Riemann spheres) glued together along the
holes. One then takes a sum over charge labellings of the internal
hole boundaries of the product of $N_{\lambda_1\lambda_2\lambda_3}$, one for
each pair of pants. For example, in the source-free case we have
\be
\dim{\cal
  H}_{g,0,k}=\sum_{\lambda=0}^{pq-1}~\sum_{\stackrel{\scriptstyle
\stackrel{\scriptstyle\lambda_1,\dots,\lambda_g}{\scriptstyle
\mu_1,\dots,\mu_g}}{\scriptstyle\nu_1,\dots,\nu_g}}N_{\lambda
\lambda_1}^{~~~\mu_1}\,N_{\mu_1\lambda_1}^{~~~~\nu_1}\,
N_{\nu_1\lambda_2}^{~~~~\mu_2}\,N_{\mu_2\lambda_2}^{~~~~\nu_2}
{}~\cdots~N_{\nu_{g-1}\lambda_g}^{~~~~~~\mu_g}\,
N_{\mu_g\lambda_g}^{~~~~\lambda} \ .
\label{Verlindenopunct}\ee
By using (\ref{Verlinde3pt}) and unitarity (\ref{Sunitary}) of the
$\Smatrix$-matrix, we then arrive at (\ref{Verlindegen}). It is in this way
that we may interpret the Verlinde formula dynamically as the
statement of overall charge conservation for arbitrary,
non-perturbative linking processes involving
charged matter in the bulk quantum field theory. In other words, the
formula (\ref{Verlindegen}) expresses the construction of the
three-dimensional wavefunctions (\ref{Xiexcited}), which carry
information about the physical spectrum of the induced string
theory, from different gluings of three-manifolds with boundaries by
pants.

\subsection{Ishibashi Wavefunctionals}

In this subsection we will construct a basis of wavefunctionals which
generate the three-dimensional counterparts of closed string, brane
boundary states of the rational circle. The basic idea is that the
diagonal, modular invariant torus partition function~\cite{TM_18,FS_4}
\be
Z^{\rm c}(\tau,\overline{\tau}\,)
=\sqrt k~\sum_{\lambda=0}^{pq-1}\Psi_\lambda(\tau)\otimes
\Psi^\dag_{\bar\lambda}(\,\overline{\tau}\,)
\label{toruspartfn}\ee
encodes information about the closed string Hilbert space ${\cal
  H}^{\rm c}$ in radial quantization of the two-dimensional
  $\sigma$-model on the infinite cylinder
  $\Sigma=\real\times\sphere^1$. It can be regarded as an
  element of the infinite-dimensional vector space
\be
{\cal H}^{\rm c}=\bigoplus_{\lambda=0}^{pq-1}\,[\phi_\lambda]\otimes
[\bar\phi_{\bar\lambda}] \ ,
\label{Hilbertclosed}\ee
where the conformal block $[\phi_\lambda]$ is regarded now (under the
operator-state correspondence) as the irreducible Virasoro module
built on the Fock state $|Q\rangle$ corresponding to the primary field
$\phi_\lambda=\e^{i\,Q\,\varphi}$. This identification makes use of
the inner products on $[\phi_\lambda]$ and $[\bar\phi_{\bar\lambda}]$
in the usual way.

With this in mind, let us consider the wavefunctionals
(\ref{Xiexcited}) of the topological membrane with
$\Sigma=\sphere_0^2$ the punctured sphere. Their charges $Q_i$
label the irreducible representations of the cyclic group $\zed_{pq}$,
and so they act on (one-dimensional) vector spaces spanned by the Fock
vacuum states $|Q_i\rangle$. Via application of the corresponding Virasoro
descendent fields, these vector spaces can be extended to the
modules $[\phi_{\lambda_i}]$. In other words, the wavefunctional
(\ref{Xiexcited}) is an operator acting on the product of the corresponding
representation spaces, so that
\be
\Xi^{\sphere_0^2}\left[A_z,A_\bz\,;\left\{\matrix{z_1&\cdots&z_s\cr
Q_1&\cdots&Q_s\cr}\right\};0\right]~\in~\bigotimes_{i=1}^s\,
\left[\phi_{\lambda_i}\right]^*\otimes\left[\phi_{\lambda_i}\right] \
{}.
\label{Xioprep}\ee
In the next section we will see how the membrane amplitudes also
incorporate the string descendent fields into the actions of the
operators in (\ref{Xioprep}).

Given this interpretation, we will now study the one-point operators
(\ref{Xioprep}) (with $s=1$) in some detail. The structure of a
vertical Wilson line operator $W_Q[A]$ corresponding to the bulk
propagation of a non-self-linking charge $Q$ from the right-moving
sector $\Sigma_0$ to the left-moving sector $\Sigma_1$ of the string
worldsheet depends crucially on what discrete symmetries we require of
the topologically massive gauge theory. Let us first consider the
simplest case whereby the bulk quantum field theory possesses the full
$\sf PCT$ invariance. Then from (\ref{T2}), and the fact that
the time inversion operation $\sf T$ reverses the orientation of the
contour $C$ defining the external particle worldline, it follows that
the $\sf PCT$ involution of the quantum field theory acts on vertical
Wilson lines as
\be
{\sf PCT}\,:\,W_Q[A]~\longmapsto~W_Q[A] \ .
\label{PTWQA}\ee
This implies that charge is conserved for external particles which
propagate along a Wilson line from $\Sigma_0$ to $\Sigma_1$,
i.e. $\bar Q=Q=m$. In this case, the charge non-conserving monopole
induced processes are suppressed, ${\cal
  W}_{\sphere_0^2\times[0,1]}=1$, and from (\ref{Xioprep}) it
follows that the one-punctured state (\ref{Xiexcited}) may be
regarded as an operator
\be
\Xi^{\sphere_0^2}\left[A_z,A_\bz\,;\left\{\matrix{z\cr
Q\cr}\right\};0\right]\,:\,[\phi_\lambda]~\longrightarrow~[\phi_\lambda] \ .
\label{Xiopdiag}\ee

By gluing together the left and right moving sectors as we did in
(\ref{PsiWilsonnonchiral}), we can then form the vacuum
wavefunctionals corresponding to $\sf PCT$-invariant states of the
membrane with the appropriate bulk propagation of external charges as
\be
\Phi^{\rm D}_m\left[A_z,A_\bz\,;z,\bz\,\right]=
\Xi^{\sphere_0^2}\left[A_z,A_\bz\,;\left\{\matrix{z\cr
Q\cr}\right\};0\right]\otimes\Xi^{\sphere_0^2}\left[A_z,A_\bz\,;
\left\{\matrix{\bz\cr\bar Q=Q\cr}\right\};0\right]^\dag \ .
\label{PsimD}\ee
Here we have made the identification $\Xi=\Xi_0\equiv\Xi_1$ as is
dictated by the orbifold symmetry (with respect to $\sf P$) which
identifies left and right movers $\Sigma_0\equiv\Sigma_1$. The
operator (\ref{PsimD}) acts only in the diagonal, left-right symmetric
product
$[\phi_\lambda]\otimes[\phi_{\bar\lambda=\lambda}]\subset{\cal H}^{\rm
  c}$, and as such is proportional to the orthogonal projection
$P_\lambda=\sum_l|\lambda,l\,\rangle\langle\bar\lambda=\lambda,l\,|$ onto
this subspace, with $|\lambda,l\,\rangle$, $l\in\zed_+$ the elements of the
corresponding number basis. By choosing the proportionality constant to be
one and using the inner product on $[\phi_{\bar\lambda}]$, it follows
that the homomorphism (\ref{PsimD}) on
$[\phi_{\bar\lambda}]\to[\phi_\lambda]$ is in a one-to-one
correspondence with the coherent state
\be
|m\rangle\!\rangle^{\rm D}=\sum_{l=0}^\infty|\lambda,l\,\rangle
\otimes U_{\sf P}\,U_{\sf C}|\lambda,l\,\rangle \ ,
\label{DIshibashi}\ee
where the anti-unitary operators $U_{\sf P}$ and $U_{\sf C}$
respectively implement the parity and charge conjugation
automorphisms on the right-moving Hilbert space with $U_{\sf
  P}\,U_{\sf C}|\lambda,l\,\rangle\in[\phi_\lambda]$. This vector is
recognized as the Dirichlet Ishibashi state for the $c=1$ Gaussian
model~\cite{I_1}, of Kaluza-Klein momentum $m$ about the compactified
direction. The set of states $|m\rangle\!\rangle^{\rm D}$,
$m\in\zed_{pq/2}$ span the space of boundary states of the rational
circle corresponding to Dirichlet boundary conditions on the open
string embedding fields. We shall therefore refer to (\ref{PsimD}) as
a Dirichlet Ishibashi wavefunctional. It is the unique vacuum state of
the topologically massive gauge theory which carries definite $U(1)$
charge $Q=m$ and which respects the $\sf PCT$ symmetry.

The analysis is similar in the case where only the $\sf PT$
sub-invariance of the quantum field theory is assumed. From (\ref{PCT}) it
follows that the action on vertical Wilson lines is now
\be
{\sf PT}\,:\,W_Q[A]~\longmapsto~W_{-Q}[A] \ ,
\label{PCTWilson}\ee
and there is a vortex operator of charge (\ref{DeltaQ}) which ruins
charge conservation in propagation along a Wilson line from $\Sigma_0$
to $\Sigma_1$, i.e. $\bar Q=-Q=-kn/4$. Again the one-punctured state
is the operator (\ref{Xiopdiag}) but, because of the non-trivial
monopole process, instead of (\ref{PsimD}) one must now form the vacuum
functional
\bea
\Phi_n^{\rm N}\left[A_z,A_\bz\,;z,\bz\,\right]&=&
{\cal W}_{\sphere_0^2\times[0,1]}(\Delta Q)~
\Xi^{\sphere_0^2}\left[A_z,A_\bz\,;\left\{\matrix{z\cr
Q\cr}\right\};0\right]\nn\\&&\otimes~\Xi^{\sphere_0^2}\left[A_z,A_\bz\,;
\left\{\matrix{\bz\cr\bar Q=-Q\cr}\right\};0\right]^\dag \ .
\label{PsinN}\eea
It coincides with the orthogonal projection onto the subspace
$[\phi_\lambda]\otimes[\phi_{\bar\lambda=-\lambda}]\subset{\cal
  H}^{\rm c}$, and therefore with the closed string coherent states
\be
|kn/4\,\rangle\!\rangle^{\rm N}=\sum_{l=0}^\infty|\lambda,l\,
\rangle\otimes U_{\sf P}|\lambda,l\,\rangle \ ,
\label{NIshibashi}\ee
where $U_{\sf P}|\lambda,l\,\rangle\in[\phi_{-\lambda}]$. These are just the
Neumann Ishibashi boundary states of the rational
circle, of vanishing total momentum along the compact dimension, and
hence we will call (\ref{PsinN}) the Neumann Ishibashi wavefunctionals
of the topological membrane. They are the unique vacuum states of
  definite charge which are invariant under three-dimensional $\sf
  PT$ transformations. The anticipated properties of closed string
Ishibashi states are thereby a consequence of boundary gauge
invariance of the bulk wavefunctions in three-dimensions.

\subsection{Boundary Couplings and Wilson Lines\label{BCOrb}}

For the remainder of this section we will analyse generic boundary
wavefunctionals which are expansions in the Ishibashi wavefunctionals
(\ref{PsimD}) and (\ref{PsinN}) of the form
\be
\psi^{~}_{\rm B}=\sum_m\Bmatrix_{\rm D}^m~\Phi_m^{\rm D}+
\sum_n\Bmatrix_{\rm N}^n~\Phi_n^{\rm N} \ .
\label{PsiBDN}\ee
We wish to determine what sorts of combinations (\ref{PsiBDN})
correspond to three-dimensional states of D-branes. In boundary
conformal field theory, such states would be constrained by various
sewing and locality conditions, the most important of which is the
celebrated Cardy condition~\cite{C_1}. The constants $\Bmatrix$ in
(\ref{PsiBDN}) characterize the boundary condition of the corresponding
two-dimensional state. We will only deal with those boundary
conditions that preserve the original $U(1)$ gauge symmetry of the
topological membrane. Then the sums in (\ref{PsiBDN}) run through all
$m,n\in\zed_{pq/2}$ for the case of the rational
circle. As we will see in the following, the three-dimensional
formalism naturally selects the ``fundamental'' branes that satisfy
the Cardy relations.

The key to this analysis will be a study of the behaviour of Wilson
line correlators of the bulk three-dimensional gauge theory after
application of the pertinent orbifold involutions. In this subsection
we shall make some general remarks about the explicit
construction of conformal boundary conditions from the membrane
perspective. For brevity we consider only the $\sf PT$ orbifold of the
topologically massive gauge theory and work to lowest order in string
perturbation theory. Consider a two-dimensional
boundary conformal field theory with a set of boundary conditions
labelled by $\alpha,\beta,\dots$. The basic
observation here is that a D-brane vertex operator insertion
corresponds to a change in boundary conditions from $\alpha$
to $\beta$ and that in terms of the full membrane picture it
is interpreted as a Wilson line insertion on the boundary when the
theory is subjected to Dirichlet boundary conditions,
as depicted in fig.~\r{fig.cplane}. The simplest counterpart of this
insertion in the full three-dimensional membrane corresponds to a
configuration with two Wilson lines and is pictured in fig.~\r{fig.orb2w}.

\fig{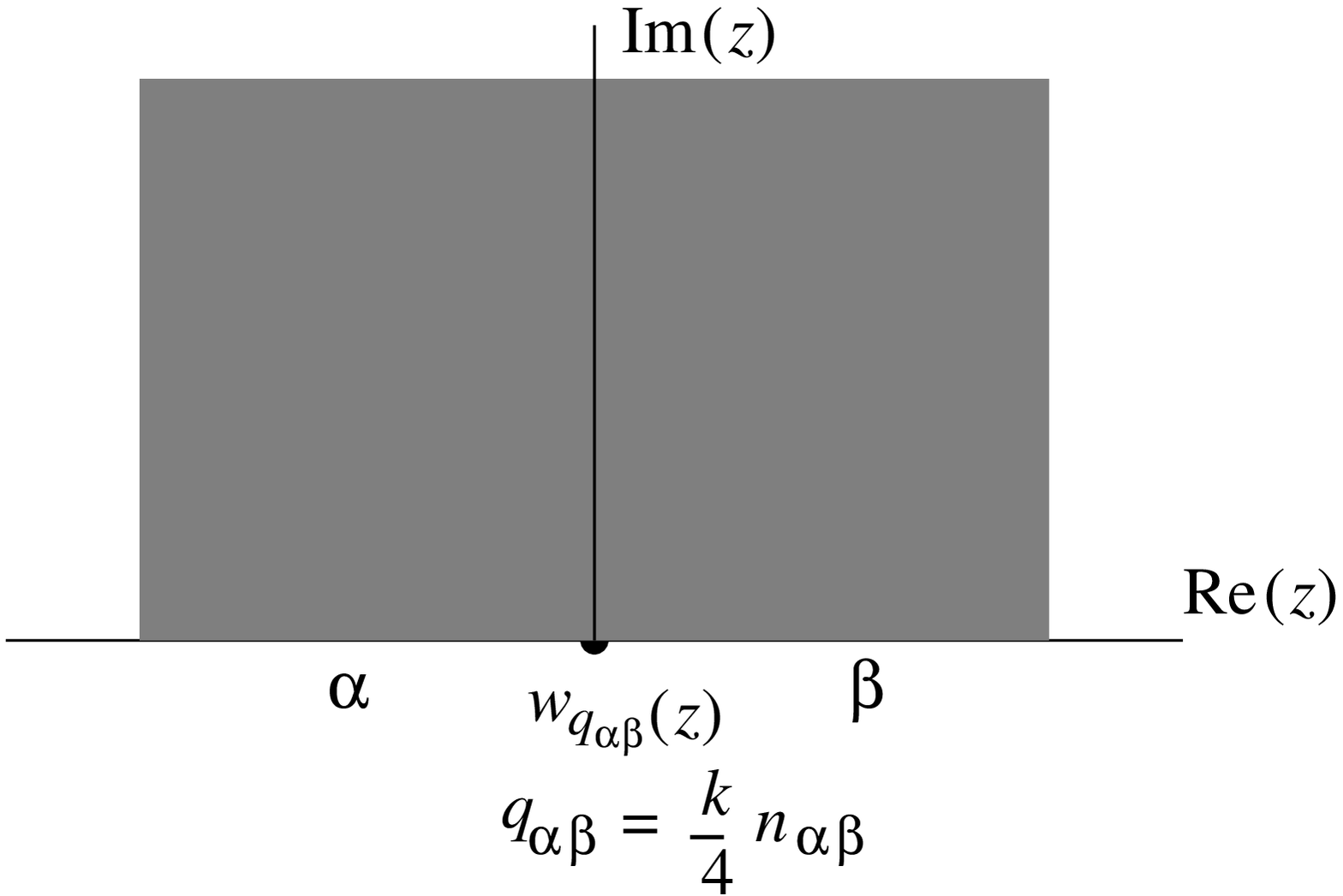}{A Wilson line insertion playing the role of a
  boundary vertex operator.}{fig.cplane}

\fig{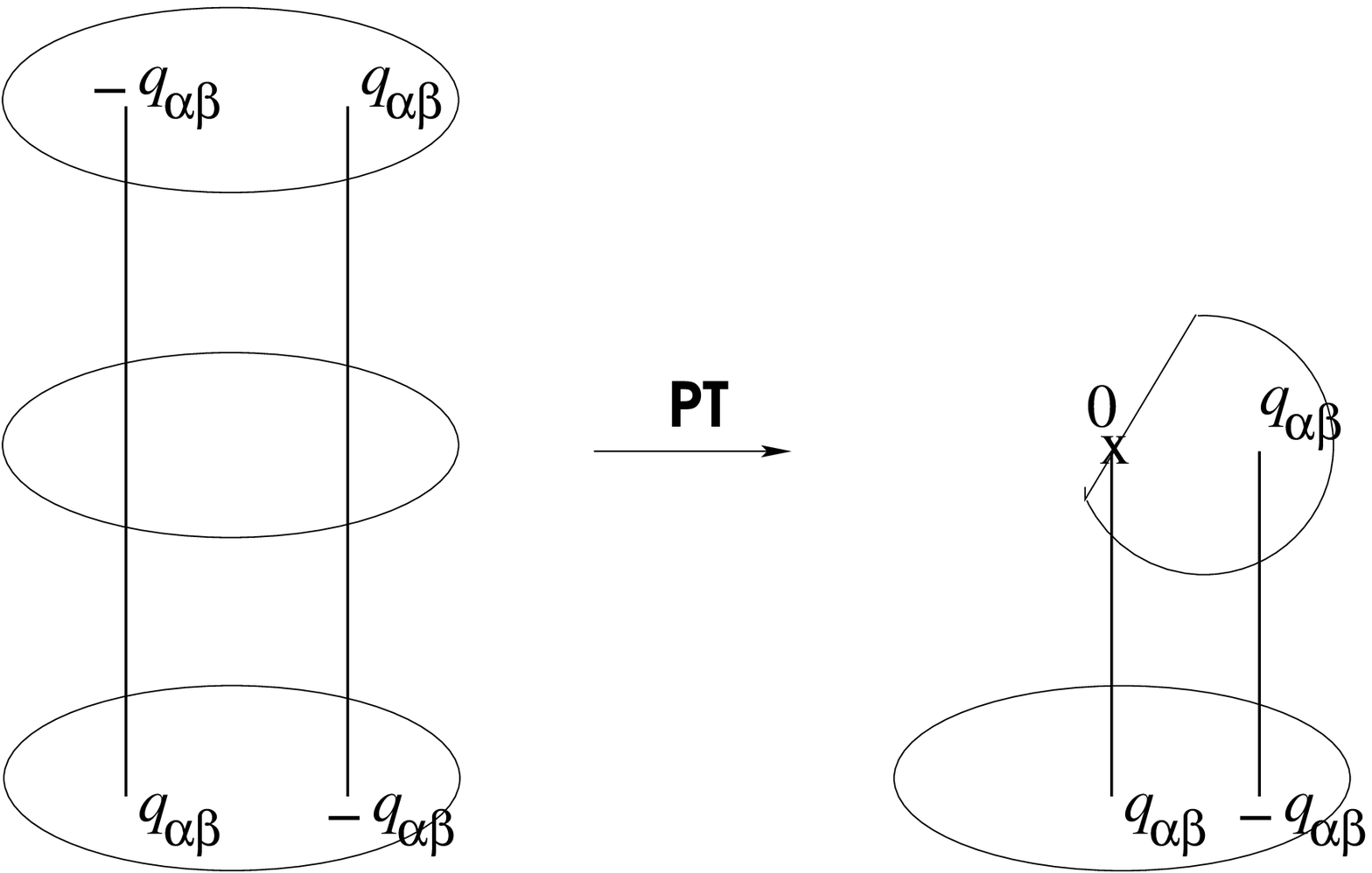}{The $\sf PT$ orbifold of the topological membrane in
  the presence of two Wilson lines with one boundary vertex
  insertion. The charges propagating along the Wilson lines are
  constrained by bulk charge conservation.}{fig.orb2w}

In the absence of boundary terms in the Schr\"odinger
wavefunctionals, the $\sf PT$ orbifold does not allow non-trivial
Wilson line insertions to live on the boundary of the string
worldsheet~\cite{TM_16}. However, they are now possible due to the
boundary terms that we have derived in the orbifold wavefunctions
which induce the vertex operators (\ref{Dvertexop}). The field $Y_{\rm
  D}$ must in this way have a \textit{discontinuity} at the insertion
point of the Wilson line which generates some charge
$q_{\alpha\beta}$. The full three-dimensional theory must generate a
mirror charge at that insertion point to screen it and its pair
somewhere in the bulk of $\Sigma^{\rm o}$. Then the new pair
``propagates'' through the bulk to the other boundary $\Sigma_0\equiv\Sigma_1$,
where the charges emerge with opposite signs due to the monopole and
linking processes described earlier (recall that here the orbifold charge
spectrum is of the form $kn/4$).

A natural assumption is that the discontinuity is generated
by a magnetic flux which is responsible for the appearence of charge
at the insertion point. As a toy model we can thereby take
\be
Y_{\rm D}(x)=2\pi n_\alpha+2\pi n_{\alpha\beta}\,\Theta(x-x_0) \ ,
\lb{Ytoy1}
\ee
where $x$ is the coordinate of the boundary $\partial\Sigma^{\rm o}$
and $x_0$ is the point of insertion on the boundary. Here
$\Theta$ denotes the usual Heaviside function, and
$n_{\alpha\beta}=n_\beta-n_\alpha$ with $n_\alpha$ and $n_\beta$
integers. The components of the corresponding current on
$\partial\Sigma^{\rm o}$ are then given by
\bea
j^{\parallel}(x)&=&\frac{k^2}{4\pi}\,n_{\alpha\beta}\,\delta(x-x_0)
\ , \nn\\\rho(x)&=&i\,k\,n_{\alpha\beta}\,\partial^{~}_\parallel
\delta(x-x_0) \ ,
\eea
and integrating the bulk Gauss law (\ref{GaussEB}) locally in a
neighbourhood $\sigma_{x_0}\subset\Sigma^{\rm o}$ of the insertion point
yields the total magnetic flux
\be
\int\limits_{\sigma_{x_0}}d^2z~B=\frac{4\pi}{k}\,\int
\limits_{\sigma_{x_0}}d^2z~\rho=2\pi n_{\alpha\beta} \ .
\ee
As required, the total flux is an integer multiple of $2\pi$, and the
integers $n_\alpha$ and $n_\beta$ are related to the decomposition of
the boundary states. Note again that the total charge in the bulk
theory vanishes, and only at the orbifold fixed point does a
non-trivial charge emerge. This same argument holds for the infinite
strip with a Wilson line insertion on one of the two boundaries.

However, the above picture is not quite complete, because here we are
working with compact manifolds, meaning that after orbifolding our
boundaries are always closed. In other words, instead of the complex
plane we should consider the parametrization of a disk, and instead of
an infinite strip we have an annulus. Thus we must consider at least
two vertex insertions, as depicted in fig.~\r{fig.disk}, such that
\be
Y_{\rm D}(x)=2\pi n_\alpha+2\pi n_{\alpha\beta}\,\Theta(x-x_1)-2\pi
n_{\alpha\beta}\,\Theta(x-x_2)
\ee
in order to induce the appropriate changes of boundary conditions. The
simplest counterpart in the full three-dimensional membrane
corresponds to a configuration with two Wilson lines ending at the
boundaries as illustrated in fig.~\r{fig.orb2wb}. Note that in the
absence of Wilson line insertions at the boundary, the boundary
condition is nonetheless fixed by a constant value for the field
$Y_{\rm D}$ on $\partial\Sigma^{\rm o}$. This same picture is valid
for the annulus diagram with each of the two boundaries treated
independently.

\fig{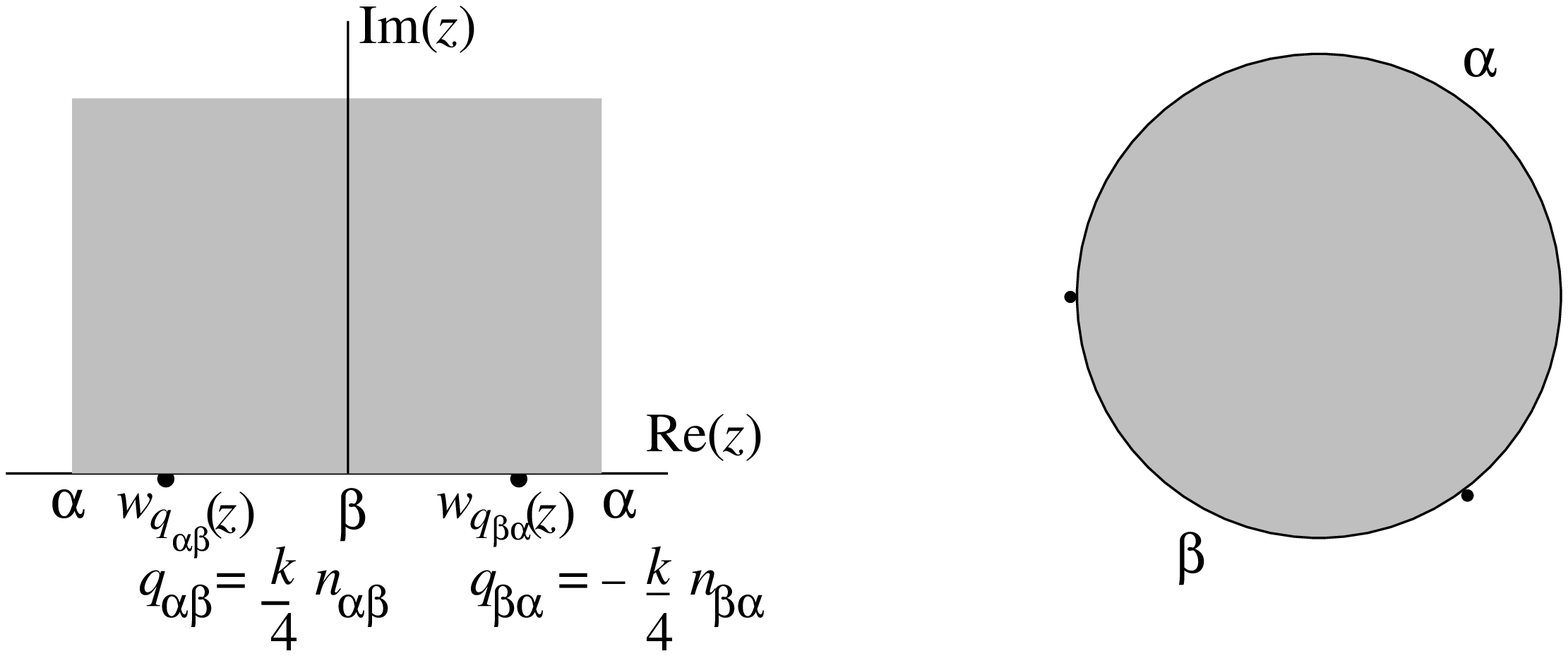}{Boundary conditions on a disk: Two-dimensional
  perspective.}{fig.disk}

\fig{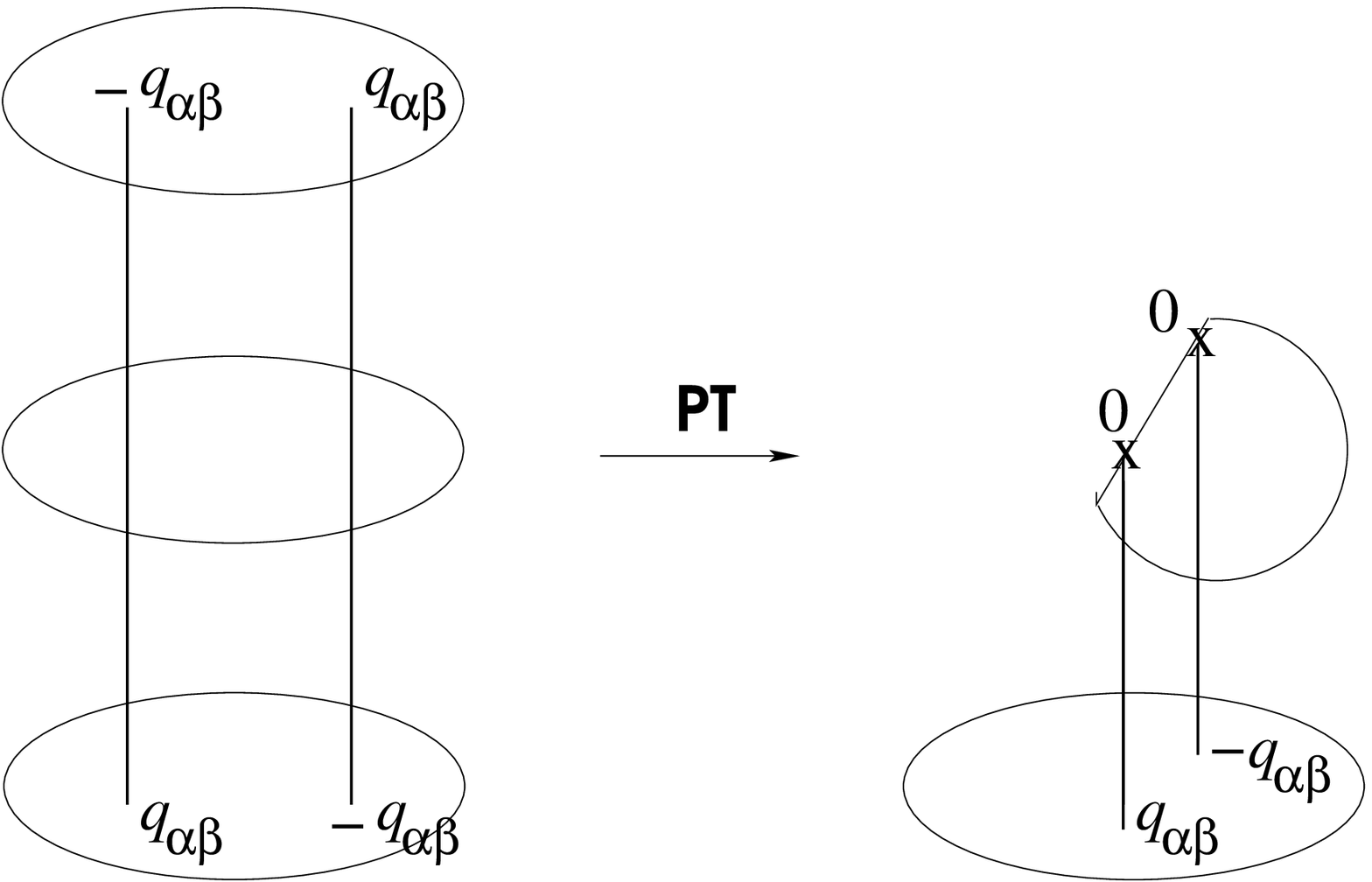}{Boundary conditions on a disk: Three-dimensional
  perspective.}{fig.orb2wb}

To quantify these arguments somewhat, let us consider the charge
insertions at the boundary of $\Sigma^{\rm o}$ as they appear in the
punctured wavefunctionals~(\r{Xiexcited}) and rewrite them as
boundary integrals over an effective charge distribution ${\cal
  Q}(x-x_0)$ which represents a discontinuity at the charge
insertion point in terms of the Heaviside function, as explained
above. Then in the presence of a D-brane vertex~(\r{Dvertexop}) we
get the total exponential factor
\be
{\cal V}_{\rm D}~\exp\Bigl\{i\,Q\,\varphi(x_0)\Bigr\}=
\exp\left\{~\oint\limits_{\partial\Sigma^{\rm o}}dx~
\left[{\cal Q}(x-x_0)-\frac{k}{4\pi}\,Y_{\rm D}(x)\right]
\,\partial^\perp\varphi(x)\right\} \ .
\ee
By shifting the brane collective coordinate as $Y_{\rm D}(x)\to Y_{\rm
  D}(x)+(4\pi/k)\,{\cal Q}(x-x_0)$, we recover the standard D-brane
vertex operator (\ref{Dvertexop}) but now $Y_{\rm D}(x)$ has a
discontinuity at $x=x_0$ as in~(\r{Ytoy1}). The same constructions can
be carried out for multiple Wilson line insertions.

\subsection{Orbifold Correlation Functions}

{}From the analysis of the previous subsection we may deduce an
important property of the computation of correlators of the orbifold
gauge theory, which may be regarded as a three-dimensional version of
the bulk-boundary correspondence of two-dimensional conformal field
theory~\cite{CL_1}. This correspondence is based on the fact that each
bulk point of an open surface $\Sigma^{\rm o}=\Sigma/\zed_2$ has {\it
  two} (mirror) pre-images on its Schottky double $\Sigma$, and it
states that the $s$-point correlators on $\Sigma^{\rm o}$ are in a one-to-one
correspondence with the chiral $2s$-point correlation functions on
$\Sigma$. The interaction of local fields with the
boundary of $\Sigma^{\rm o}$ (in the form of boundary conditions on
$\partial\Sigma^{\rm o}$) is then simulated by the interaction between
mirror images of the same holomorphic field on $\Sigma$, carrying
conjugate primary charges $\lambda,\bar\lambda=\pm\,\lambda$. The basic
example of this correspondence in the membrane picture is provided by
the orbifold partition function (\ref{Zodef}) at genus $g$ obtained from
(\ref{partfnTM}), which reads
\be
Z^{\rm o}(\Gamma)=k^{g/4}\,\sum_{\stackrel{\scriptstyle\vec\lambda
\in(\zed_{pq/2})^g}{\scriptstyle\vec{\bar\lambda}=\pm\,\vec\lambda}}
\Psi_{\vec\lambda}^{\rm orb}(\Gamma) \ ,
\label{partfnTMorb}\ee
where $\Psi_{\vec\lambda}^{\rm orb}(\Gamma)$ are the reduced
topological wavefunctions whose quantum numbers $\vec\lambda$ are
either purely winding or momentum integers according to which of the
$\sf PT$ or $\sf PCT$ orbifolds of the quantum field theory has been
taken.

More generally, let us assume that the boundary of $\Sigma^{\rm o}$
consists of $b$ connected components $C_\alpha'\cong\sphere^1$,
$\alpha=1,\dots,b$, so that
\be
\partial\Sigma^{\rm o}=\coprod_{\alpha=1}^bC_\alpha' \ .
\label{Sigmaocomps}\ee
The pre-image of each $C_\alpha'$ on the double cover $\Sigma$ is a
$\zed_2$-invariant, equatorial loop. It therefore corresponds to a
Wilson loop on a chosen time slice in the covering cylinder
$\Sigma\times[0,1]$, which becomes a circle of singular points in the
corresponding three-dimensional orbifold~\cite{H_1,FS_1,FS_2}. This
fact follows immediately from comparison of the orbifold wavefunctionals
(\ref{Psi0orb}) and (\ref{Psi12orb}), and from the remarks made in the
last subsection. In particular, as each component $C_\alpha'$
corresponds to the bulk propagation of an external $U(1)$ particle, it
carries an integral charge $Q_\alpha'$, again dependent on the type of
orbifold taken.

A generic vertical Wilson line correlator in the orbifold theory with
the boundary values $Q_1',\dots,Q_b'$ of the field $Y=Y_{\rm D}$ or
$Y_{\rm N}$ on $\partial\Sigma^{\rm o}$ may then be computed by using
those of section~\ref{WilsonCorr} through the prescription
\bea
\left\langle\,\prod_{i=1}^sW_{Q_i}\right\rangle_{\rm orb}^{Q_1'
\cdots Q_b'}&=&\prod_{i,j=1}^s\e^{\frac ik\,Q_iQ_j\,\bigl(
\theta_{ij}(1/2)-\theta_{ij}(0)\bigr)}~\int[DA_z~DA_\bz\,]~
\e^{2iS_{\rm TMGT}^{\rm orb}[A]}\nn\\&&\times\,\Xi_{1/2}^{\rm orb}
\left[A_z,A_\bz\,;\left\{{\matrix{\bz_1&\cdots&\bz_s\cr
\bar Q_1=\pm\,Q_1&\cdots&\bar Q_s=\pm\,
Q_s\cr}}\right\};\,Y\Bigm|_{C_\alpha'}=4\pi
\lambda_\alpha'\right]^\dag\nn\\&&\times\,{\cal W}_{\Sigma\times[0,1]}
\left(\Delta\vec Q_{\vec\lambda}\,\right)~\Xi_0^{\rm orb}\left[A_z,A_\bz\,;
\left\{{\matrix{z_1&\cdots&z_s\cr Q_1&\cdots&Q_s\cr}}\right\};0\right] \ .
\label{WQicorrorb1}\eea
This is of course just the orbifold inner product
$\langle\Xi_{1/2}|\Xi_0\rangle^{~}_{\rm orb}$ defined in (\ref{Zodef})
on the wavefunctional (\ref{PsiWilsonnonchiral}) with the prescribed
boundary values of the external currents. By using a finite
temperature correspondence completely analogous to that of
(\ref{tempmembrane},\ref{tempmembranefinal}) in the presence of
Polyakov loops, we may evaluate (\ref{WQicorrorb1}) as the ordinary
Wilson line correlator
\be
\left\langle\,\prod_{i=1}^sW_{Q_i}\right\rangle_{\rm orb}^{Q_1'
\cdots Q_b'}=\left\langle\,\prod_{\alpha=1}^b
W_{Q_\alpha'}~\prod_{i=1}^s{\cal W}_{\Sigma\times
[0,1]}(\Delta Q_i)~W_{Q_i}\,W_{\bar Q_i=\pm\,
Q_i}\right\rangle_{\Sigma\times\sphere^1} \ ,
\label{WQicorrorb2}\ee
where it is understood that the weight ${\cal
  W}_{\Sigma\times[0,1]}(\Delta Q_i)$ is unity whenever the bulk
  monopole processes associated with particle $i$ are suppressed,
  i.e. $\bar Q_i=Q_i$. While this correspondence doesn't completely
  determine the functional dependence of the correlators on
  $\Sigma^{\rm o}$, it suffices to determine the relationships between
  the conformal blocks corresponding to the left and right hand sides of
(\ref{WQicorrorb2}). We have also assumed that the Wilson
line insertions $W_{Q_i}$ only pierce the interior of the open string
worldsheet $\Sigma^{\rm o}$ as illustrated in fig.~\ref{Wilsonorb}, so
that the left-hand side of (\ref{WQicorrorb2}) corresponds to a bulk
insertion of primary fields in the boundary conformal field
theory. In order to properly account for the topological dependences
  of the correlation functions, one would need to further specify an
  appropriate homological truncation to the Lagrangian subspace
  (\ref{kericalL}). This will not be required in the following.

\fig{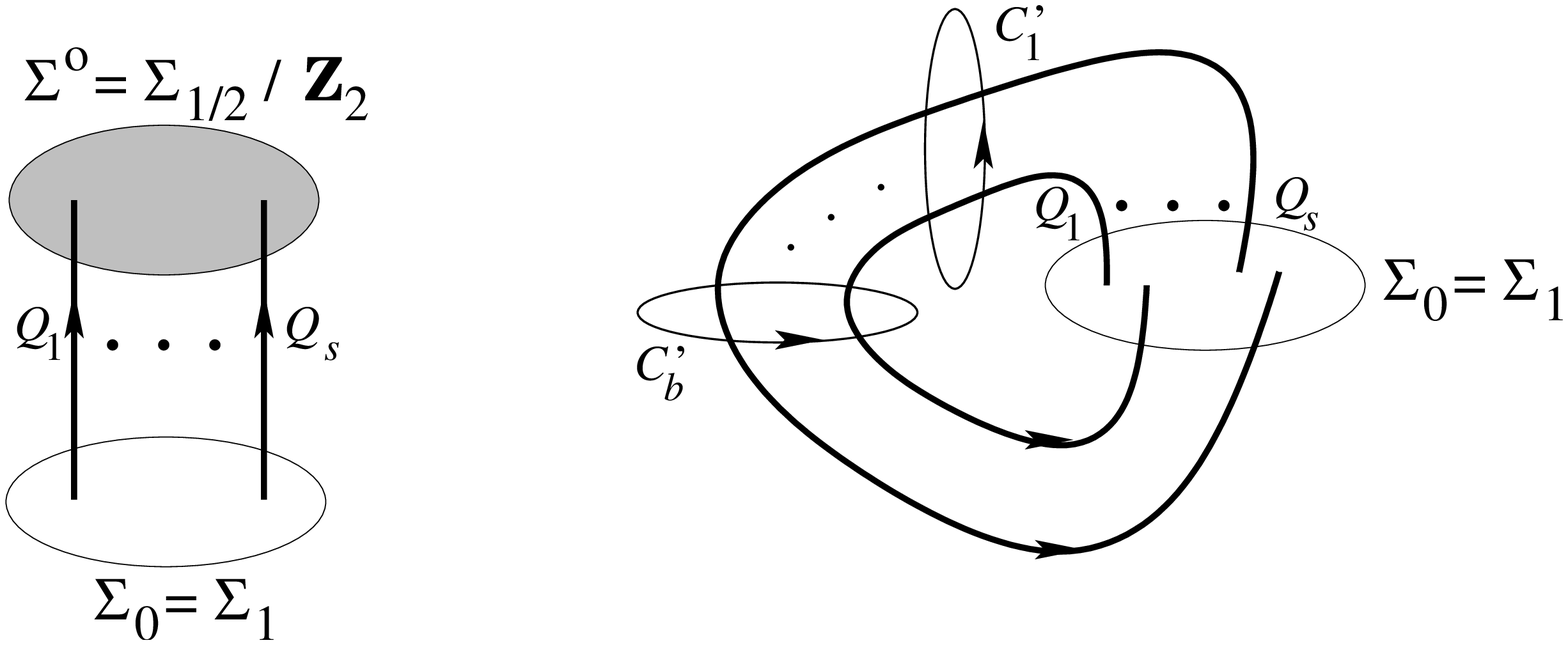}{The two equivalent representations of the orbifold
    Wilson line correlator which defines a bulk $s$-point function on
    the open string worldsheet. In the second picture the holomorphic
    and antiholomorphic sectors of the covering closed string
    worldsheet are identified as in fig.~\ref{Wilsonlines}(b), with
    the two ends of each Wilson line interpreted as independent chiral
    insertions on $\Sigma$. Thus each orbifold Wilson line is
    associated with two Polyakov loops carrying external charges $Q_i$
    and $\bar Q_i=\pm\,Q_i$ (not shown explicitly here), which are
    constrained by the discrete symmetries which are gauged to perform
    the orbifold of the topologically massive gauge theory.}{Wilsonorb}

As is seen from (\ref{Psi12orb}) and discussed above, the equality
(\ref{WQicorrorb2}) can be understood from the fact that the
branch point locus of the orbifold is equivalent to point-like
insertions of curvature singularities for the gauge field in the
quantum theory~\cite{H_1}. In addition, the charges of these closed
Wilson loops have the very natural interpretation as boundary conditions on the
open string worldsheet. This provides a very simple and elegant
derivation of the well-known result that the allowed conformal
boundary conditions correspond to specific integral reductions of the
primary charges of the rational circle, i.e. that the boundary
conditions are in a one-to-one correspondence with the primary fields
of the chiral conformal field theory. The equality
(\ref{WQicorrorb2}) now incorporates the worldsheet boundary effects
through the linkings of the vertical Wilson lines with the closed
boundary loops, as depicted in fig.~\ref{Wilsonorb}. The case of the
sphere $\Sigma=\sphere^2$ with orbifold the disk $\Sigma^{\rm o}=\disc^2$
is depicted in fig.~\ref{fig.sphereorb}. In the presence of
Wilson lines in the full three-dimensional membrane, each sphere
living at each fixed time slice will be pierced and each of these piercings
constitutes a new boundary. Consider two Wilson lines of charges $q$
and $q_c$ piercing $\Sigma_{1/2}$ at points $z$ and $z_c$ which are
related by the action of the orbifold group
generator~\cite{TM_16}. The piercing is considered to live in the bulk
of the disk as pictured in fig.~\r{fig.sphereorb}.

\fig{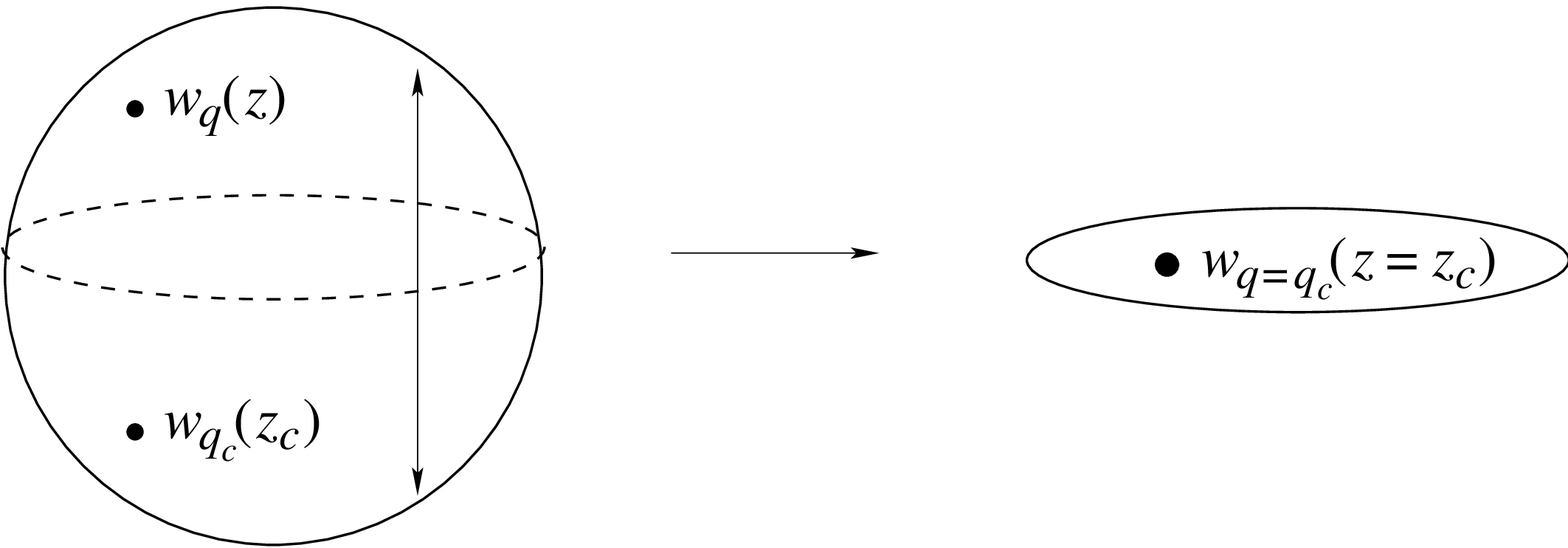}{Orbifold of $\Sigma_{1/2}=\sphere^2$ under
${\mathbb{Z}}_2$, whose action yields the open string worldsheet
$\Sigma^{\rm o}=\Sigma_{1/2}/\zed_2=\disc^2$, with Wilson line
insertions.}{fig.sphereorb}

As in section~\ref{VerlindeForm}, the linking of external particle $i$
with the boundary component $C_\alpha'$ of (\ref{Sigmaocomps})
produces the statistical phase factor $\e^{4\pi
  i\,Q_iQ_\alpha'/k}=\Smatrix_{\lambda_i\lambda_\alpha'}/\Smatrix_{0
    \lambda_\alpha'}$, so that in this simple case the boundary
  effects may be unravelled to write the orbifold correlator
  (\ref{WQicorrorb2}) as
\be
\left\langle\,\prod_{i=1}^sW_{Q_i}\right\rangle_{\rm orb}^{Q_1'
\cdots Q_b'}=\prod_{\alpha=1}^b\frac1{
\Smatrix_{0\lambda_\alpha'}}\,\left\langle\,\prod_{i=1}^s
\Smatrix_{\lambda_i\lambda_\alpha'}~{\cal W}_{\Sigma\times
[0,1]}(\Delta Q_i)~W_{Q_i}\,W_{\bar Q_i=\pm\,Q_i}\right
\rangle_{\Sigma\times\sphere^1} \ .
\label{WQiunravel}\ee
Let us stress again that in (\ref{WQiunravel}) the allowed particle
charges $Q_i,\bar Q_i=\pm\,Q_i$ and $Q_\alpha'$ are prescribed by the
discrete $\sf PT$ or $\sf PCT$ symmetries in three-dimensions. This
formula is the key identity which will allow us to completely specify
the membrane states of D-branes.

\subsection{Fundamental Wavefunctionals\label{FundWave}}

We are now ready to analyse the expansion (\ref{PsiBDN}) into
Ishibashi wavefunctionals. Branes in the $\sphere^1$ background are
completely characterized by their couplings to closed string modes,
and these couplings are encoded in the expansion coefficients
$\Bmatrix$ of (\ref{PsiBDN}). Not all linear combinations of Ishibashi
states can couple to the bulk conformal field theory. To determine the
appropriate couplings, we will compute the one-point functions of bulk
fields on the interior of the disk $\disc^2$.

Consider the propagation of a charge $Q$ from the Schottky double
$\sphere^2$ to the orbifold $\disc^2$ whose boundary carries a charge
$Q'$. According to the general relation (\ref{WQiunravel}), the
orbifold Wilson line correlator can be represented in terms of a bulk
Polyakov loop correlator in the finite temperature gauge theory as
\be
\Bigl\langle W_Q\Bigr\rangle^{Q'}_{\rm orb}=\frac{\Smatrix_{\lambda
\lambda'}}{\Smatrix_{0\lambda'}}\,\Bigl\langle
{\cal W}_{\sphere^2\times[0,1]}(\Delta Q)~W_Q\,W_{\bar Q=\pm\,Q}
\Bigr\rangle_{\sphere^2\times\sphere^1} \ .
\label{disc1pt}\ee
According to (\ref{WQicorrorb1}), the correlator on the right-hand side
of (\ref{disc1pt}) coincides with the membrane inner product
(\ref{Zcdef}) on Ishibashi wavefunctionals as
\be
\Bigl\langle W_Q\Bigr\rangle^{Q'}_{\rm orb}=\frac{\Smatrix_{\lambda
\lambda'}}{\Smatrix_{0\lambda'}}\,\Bigl\langle1
\Bigm|\Phi_\lambda^{\rm B}\Bigr\rangle \ ,
\label{disc1ptIsh}\ee
where ${\rm B}={\rm D},{\rm N}$ labels the particular type of Ishibashi
wavefunctional (\ref{DIshibashi}) or (\ref{NIshibashi}) obtained from
the given $\sf PCT$ or $\sf PT$ symmetry of the three-dimensional
gauge theory. These latter states are defined on the punctured sphere
$\sphere^2_0$, whereby a fictitious screening charge is present at the
puncture in order to render the right-hand side of (\ref{disc1ptIsh})
non-vanishing, as explained in section~\ref{BCOrb}. Such screenings
will always be implicitly assumed in the following.

Consider now a generic membrane state $\psi_{\lambda'}^{\rm B}$ of the
geometry $\sphere_0^2\times[0,1]$ which is labelled by the boundary
charge $Q'$, and which is invariant under the given orbifold
involution, i.e. an expansion in the appropriate Ishibashi
wavefunctionals as
\be
\psi_{\lambda'}^{\rm B}=\sum_{\stackrel{\scriptstyle\mu\in\zed_{pq/2}}
{\scriptstyle\bar\mu=\pm\,\mu}}\Bmatrix_{\lambda'}^{{\rm B}~\mu}~
\Phi_\mu^{\rm B} \ .
\label{genericIsh}\ee
We wish to consider in particular those wavefunctionals
$\psi_{\lambda'}^{\rm B}$ which are ``fundamental'', in the sense that
their membrane inner products are determined by the trivial $Q=0$
Wilson line in the orbifold theory with boundary condition $Q'$. This
is achieved via the insertion of a complete set of orbifold Wilson
lines, and by using (\ref{disc1ptIsh}) we find
\be
\Bigl\langle1\Bigm|\psi_{\lambda'}^{\rm B}\Bigr\rangle~=~
\left\langle\,\sum_{Q\,:\,
\bar Q=\pm\,Q}W_Q\right\rangle^{Q'}_{\rm orb}~=~
\sum_{\stackrel{\scriptstyle\lambda\in\zed_{pq/2}}
{\scriptstyle\bar\lambda=\pm\,\lambda}}\frac{\Smatrix_{\lambda
\lambda'}}{\Smatrix_{0\lambda'}}\,\Bigl\langle1
\Bigm|\Phi_\lambda^{\rm B}\Bigr\rangle \ .
\label{psilambdacompl}\ee
{}From (\ref{psilambdacompl}) it follows that the coupling coefficients
of (\ref{genericIsh}) are then given by
\be
\Bmatrix_{\lambda'}^{{\rm B}~\lambda}=\frac{\Smatrix_{\lambda\lambda'}}
{\Smatrix_{0\lambda'}} \ ,
\label{CardyB}\ee
and we have thereby arrived at a remarkably simple derivation of the
celebrated Cardy solution (up to an irrelevant normalization by
$\sqrt{\Smatrix_{0\lambda'}}\,$) of the sewing constraints in boundary
conformal field theory~\cite{C_1}. As before, a more general argument uses an
appropriate surgery prescription~\cite{FS_2}.

The couplings (\ref{CardyB}) (trivially) obey the factorization
constraints
\be
\Bmatrix_\mu^{{\rm B}~\lambda}~\Bmatrix_\nu^{{\rm B}~\lambda}=
\sum_{\stackrel{\scriptstyle\lambda'\in\zed_{pq/2}}
{\scriptstyle\bar\lambda'=\pm\,\lambda'}}N_{\mu\nu}^{~~\lambda'}~
\Fmatrix_{\lambda'1}\left[\matrix{\mu&\nu\cr\mu&\nu\cr}\right]~
\Bmatrix_{\lambda'}^{{\rm B}~\lambda} \ ,
\label{factconstr}\ee
where $\Fmatrix$ are the fusing matrices of the rational circle which
are given by
\bea
\Fmatrix_{\lambda\lambda'}\left[\matrix{\mu&\nu\cr\rho&\sigma\cr}
\right]&=&\delta_{[\mu+\nu+\sigma-\rho]}~\delta_{[\nu+\sigma-\lambda]}
{}~\delta_{[\mu+\nu-\lambda']}\nn\\&&\times\,\exp\left\{
\frac{\pi i\,(\mu+\sigma+1)}{pq}\,\left[\nu\Bigl(\mu+\nu+\sigma
-[\mu+\nu+\sigma]\Bigr)\right.\right.\nn\\&&+
\left.\left.(\nu+\sigma)\Bigl(\mu+2\nu+\sigma-
[\mu+\nu]-[\nu+\sigma]\Bigr)\right]\right\} \ .
\label{fusing}\eea
They may be defined in terms of the interactions between Wilson lines
in the bulk three-dimensional gauge theory, as depicted schematically
in fig.~\ref{fusion}. The classifying algebra (\ref{factconstr})
thereby completely determines the brane moduli in the form
(\ref{CardyB}), whose physical interpretation now depends on the type
of orbifold taken.

\fig{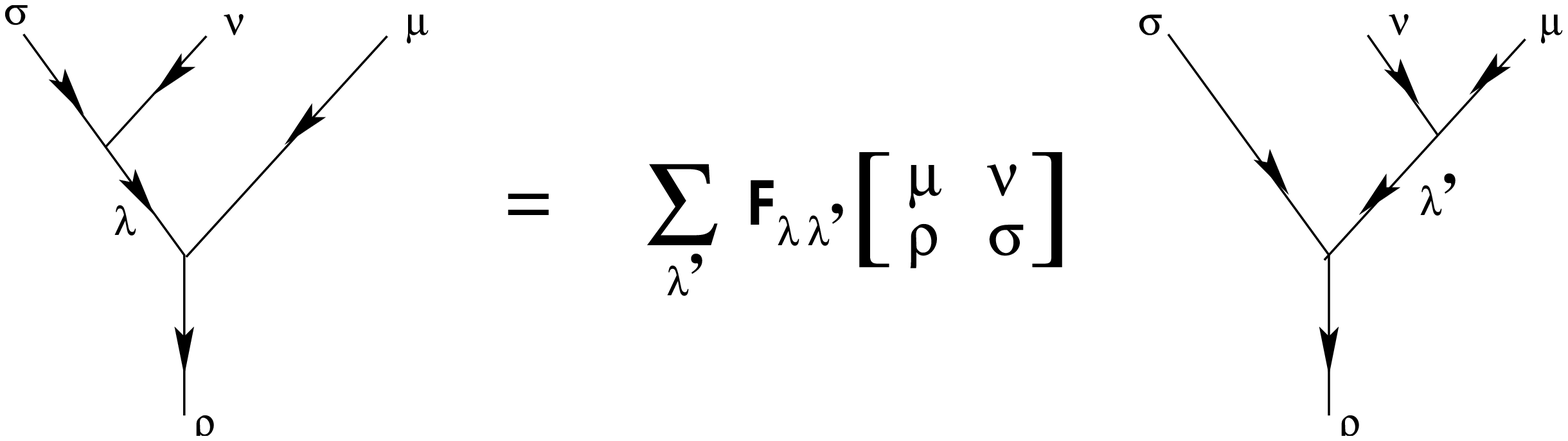}{The fusing matrices $\Fmatrix$ are defined through
  the worldline junctions associated with the charge conserving
  interactions of four charged particles in the bulk labelled by
  $\mu,\nu,\rho,\sigma$. They represent
  a cross channel duality in the intersections of the corresponding
  Wilson lines and are related to the statistical exchange phases that
  arise from the linkings of the charged particle trajectories.}{fusion}

For ${\rm B}={\rm D}$, by using (\ref{Qm0}), (\ref{CScouplingid}),
(\ref{k2pq}) and (\ref{genericIsh}) we thereby find that the most
general fundamental, $U(1)$ gauge symmetry preserving and $\sf
PCT$-invariant membrane wavefunctionals are given by
\be
\psi_a[A_z,A_\bz\,;z,\bz\,]=\sum_{m=0}^{pq/2-1}\e^{i\,ma/R}~
\Phi_m^{\rm D}[A_z,A_\bz\,;z,\bz\,] \ ,
\label{genDwave}\ee
where
\be
a=\frac{2\pi lqR}p \ , ~~ l\in\zed_{pq/2} \ .
\label{Dbranepos}\ee
The state (\ref{genDwave}) can be given the following physical
interpretation. Let us compute the inner product of (\ref{genDwave})
with the action of an unknotted vertical Wilson line $W_{m'}[A]$
corresponding to the propagation of a charge $m'\in\zed_{pq/2}$ from
$\Sigma_0$ to $\Sigma_1$. By using the explicit forms (\ref{Xiexcited}) of the
Ishibashi wavefunctionals (\ref{PsimD}), and the fact that the effect
of the Wilson line $W_{m'}[A]$ is to induce the $\sphere^1$-valued operators
$\phi_{m'}=\e^{i\,m'\,\varphi}$ on the string worldsheet, we find
\bea
\left\langle1\left|W_{m'}\,\psi_a\right.\right
\rangle&=&\sum_{m=0}^{pq/2-1}\e^{i\,ma/R}~
\left\langle1\left|W_{m'}\,\Phi_m^{\rm D}\right.\right
\rangle\nn\\&=&\sum_{m=0}^{pq/2-1}\e^{i\,ma/R}~
\left\langle1\left|\Phi_{[m+m']}^{\rm D}\right.\right
\rangle~=~\e^{-i\,m'a/R}~\left\langle1\left|\psi_a\right.\right\rangle \ .
\label{ainterpret}\eea
We interpret (\ref{ainterpret}) to mean that as a particle propagates
through the bulk of the membrane in the presence of the particle state
described by (\ref{genDwave}), it undergoes a non-local interaction
with this particle defect and acquires a statistical phase factor
$\e^{-4\pi i\,m'm/k}$ from linking. In the worldsheet picture, this
defect clearly corresponds to a D-brane situated
at the point (\ref{Dbranepos}). Note that in this case the brane
moduli (\ref{CardyB}) are $p^{\rm th}$ roots of unity, and so the
positions of the D-branes are restricted to lie at the vertices of a
regular $p$-gon.

For the case ${\rm B}={\rm N}$, from (\ref{Q0n}) it follows that the
most general fundamental, gauge symmetry preserving and $\sf
PT$-invariant membrane states are
\be
\psi_w[A_z,A_\bz\,;z,\bz\,]=\sum_{n=0}^{pq/2-1}\e^{i\,wnR}~
\Phi_n^{\rm N}[A_z,A_\bz\,;z,\bz\,] \ .
\label{genNwave}\ee
They describe Neumann branes which carry Wilson lines
\be
w=\frac{2\pi lq}{pR} \ , ~~ l\in\zed_{pq/2} \ .
\label{NbraneWilson}\ee
In this case the target $\sphere^1$ Wilson line is induced by those of
charged particles in the bulk of the topological membrane.

Let us stress again that the particular coupling coefficients (\ref{CardyB}),
which correspond to the statistical exchange phase between two Hopf
linked charged particles, arise from the assumption that the brane
states (\ref{genericIsh}) are compatible with the full $U(1)$ gauge
symmetry of the topological membrane. This property was captured by the first
equality of (\ref{psilambdacompl}), which took into account the sum
over all $U(1)$ charges. It is possible in this simple case to find
other expansions (\ref{genericIsh}) which preserve a smaller symmetry
group and still obey the necessary conformal factorization constraints
(\ref{factconstr}). These rational boundary symmetries would then
restrict the D-brane moduli to a smaller subset of those found
above. However, these branes do not constitute very natural
objects from the point of view of the underlying topologically
massive gauge theory, and within the present framework we may use
gauge invariance as a guiding principle to conclude that the special
branes found above account for {\it all} the relevant branes of the
topological membrane.

\subsection{The Cardy Condition\label{Cardy}}

In the previous subsection we saw that the closed string coupling
coefficients of membrane wavefunctionals were determined unambiguously
by the requirement of gauge invariance. We recall that a similar
principle was used in the derivation of the Verlinde formula (see
fig.~\ref{Verlinde}). This suggests that there could be a natural
relationship between them. In boundary conformal field theory this
connection is known as the Cardy condition~\cite{C_1}. Here we shall
investigate how it arises as a natural feature of the membrane
formulation of the open string sector, and thereby provide a nice
consistency check of the present formalism.

For this, we will study the one-loop, open string annulus
amplitude. The Schottky double of the annulus is the torus of purely
imaginary modulus $\tau=i\,\tau_2$, and the Neumann partition function
is obtained from (\ref{toruspartfn}) by the orbifold with respect to
the $\sf PCT$ symmetry of the three-dimensional gauge theory. This
results in the genus~1 annulus amplitude~\cite{TM_18}
\be
Z^{\rm o}(\tau_2)=k^{1/4}\,\sum_{n=0}^{pq/2-1}\Psi_n^{\rm orb}(i\,\tau_2) \ ,
\label{annulusampl}\ee
where
\be
\Psi_n^{\rm orb}(i\,\tau_2)=\frac1{2(k\tau_2)^{1/4}\,\Bigl|
\eta(i\,\tau_2)\Bigr|}~\sum_{r=-\infty}^\infty\exp\left\{-
\frac{2\pi\tau_2}k\,\left(rp+\frac{kn}4\right)^2\right\} \ .
\label{annuluswave}\ee
The orbifold partition function (\ref{annulusampl}) is a vector in the
open string Hilbert space
\be
{\cal H}^{\rm o}=\bigoplus_{n=0}^{pq/2-1}\left[\phi_n^{\rm orb}\,
\right] \ .
\label{Hilbertopen}\ee
Now let us consider the annulus with prescribed charges $n_1$ and $n_2$
on its two boundary circles. By using bulk charge conservation and
the fusion coefficients (\ref{fusioncoeffs}), we may write the annulus
partition function restricted to this sector as
\be
Z^{\rm o}_{\lambda_1,\lambda_2}(\tau_2)=k^{1/4}\,\sum_{\lambda=0}^{pq/2-1}
N_{\lambda_1\lambda_2}^{~~~~\lambda}~\Psi^{\rm orb}_\lambda(i\,
\tau_2) \ .
\label{annulusamplsec}\ee

On the other hand, the Dirichlet cylinder amplitude for closed string
propagation in a time span $2\pi\tilde\tau_2$ is obtained from
(\ref{toruspartfn}) via a $\sf PT$ orbifold in three-dimensions and may be
written as~\cite{TM_18}
\be
\tilde Z^{\rm o}(\tilde\tau_2)=k^{1/4}\,\sum_{m=0}^{pq/2-1}
\tilde\Psi_m^{\rm orb}(i\,\tilde\tau_2) \ ,
\label{cylampl}\ee
where
\be
\tilde\Psi_m^{\rm orb}(i\,\tilde\tau_2)=\frac{\left(k\,\tilde
\tau_2\right)^{1/4}}{2\,\Bigl|\eta(i\,\tilde\tau_2)\Bigr|}~
\sum_{r'=-\infty}^\infty\exp\left\{-\frac{2\pi\tilde\tau_2}k\,
\Bigl(r'p+m\Bigr)^2\right\} \ .
\label{cylwave}\ee
Worldsheet duality is the statement that the partition functions
(\ref{annulusampl}) and (\ref{cylampl}) coincide after a modular
transformation,
\be
\tilde Z^{\rm o}(\tilde\tau_2=2/\tau_2)~=~Z^{\rm o}(\tau_2) \ ,
\label{worldsheetduality}\ee
which may be derived explicitly by using the Poisson resummation
formula (\ref{Poisson})~\cite{TM_18}. In the next section we will give
a dynamical interpretation of this duality in the finite temperature gauge
theory. When restricted to a fixed boundary charge
sector as in (\ref{annulusamplsec}), application of the definition
(\ref{Psimodular}) to (\ref{cylampl}) reproduces the Verlinde
diagonalization formula (\ref{Verlinde3pt}). This is thereby a three
dimensional version of the open string derivation of the Verlinde
formula through the Cardy relations~\cite{C_1}.

Formally, the Cardy condition arises because the fixed boundary charge
restriction of the partition function (\ref{cylampl}) is, by
definition, the membrane inner product
\be
\tilde Z^{\rm o}_{\lambda,\lambda'}(\tilde\tau_2)=\left.\left
\langle\psi_{\lambda'}^{\rm D}\,\right|\,\psi_\lambda^{\rm D}
\right\rangle
\label{tildeZoinnprod}\ee
of fundamental wavefunctionals (\ref{genericIsh}). By using the
expansion coefficients (\ref{CardyB}) and comparing
(\ref{tildeZoinnprod}) with (\ref{annulusamplsec}) after a modular
transformation, we arrive at the Verlinde formula. Alternatively, the
three dimensional dynamical description of this condition follows from
surgery~\cite{FS_2}, analogously to the derivation depicted in
fig.~\ref{Verlinde}. In the present case, we use the fact, discussed
earlier, that the annulus graph can be interpreted in three dimensions
as the correlator of two Wilson lines carrying the prescribed boundary
charges $Q_1$ and $Q_2$. We may thereby represent (\ref{annulusamplsec})
as the $\sphere^3$ correlator of three Wilson loops of charges $Q_1$,
$Q_2$ and $Q_3=0$ linked with a trivial $Q=0$ loop, as in the first
equality of fig.~\ref{Verlinde}. By rewriting the trivial loops as
sums over complete sets of Polyakov loops, we again arrive at the
anticipated result above.

\setcounter{equation}{0}
\section{BRANE TENSIONS\lb{Daction}}

In this final section we shall describe what is perhaps the most
important aspect of the three-dimensional representation of D-branes,
that the topological membrane approach yields the correct tension
of a D-brane. This will require two additional steps beyond what we
have developed thus far. First, we shall have to properly introduce
and study the mass scales of the bulk three-dimensional theory and
discuss how they generate those of the induced worldsheet
theory. Second, we will introduce the excited eigenstates of the gauge
theory Hamiltonian and study the Hilbert space of physical states
beyond the topological vacuum sector. With these developments we will
show that the appropriate D-brane tension arises both as a dynamical
property of the topological membrane and as an intrinsic quantity set
by the mass scales of the bulk theory. As we will see, the careful
incorporation of the bulk mass scales is absolutely crucial for the
appropriate appearence of the D-brane mass, as they are required to
remove the genuine quantum field theoretic divergences which arise
above the vacuum sector in the three-dimensional formulation. As
additional bonuses, we shall discover the proper formulation of
descendent string states in three-dimensions and also acquire a new
perspective on worldsheet modular duality.

\subsection{Dilaton Coupling}

In our derivation of the D-brane tension in three dimensions, we shall
first describe how the appropriate factors of the string coupling
constant $g_s$ appear. For this, we need to describe how to
incorporate the string dilaton field into the topological membrane
formalism. This problem was addressed in~\cite{TM_14} and consists of
examining the conformal coupling of topologically massive gauge theory
to topologically massive gravity through the action
\bea
S_{\rm CTMGT}[A,\omega;D]&=&\int\limits_Md^3x~\sqrt{-g}~\left[\kappa\,D^2\,
R(\omega)+8\kappa\,\partial_\mu D\,\partial^\mu D-\frac1{4\gamma\,D^2}\,
F_{\mu\nu}F^{\mu\nu}\right]\nn\\&&-\,8\kappa\,\oint\limits_{\partial M}
D\,\partial^{~}_\perp D+S_{\rm CS}[A,\omega]
\label{SCTMGT}\eea
which is defined on a three-manifold $M$. Here $\kappa$ is the
three-dimensional Planck mass, $R(\omega)$ is the curvature of the
torsion-free, $SO(2,1)$ Lie algebra valued spin-connection
$\omega_\mu^a$ of the frame bundle of $M$, and $D$ is a dimensionless
scalar field in three spacetime dimensions. The first term in
(\ref{SCTMGT}) is a modification of the Einstein-Hilbert action, while
\be
S_{\rm CS}[A,\omega]=\int\limits_Md^3x~\left[\frac k{8\pi}\,
\epsilon^{\mu\nu\lambda}\,A_\mu\,\partial_\nu A_\lambda+
\frac{k'}{8\pi}\,\epsilon^{\mu\nu\lambda}\,\left(\omega_\mu^a\,
\partial_\nu\omega_\lambda^a+\frac23\,\epsilon^{abc}\,\omega_\mu^a\,
\omega_\nu^b\,\omega_\lambda^c\right)\right]
\label{SCSgaugegrav}\ee
is the sum of the gauge and gravitational Chern-Simons actions.

The action (\ref{SCTMGT}) is invariant under three-dimensional
conformal transformations~\cite{TM_14}. In the naive vacuum $\langle
D^2\rangle=0$, the spectrum of the corresponding quantum field theory
contains a massless scalar particle but no graviton or gauge degrees
of freedom, so that in this phase the model is equivalent to the pure
topological field theory defined by the Chern-Simons actions
(\ref{SCSgaugegrav}). On the other hand, in the phase where the scalar
field $D^2$ has a non-vanishing vacuum expectation value, there is a
propagating graviton mode with mass $8\pi\kappa\,\langle
D^2\rangle/k'$ and a photon with topological mass
\be
\mu=\frac{\gamma k\,\left\langle D^2\right\rangle}{4\pi} \ .
\label{photonmass}\ee
The zero-point quantum fluctuations of the field $D$ thereby set the
mass scales of the bulk theory.

For the membrane geometry $M=\Sigma\times[0,1]$, in addition to the
usual conformal $\sigma$-model of central charge $c=1$ living at the
boundary $\Sigma$, the gravitational sector of the three-dimensional
theory (\ref{SCTMGT}) induces Liouville theory coupled to a worldsheet
dilaton field on $\Sigma$~\cite{TM_14}. After the usual boundary
identifications and an appropriate rescaling of the fields, the
effective two-dimensional gravity action is given by
\be
S_{\rm L}[D,\phi]=\int\limits_\Sigma d^2z~\left[-\frac1{4\pi}\,
\left(\ln D^4+\phi\right)\,R^{(2)}-2\kappa\,D\,\partial_\perp^{~}D+
\frac1{16\pi}\,\partial_\bz\phi\,\partial_z\phi+\Lambda_\Sigma
{}~\e^{-\phi}\right] \ ,
\label{SLDphi}\ee
where $\phi$ is the Liouville field, $R^{(2)}$ is the usual
two-dimensional curvature of the worldsheet $\Sigma$, and $\Lambda_\Sigma$ is
the worldsheet cosmological constant which is induced by the dreibein
condensate and topological graviton mass of the gravitational sector
of the original three-dimensional theory (\ref{SCTMGT}), and by the
vacuum expectation value of the scalar field $D^2$. From (\ref{SLDphi})
it is clear that $D$ is the three-dimensional version of the string
dilaton field, and that the string coupling constant, like the three
dimensional mass scales, is dynamically generated by the vacuum
expectation value
\be
g_s=\left\langle D^4\right\rangle \ .
\label{gsdilaton}\ee
Then, from (\ref{SLDphi}) it follows that the contribution to the
string statistical sum from a Riemann surface $\Sigma$ of genus $g$
with $b$ boundaries and $c$ crosscaps will contain the factors
\be
\left\langle\exp\left\{-\frac1{4\pi}\,\int\limits_\Sigma d^2z~
\left(\ln D^4\right)\,R^{(2)}\right\}\right\rangle~=~(g_s)^{-\chi(\Sigma)} \ ,
\label{dilatoncontr}\ee
where
\be
\chi(\Sigma)=2-2g-b-c=\frac1{4\pi}\,\int\limits_\Sigma d^2z~R^{(2)}
\label{Eulernumber}\ee
is the Euler number of $\Sigma$.

If $\Sigma$ is a closed surface, then (\ref{dilatoncontr}) gives the
contribution to the closed string sector of the worldsheet theory. The
crucial point now is that the open string sector can be obtained as a
$\zed_2$-orbifold of the closed string sector. In the usual way, the
original action $S_\Sigma$ defined on the closed surface $\Sigma$ is
twice that of its orbifold, $S_\Sigma=2S^{\rm
  o}_{\Sigma/\zed_2}$~\cite{SBH_05}. Thus the {\it open} string
contribution is given by $(g_s)^{-\chi(\Sigma)/2}$. Heuristically,
this feature manifests itself within the three dimensional picture in the
membrane scattering amplitudes $\langle\Psi_1|\Psi_0\rangle$ between
states $\Psi_0$ and $\Psi_1$ inserted at the initial and final
surfaces $\Sigma_0$ and $\Sigma_1$ in the closed string picture (see
fig.~\ref{fig.orbi}). Inserting a complete set of wavefunctionals
$\Psi_{1/2}$ at the orbifold fixed point $t=1/2$ in the bulk of the
topological membrane determines this amplitude as~\cite{TM_18}
\be
\left\langle\Psi_1\left|\Psi_0\right.\right\rangle=\int\limits_{\Psi_{1/2}}
\!\!\!\!\!\!\!\!\!\sum\,
\left\langle\Psi_1\left|\Psi_{1/2}\right.\right\rangle\left\langle
\left.\Psi_{1/2}\right|\Psi_0\right\rangle \ .
\label{amplcomplset}\ee
Under the orbifold operation, the initial and final surfaces are
identified, $\Sigma_0\equiv\Sigma_1$, and the only wavefunction that
can live at $\Sigma_{1/2}$ is the identity character state
$|\Psi_{\vec\lambda=\vec0}\rangle=|1\rangle$, so that the right-hand
side of (\ref{amplcomplset}) reduces to
$|\langle1|\Psi_0\rangle|^2$. Thus the decomposition
(\ref{amplcomplset}) heuristically means that closed strings come in a
double volume of open strings, as usual. It follows that the open
string membrane amplitudes should be identified with the square roots
of closed string ones,
\be
\left\langle\Psi_1\left|\Psi_0\right.\right\rangle_{\rm orb}=
\sqrt{\left\langle\Psi_1\left|\Psi_0\right.\right\rangle} \ .
\label{orbsqrtclosed}\ee
For example, while the leading order closed string diagram $\sphere^2$
varies with the string coupling constant as $1/(g_s)^2$, the disk
amplitude $\disc^2$ behaves like $1/g_s$, which is the standard coupling
dependence of the D-brane tension at tree-level in string perturbation
theory. These arguments can also be used to account for the missing
square root dependences of the Wilson line correlators
(\ref{WQiunravel}).

\subsection{D-Brane Tension\label{Tension}}

To derive a formula for the brane tension one can proceed in one of
two ways. The first one, which we shall begin developing in
section~\ref{Excited}, computes the three-dimensional version of the
one-point tadpole insertion on a disk, i.e. the Wilson line correlator
(\ref{disc1pt}) in the presence of a monopole. This uses the fact
that, in the worldsheet theory, the mass is measured by the one-point
function with the graviton vertex operator and it requires a
development of excited string states in the topological membrane. The
second approach, which we will pursue in this subsection and the next,
is to use our bulk construction of the brane coordinates $Y_{\rm N}$
and $Y_{\rm D}$ to derive the Dirac-Born-Infeld action from the
membrane partition function. This exploits the fact that the tension
appears as the boundary entropy factor in the overall normalization of
the open string partition function.

To make the derivation as explicit and general as possible, we will
consider, in this subsection and the next only, a more general
topologically massive gauge theory with structure group $U(1)^d$,
abelian gauge fields $A^I$, and corresponding field strengths $F^I$,
where $I=1,\dots,d$. The source-free action is given by
\be
S^{(d)}_{\rm TMGT}[A;D]=\int\limits_0^1dt~\int\limits_\Sigma d^2z~
\left[-\frac{\sqrt{-g}}{4\gamma\,D^2}\,F^I_{\mu\nu}F_I^{\mu\nu}-
\frac2\pi\,K_{IJ}\,\epsilon^{\mu\nu\lambda}\,A_\mu^I\,
\partial_\nu A_\lambda^J\right] \ .
\label{STMGTd}\ee
The inclusion of external sources, as in (\ref{STMGT}), will be
described in the next subsection. The Chern-Simons coefficient
(\ref{CScouplingid}) is now modified to the rank~2 constant tensor
\be
K_{IJ}=\frac1{\alpha'}\,\Bigl(G_{IJ}+i\,B_{IJ}\Bigr) \ ,
\label{KIJGB}\ee
where $G_{IJ}$ is a symmetric real matrix which is interpreted as the
target space graviton condensate, while $B_{IJ}$ is an antisymmetric
real matrix which is interpreted as a constant background NS--NS
two-form field. Note that in complex worldsheet coordinates, the
antisymmetric part of the Chern-Simon coefficient matrix is purely
imaginary, owing to the Euclidean signature of the target space. The
previous results derived in section~\ref{Verts} generalize
straightforwardly with the inclusion of the symmetric part
$G_{IJ}$. On the other hand, the terms containing $B_{IJ}$ are total
derivatives and thereby contribute only at worldsheet boundaries. The
bulk gauge theory (\ref{STMGTd}) induces the two-dimensional $c=d$
conformal $\sigma$-model with target space the $d$-dimensional torus
$\torus^d$ of metric $G_{IJ}$. The closed string conformal field
theory is characterized by a Narain lattice
$\Gamma(K)\subset\real^{d,d}$ with rational-valued moduli.

We shall now present a careful evaluation of the partition functions
associated with the quantum field theory defined by (\ref{STMGTd}),
keeping track in particular of the normalization factors that were
neglected in the analysis of section~\ref{OrbPart}. For closed
strings, after identifying both boundaries $\Sigma_0$ and $\Sigma_1$
with opposite orientations~\cite{TM_06,TM_14,TM_15,TM_16,TM_17,TM_18}, we
find that the generalization of (\ref{Zcdef},\ref{Zcfact}) is given by
the path integral
\be
\ba{rcl}
Z^{\rm c}&=&\displaystyle(g_s)^{-\chi(\Sigma)}\,{\mathcal{N}}\\[2mm]&&
\displaystyle\times\,
\prod_{I=1}^d\,\int\left[D\bar A^I_z~D\bar A^I_\bz\right]~
\int\left[D\varphi^I\right]~
\exp\left\{-\frac{K_{IJ}}{8\pi}\,\int\limits_{\Sigma}d^2z~
\Bigl[2\,\left(\bar A_z^I-\partial_z\varphi_0^I\right)\left(\bar A_\bz^J-
\partial_\bz\varphi_1^J\right)\Bigr.\right.\\[2mm]&&+\displaystyle
\Biggl.\Bigl.\partial_z\left(\varphi_0^I-
\varphi_1^I\right)\partial_\bz\left(\varphi_0^J-\varphi_1^J
\right)\Bigr]\Biggr\} \ .
\ea
\label{Zcds}\ee
Here we have identified the gauge field degrees of freedom $\bar A^I$ on
both boundaries, and assumed that the path integral over the Liouville
field is performed with its contribution absorbed in an
appropriate overall normalization constant included in
$\mathcal{N}$. We have also frozen the dilaton field $D$ at its
classical value, as will be discussed further in
section~\ref{Excited}. The chiral scalar fields $\varphi_0^I$ and
$\varphi_1^I$ are induced by $U(1)^d$ gauge invariance of the
corresponding Schr\"odinger wavefunctionals on the surfaces $\Sigma_0$ and
$\Sigma_1$, respectively, which combine into the non-chiral fields
$\varphi^I=\varphi^I_0-\varphi^I_1$ on $\Sigma$.

In keeping with the discussion of the previous subsection, we will fix
$\cal N$ such that the closed string partition function (\ref{Zcds})
contributes the usual perturbative weight $(g_s)^{-\chi(\Sigma)}$,
along with the usual topological contributions as in (\ref{partfnTM})
which to avoid clutter we will not display explicitly. The functional
Gaussian integrations over $\bar A^I$ and $\varphi^I$ in (\ref{Zcds}) may
be done explicitly as described in section~\ref{OrbPart}, and by
exploiting the fact that the fields are defined
on a compact surface to expand them in a countable basis of
functions on $\Sigma$. In genus $g=0$ this is the Laurent
basis on the complex plane which is labelled by $\zed$, for
$g=1$ it is the Fourier basis on $\sphere^1\times\sphere^1$ which is
indexed by $\zed^2$, while for $g\geq2$ it is given by the
Krichever-Novikov basis which is labelled by $\zed$ for
$g$ even and by $\zed+1/2$ for $g$ odd. Infinite constants are then
regulated by using the standard zeta-function regularizations
\be
\prod_{n=1}^\infty\,c=c^{-1/2} \ , ~~ \prod_{n+1/2=1}^\infty\,c=1 \ .
\label{zetaregc}\ee
The $\bar A^I$ integrals are performed after an appropriate shift of
the fields $(\bar A_z^I,\bar A_\bz^I)\to(\bar A_z^I-\partial_z\varphi^I_0,
\bar A^I_\bz-\partial_\bz\varphi^I_1)$. In this way the contribution to
the normalization constant comes from two complex functional
Gaussian integrations and results in
\be
{\mathcal N}=\left(\frac{\sqrt{2}}{8\pi^2}\,\Bigl|\det(K_{IJ})\Bigr|
\,\right)^{\nu_g} \ ,
\label{calNfix}\ee
where the index $\nu_g$ depends on the genus of the surface and is
given by $\nu_0=1$, $\nu_1=2$, while $\nu_g=1$ for $g\geq2$ even and
$\nu_g=0$ for $g>2$ odd.

In the case of open strings, the worldsheet $\Sigma_1$
is, under the given orbifold involution, identified with $\Sigma_0$
and the wavefunction at $\Sigma_{1/2}$ consists only of terms on the
boundary $\partial\Sigma^{\rm o}$. This point was already used in
(\ref{WQicorrorb2}) and will be further elucidated in the next
subsection. Then we have only the integral over one of the boundaries,
say $\Sigma=\Sigma_0$, and the orbifold of (\ref{Zcds}) yields the
mass formula
\be
\ba{rcl}
{\cal M}^2&=&\alpha'\,\displaystyle(g_s)^{-\chi(\Sigma)}\,{\mathcal{N}}~
\left(\frac{\det(G_{IJ}/\alpha')}{\sqrt{32\pi^2}}\right)^{\nu_g/2}\\[2mm]&&
\displaystyle\times\,\prod_{I=1}^d\,\int\left[D\bar A_z^I~
D\bar A_\bz^I\right]~\int\left[D\varphi^I\right]~\exp\left\{-
\frac{1}{4\pi\alpha'}\,\int\limits_{\Sigma}d^2z~\Bigl[G_{IJ}\,\bar A_z^I\,
\partial_\bz\varphi^J\Bigr.\right.\\[2mm]&&\displaystyle-\Biggl.\Bigl.
G_{IJ}\,\partial_z\varphi^I\,\bar A_\bz^J
+\alpha'\,K_{IJ}\left(\bar A_z^I-
\partial_z\varphi^I\right)\left(\bar A_\bz^J-\partial_\bz\varphi^J
\right)\Bigr]\Biggr\} \ ,
\ea
\label{effSgen}\ee
where the overall metric determinant factor in (\ref{effSgen}) comes from the
regularized integrations enforcing the functional Dirac
delta-functions which impose the boundary
conditions~(\r{Psi12orb2D},\r{Psi12orb2N}). Remembering the orbifold
correspondence rule (\ref{orbsqrtclosed}), we have identified
(\ref{effSgen}) as the {\it square} of the orbifold contribution to
the normalization of (\ref{Zodef}) determining the brane tension. As
before, to determine the precise topological dependence here requires
an appropriate truncation to the Lagrangian subspace
(\ref{kericalL}). This is not required to determine the dependence of
the mass formula on the couplings. Note that on the boundary
$\partial\Sigma^{\rm o}$ only the antisymmetric part $B_{IJ}$ of $K_{IJ}$
is present, and generally its symmetric part $G_{IJ}$ acts as a metric
for the raising and lowering of the $U(1)^d$ group indices (target
space directions in the induced string theory). As with the
topological contributions, we suppress these boundary terms in all
formulas of this subsection.

Let us first consider the $\sf PT$ orbifold of the three-dimensional
gauge theory (\ref{STMGTd}), corresponding to a $d$-brane wrapping
$\torus^d$. Integration over the gauge field components in
(\ref{effSgen}) as described in section~\ref{OrbPart} then yields
\bea
{\cal M}^2&=&\alpha'\,(g_s)^{-\chi(\Sigma)}\,{\mathcal{N}}~
\left(\frac{\pi}{\sqrt2}\,\frac{\det(G_{IJ}/\alpha')}{\Bigl|\det(K_{IJ})\Bigr|}
\,\right)^{\nu_g/2}
\nn\\&&\times\,\prod_{I=1}^d\,\int\left[D\varphi^I\right]~
\exp\left\{-\frac{\tilde{K}_{IJ}}{4\pi}\,\int\limits_{\Sigma}d^2z~
\partial_z\varphi^I\,\partial_\bz\varphi^J\right\} \ ,
\label{effSgenPT}\eea
where
\be
\tilde{K}_{IJ}=\frac1{(\alpha')^2}\,G_{II'}\,\left(K^{-1}\right)^{I'J'}\,
G_{J'J}\equiv\frac1{\alpha'}\,\Bigl(\tilde{G}_{IJ}+i\,\tilde{B}_{IJ}
\Bigr)
\lb{tK}
\ee
is the T-dual background to the Chern-Simons coupling matrix
(\ref{KIJGB}), with symmetric and antisymmetric parts $\tilde{G}_{IJ}$
and $\tilde{B}_{IJ}$, respectively. Carrying out the remaining path
integration in (\ref{effSgenPT}) and inserting (\ref{calNfix}) then
yields finally
\be
{\cal M}^2=\alpha'\,(g_s)^{-\chi(\Sigma)}\,\left(
\frac{\Bigl|\det(K_{IJ})\Bigr|}{\sqrt{4\pi\det(G_{IJ}/\alpha')}}\,
\right)^{\nu_g} \ .
\lb{effS}\ee
The effective coupling (\ref{effS}) is the familiar
Born-Infeld action for a flat background $K_{IJ}$ times the right
power of the string coupling constant $g_s$, giving the familiar
contribution to the D-brane tension on $\torus^d$~\cite{TD,POL_book}. Note that
it is determined in part by the effective $\sigma$-model action of
(\ref{effSgenPT}) in the T-dual background fields, as expected.

For the $\sf PCT$ orbifold of the gauge theory (\ref{STMGTd}), the
terms involving solely the metric $G_{IJ}$ disappear. Working through
as above one finds that the final result of the open string path
integral is ${\cal M}^2\propto\det(G_{IJ})^{-\nu_g/2}$, yielding the
correct dependence of the tension appropriate for Neumann boundary
conditions along each of the directions of $\torus^d$~\cite{TD,POL_book}. Thus
from the topological membrane approach we naturally obtain the anticipated
spacetime supergravity field and string coupling dependences of the
tensions for both Neumann and Dirichlet branes. Note that this
derivation is completely independent of the induced boundary vertex
operators that we obtained in section~\ref{OrbPart}, and is simply a
consequence of the orbifold properties of the topological membrane.

\subsection{Effective Actions}

We will now carry the analysis of the previous subsection one step
further and incorporate sources into the orbifold partition functions,
as we did in section~\ref{Verts}. For this, it suffices to consider
the following construction. At time $t=1/2$ the wavefunctional
(\ref{Psi12orbfact}) simply sets boundary conditions on the fields
$\varphi$ and produces the appropriate boundary vertex operator insertions of
section~\ref{OrbPart}, while at $t=0$ (identified by the orbifold
involution with $t=1$) the wavefunctional (\ref{Psi0orb}) produces the
actual effective $\sigma$-model with a constant background
$\tilde{K}_{IJ}$, as is apparent
from~(\ref{effSgenPT})--(\r{effS}). So let us identify each point on
$\Sigma^{\rm o}=\Sigma_{1/2}/{\mathbb{Z}}_2$ with two points on
$\Sigma_0=\Sigma$, so that the closed Riemann surface $\Sigma$ constitutes
a double covering of $\Sigma^{\rm o}$. In this way, as described at
the end of section~\ref{OrbPart}, from a $U(1)^d\times U(1)^D$
topologically massive gauge theory one obtains the open bosonic string
partition function in constant backgrounds with D-branes and Wilson
lines,
\be
Z^{\rm o}=(g_s)^{-\chi(\Sigma)/2}\,\prod_{I=1}^{d+D}\,\int
\left[D\varphi^I\right]~\prod_{a=1}^d\delta_{\partial\Sigma^{\rm o}}
\left(\varphi_{\rm b}^a\right)~\prod_{m=d+1}^{d+D}
\delta_{\partial\Sigma^{\rm o}}\left(\partial^{~}_\perp
\varphi_{\rm b}^m\right)~\e^{-S_{\rm eff}[\varphi]} \ ,
\label{Zoeffaction}\ee
where the fields $\varphi^a$, $a=1,\ldots,d$ obey Dirichlet
boundary conditions and $\varphi^m$, $m=d+1,\ldots,d+D$ obey Neumann
boundary conditions on the boundary of the open string worldsheet
$\Sigma^{\rm o}$. As before, the boundary of $\Sigma^{\rm o}$ in the
three-dimensional membrane is situated at the branch point locus
$x^\perp=0$ of the orbifold, and we have already used the prescription
(\ref{orbsqrtclosed}). By using the shifts $\bar A^I_\bz\to
\bar A^I_\bz-4\pi\,(K^{-1})^{IJ}\,G_{JJ'}\,\tilde{Y}_\bz^{J'}$ and
$\varphi^a\to\varphi^a-Y^a_{\rm D}$ in the original orbifold path integral,
the effective action in (\ref{Zoeffaction}) is given modulo
topological contributions by
\be
\ba{rcl}
S_{\rm eff}[\varphi]&=&\displaystyle\frac{1}{8\pi\alpha'}\,\int\limits_{\Sigma}
d^2z~\left[\sqrt{h}\,h^{ij}\,\tilde{G}_{IJ}+i\,\epsilon^{ij}
\tilde{B}_{IJ}\right]\,\partial_i\varphi^I\,\partial_j\varphi^J\\[4mm]
&&+\,\displaystyle\frac{1}{4\pi}\,\oint\limits_{x^\perp=0}
\left[Y_{{\rm D},a}\,\partial^\perp\varphi_{\rm b}^a+i\,
Y_{{\rm N},m}\,\partial^\parallel\varphi_{\rm b}^m\right] \ ,
\ea
\ee
where we have used a Hodge decomposition of the sources analogous to
(\ref{HodgeY}) and a suitable source rescaling.

This is the usual string theoretic, deformed worldsheet $\sigma$-model
action in the presence of D-branes and Wilson lines. The effective
target space action is obtained in the standard way by integrating out
the fields $\varphi^a$ with Dirichlet boundary conditions~\cite{Db_02}. In
particular, by using the boundary conditions $\varphi^a_{\rm b}=Y_{\rm
  D}^a|_{\partial\Sigma}$ in the static gauge
for constant background $\varphi^m$, the orbifold partition function
(\ref{Zoeffaction}) integrates (modulo topological terms) to the
Dirac-Born-Infeld action~\cite{Db_03}
\be
S_{\rm DBI}=(g_s)^{-\chi(\Sigma)/2}\int d^D\varphi~
\Bigl|\det({\cal K}_{mn}+{\cal F}_{mn})\Bigr|^{\nu_g/2} \ ,
\lb{Sdbrane}
\ee
where
\bea
{\cal K}_{mn}(\varphi^m)&=&K_{IJ}\,
\frac{\partial \varphi^I}{\partial\varphi^m}\,
\frac{\partial \varphi^J}{\partial\varphi^n} \ , \nn\\{\cal F}_{mn}(
\varphi^m)&=&\frac{\partial Y_{{\rm N},m}}{\partial\varphi^n}-
\frac{\partial Y_{{\rm N},n}}{\partial\varphi^m} \ , \nn\\
\varphi^I&=&\left(\varphi^m\,,\,Y_{\rm D}^a\right) \ .
\eea
Since we are working with constant background $K_{IJ}$, the D-brane action
(\ref{Sdbrane}) is manifestly invariant under T-duality.

\subsection{Excited Wavefunctionals\label{Excited}}

In this subsection we will construct excited states of the topological
membrane which will enable an alternative, intrinsic derivation of the D-brane
tension within the three-dimensional formalism. At the same time, this
solves a long-standing problem in the topological membrane approach to
string theory, the proper construction of states in the
three-dimensional gauge theory which correspond to excited, descendent
string states in two-dimensions. For this, we rewrite the
non-vanishing electric field commutators in (\ref{com_EB}) in complex
coordinates as
\be
\Bigl[E_z(\vb{z})\,,\,E_\bz(\vb{z}')\Bigr]=-\frac k{4\pi}~
\delta^{(2)}(\vb{z}-\vb{z'}) \ .
\label{fnHeisen}\ee
This is a functional Heisenberg oscillator algebra which in the
Schr\"odinger picture can be represented as in
section~\ref{sec.basics} by the electric field operators
\bea
E_z&=&-i\,\frac\delta{\delta A_z}+\frac{i\,k}{8\pi}\,
A_\bz \ , \nn\\E_\bz&=&-i\,\frac\delta{\delta A_\bz}-
\frac{i\,k}{8\pi}\,A_z \ .
\label{electricops}\eea

The Hamiltonian of the dilaton-coupled topologically massive gauge
theory part of the action (\ref{SCTMGT}) is straightforward to obtain
in the classical background of the dilaton field with $\langle
D^2\rangle\neq0$. As usual, the vacuum energy density arising from
normal ordering the Hamiltonian operator using the commutation
relations (\ref{fnHeisen}) is a divergent term $\delta^{(2)}(0)$. The
normal ordered energy density is then defined such that the divergent
term drops out, $\NO h_{ij}\,E^iE^j\NO\equiv E_\bz E_z$, and the
dilaton-coupled gauge theory Hamiltonian in the temporal gauge $A_0=0$ is
\be
H_D=\frac14\,\int\limits_\Sigma d^2z~\frac1{D^2}\,\left[\gamma\,
E_\bz E_z+\frac1\gamma\,B^2\right] \ .
\label{HDdef}\ee
The equations of motion obtained by varying the classical action
(\ref{SCTMGT}) with respect to the dilaton field $D$ read
\be
\frac1{2\gamma\,D^3}\,F_{\mu\nu}F^{\mu\nu}-16\kappa\,\nabla^2D+
2\kappa\,D\,R(\omega)=0 \ ,
\label{dilatonEOM}\ee
which we note carefully uses the boundary term over $\partial M$ in
(\ref{SCTMGT}) that ensures the dilaton field theory has a classical
extremum. By using the Minkowski signature of the three-dimensional
metric (\ref{ds3}) to compute $F_{\mu\nu}F^{\mu\nu}=-\gamma^2\,E_\bz
E_z+B^2$, we can use the classical value of the dilaton field
determined from (\ref{dilatonEOM}) to express the magnetic field term
$B^2$ in (\ref{HDdef}) in terms of the field theoretic oscillator
number operator $E_\bz E_z$. We find
\be
H_D=\frac12\,\int\limits_\Sigma d^2z~\left[\frac\gamma{D^2}\,E_\bz E_z+
2\kappa\,V(D,\omega)\right] \ ,
\label{HDclassicalD}\ee
where the dilaton potential is given by
\be
V(D,\omega)=8\,D\,\nabla^2D-D^2\,R(\omega)
\label{dilatonpot}\ee
and is essentially the three-dimensional energy density of the Liouville
field theory (\ref{SLDphi}). Note that substituting $D^2$ by its vacuum
expectation value given from (\ref{dilatonEOM}) breaks the quantum
S-duality symmetry $D^2\mapsto({\rm const.})/D^2$.

{}From (\ref{0mode}) it follows that the vacuum states (\ref{Xiexcited})
are zero-modes of the field theoretic harmonic oscillator annihilation
operator $E_z$ in (\ref{electricops}),
\be
E_z\Xi\left[A_z,A_\bz\,;\left\{{\matrix{z_1&\cdots&z_s\cr Q_1&\cdots&
Q_s\cr}}\right\};0\right]=0 \ .
\label{Ezann}\ee
Physically, from (\ref{HDclassicalD}) it then follows that the dilaton
coupling to the gravitational sector induces a shift in the ground state
energy of the membrane from ${\cal E}_0=0$ to
\be
{\cal E}_0=\kappa~\Bigl\langle V(D,\omega)\Bigr\rangle \ .
\label{calE0VD}\ee
This zero-point energy vanishes in the pure
Chern-Simons limit $\kappa\to0$ of topologically massive gravity
(equivalently in the phase of the topological field theory with $\langle
D^2\rangle=0$). Geometrically, the electric field operators
(\ref{electricops}) define a connection on a principal $U(q)$-bundle
$\cal P$ over the space of gauge fields of the given (non-trivial)
complex line bundle over the Riemann surface $\Sigma$~\cite{W_1}. This
connection, according to (\ref{fnHeisen}), has constant curvature
$\frac k2=\frac pq$. By using in addition the constraint (\ref{WW}),
the condition (\ref{Ezann}) simply means that the vacuum wavefunctionals
(\ref{Xiexcited}) of the topological membrane are gauge-invariant
holomorphic sections of the bundle $\cal P$. This property is of
course tied to the fact that the vacuum sector is a topological field
theory, whose quantization yields the holomorphic Friedan-Shenker
vector bundles of the induced conformal field theory on $\Sigma$.

We will now construct the eigenstates of the Hamiltonian
(\ref{HDclassicalD}) by using these observations. The natural
gauge-invariant excited states of the topological membrane are the
Landau levels created by the field theoretic harmonic oscillator
creation operator $E_\bz$ of (\ref{electricops}). They are obtained
by acting on the wavefunctionals (\ref{Xiexcited}) with non-negative
powers $l_i$ of the operators $E_\bz(z_i)$ at the primary field
insertion points $z_i\in\Sigma$ to give
\bea
&&\Psi^s\left[A_z,A_\bz\,;\left\{{\matrix{z_1&\cdots&z_s\cr Q_1&\cdots&
Q_s\cr l_1&\cdots&l_s\cr}}\right\};0\right]\nn\\&&{~~~~}_{~~}^{~~}
\nn\\&&~~~~~~~~~~\equiv~
\prod_{i=1}^s\frac1{\sqrt{l_i!}}\,\left(\frac{4\pi\,i}{k}\,
E_\bz(z_i)\right)^{l_i}\Xi
\left[A_z,A_\bz\,;\left\{{\matrix{z_1&\cdots&z_s\cr Q_1&\cdots&
Q_s\cr}}\right\};0\right]\nn\\&&{~~~~}_{~~}^{~~}\nn\\&&~~~~~~~~~~=~
\exp\left\{\frac k{8\pi}\,\int\limits_\Sigma d^2z~A_\bz\,A_z\right\}\,
\int[D\varphi]~\exp\left\{\frac k{8\pi}\,\int\limits_\Sigma d^2z~\left(
\partial_\bz\varphi-2A_\bz\right)\,\partial_z\varphi\right\}\nn\\&&
{}~~~~~~~~~~~~~~~\times\,\prod_{i=1}^s\frac1{\sqrt{l_i!}}\,
\Bigl(A_z(z_i)-\partial_z\varphi(z_i)
\Bigr)^{l_i}~\e^{i\,Q_i\,\bigl(\varphi(z_i)+h_\varphi(z_i)
\bigr)}\ .
\label{Psisexcited}\eea
As before, the excited states $\Psi^s$ are gauge invariant up to a
projective phase, and as in (\ref{Xioprep}) in the cases that the worldsheet
is the punctured sphere, $\Sigma=\sphere_0^2$, they act on the product of
the corresponding Virasoro modules,
\be
\Psi_{\sphere_0^2}^s
\left[A_z,A_\bz\,;\left\{{\matrix{z_1&\cdots&z_s\cr Q_1&\cdots&
Q_s\cr l_1&\cdots&l_s\cr}}\right\};0\right]~\in~\bigotimes_{i=1}^s\,
\left[\phi_{\lambda_i}\right]^*\otimes\left[\phi_{\lambda_i}\right] \
{}.
\label{Psisoprep}\ee
They are the proper gauge invariant wavefunctionals that correspond to
string descendent states of higher level numbers
\be
N=\sum_{i=1}^sl_i\geq0 \ .
\label{levelnumber}\ee

By construction, the wavefunctionals (\ref{Psisexcited}) are
eigenstates of the Hamiltonian operator (\ref{HDclassicalD}),
$H_D\Psi^s={\cal E}_N\,\Psi^s$, where the excited state energies are
those of Landau levels,
\be
{\cal E}_N={\cal E}_0+\frac\mu{\left\langle D^4\right
\rangle}\,N \ ,
\label{LandaucalE}\ee
with $\mu$ the topological photon mass (\ref{photonmass}). There is
therefore an infinite, continuous degeneracy labelled by the locations
$z_i\in\Sigma$ of the insertions. From the three-dimensional
perspective, higher string states thus correspond to Landau levels
which consist of $N=\sum_il_i$ gauge invariant combinations of external
charged particles and photons, situated at the points $z_i$. As
expected, in the naive vacuum $\langle D^2\rangle=0$, the excited
states become infinitely massive and decouple from the topological
sector of the quantum field theory with ${\cal E}={\cal E}_0=0$. For a
generic dilaton coupling to topologically massive gravity, the scale
of the theory is just the photon mass $\mu$.

In what follows we shall need the inner product of two states of the
form (\ref{Psisexcited}) in the finite temperature gauge theory on
$\Sigma\times\sphere^1$. The calculation proceeds along the lines of
that in section~\ref{TopCorrs}, and in keeping with the conventions of
the rest of this section we absorb all functional determinant factors
into an appropriate bulk normalization. The inner product of two chiral
excited states is given explicitly by the path integral
\bea
&&\Tr\left\langle\left.\Psi_1^s\left\{\matrix{x_i\cr Q_i\cr l_i\cr}\right\}
  \right|\Psi_0^{s'}\left\{\matrix{x_{i'}'\cr Q'_{i'}\cr l'_{i'}\cr}
  \right\}\right\rangle~=~\int\left[DA_z~DA_\bz\right]~
\e^{iS_{\rm TMGT}^{\Sigma\times[0,1]}[A]}\,\exp\left\{\frac k{4\pi}\,
\int\limits_\Sigma d^2z~A_\bz\,A_z\right\}\nn\\&&~~~~~~~~~~~~~~~~~~~~\times\,
\int[D\varphi]~\exp\left\{\frac k{8\pi}\,\int\limits_\Sigma d^2z~\left(
\partial_\bz\varphi-2A_\bz\right)\,\partial_z\varphi\right\}
\nn\\&&~~~~~~~~~~~~~~~~~~~~\times\,
\int[D\varphi']~\exp\left\{\frac k{8\pi}\,\int\limits_\Sigma d^2z~\left(
\partial_z\varphi'-2A_z\right)\,\partial_\bz\varphi'\right\}
\nn\\&&~~~~~~~~~~~~~~~~~~~~\times\,\prod_{i=1}^s\frac1{\sqrt{l_i!}}\,
\Bigl(A_z(x_i)-\partial_z\varphi(x_i)
\Bigr)^{l_i}~\e^{-i\,Q_i\,\bigl(\varphi(x_i)+h_\varphi(x_i)\bigr)}
\nn\\&&~~~~~~~~~~~~~~~~~~~~\times\,
\prod_{i'=1}^{s'}\frac1{\sqrt{l'_{i'}!}}\,
\Bigl(A_\bz(x'_{i'})-\partial_\bz\varphi'(x'_{i'})
\Bigr)^{l'_{i'}}~\e^{i\,Q'_{i'}\,\bigl(\varphi'(x'_{i'})+
  h'_{\varphi'}(x'_{i'})\bigr)} \ .
\label{excitedinnprod}\eea
We shift $A_z\to A_z+\partial_z\varphi$, $A_\bz\to
A_\bz+\partial_\bz\varphi'$ in (\ref{excitedinnprod}) and use gauge
invariance of the topologically massive gauge theory action. By
substituting in the gauge orbit decomposition (\ref{Aigaugeorbit}) and
using the bulk normalization described in section~\ref{OrbPart}, we
arrive at
\bea
&&\Tr\left\langle\left.\Psi_1^s\left\{\matrix{x_i\cr Q_i\cr l_i\cr}\right\}
  \right|\Psi_0^{s'}\left\{\matrix{x_{i'}'\cr Q'_{i'}\cr l'_{i'}\cr}
  \right\}\right\rangle\nn\\&&~~~~~~=~\int\left[D\bar A_z~D\bar A_\bz\right]~
\exp\left\{\frac k{4\pi}\,\int\limits_\Sigma d^2z~\bar A_\bz\,\bar
A_z\right\}\,\prod_{i=1}^s\frac1{\sqrt{l_i!}}\,\bar A_z(x_i)^{l_i}\,
\prod_{i'=1}^{s'}\frac1{\sqrt{l'_{i'}!}}\,\bar A_\bz(x'_{i'})^{l'_{i'}}\nn\\&&
{}~~~~~~~~~~~~\times\,\int[D\varphi]~\int[D\varphi']~
\exp\left\{\frac k{8\pi}\,\int\limits_\Sigma d^2z~\partial_\bz
  (\varphi+\varphi')\,\partial_z(\varphi+\varphi')\right\}\nn\\&&~~~~~~~~~~~~
  \times\,\prod_{i=1}^s\e^{-i\,Q_i\,\bigl(\varphi(x_i)+h_\varphi(x_i)\bigr)}
  ~\prod_{i'=1}^{s'}\e^{i\,Q'_{i'}\,\bigl(\varphi'(x'_{i'})+
  h'_{\varphi'}(x'_{i'})\bigr)} \ .
\label{excitedshifts}\eea

We evaluate the functional integrals in (\ref{excitedshifts}) exactly
as we did in section~\ref{TopCorrs}. The new feature here is the
insertions of the gauge fields in the holomorphic and antiholomorphic
sectors. The Gaussian integrals over these insertion points require $s=s'$,
and that the collections $\{x_i;l_i\}_{i=1}^s$ and
$\{x'_i;l'_i\}_{i=1}^s$ be equal as unordered sets, i.e. up to a
permutation $\pi\in S_s$ on $s$ letters. This is of course just the
usual Wick expansion, which is weighted by the combinatorical factor
$1/s!$. Each of these integrals also introduces a factor
$l_i!\,(4\pi/k)^{l_i+1}$, and additional normalization terms from
the harmonic degrees of freedom of the gauge fields due to these extra
insertions are taken care of as described in section~\ref{Tension}.
The integrals over the insertions $\varphi(x_i)$ and $\varphi'(x'_i)$
then impose charge conservation $\sum_i(Q_i+Q'_i)=0$. The rest of the
calculation proceeds exactly as before.

In this way we find that the dimension formula (\ref{dimcalHfinal}) is
modified by these excited states to
\bea
\Tr\left\langle\left.\Psi_1^s\left\{\matrix{x_i\cr Q_i\cr l_i\cr}\right\}
  \right|\Psi_0^{s'}\left\{\matrix{x_{i'}'\cr Q'_{i'}\cr l'_{i'}\cr}
  \right\}\right\rangle&=&\left(\frac{4\pi}k\right)^{N-\nu_g}
\,(pq)^g~\delta_{ss'}
{}~\delta_{[\lambda_1+\lambda_1'+\dots+\lambda_s+\lambda_s']}
\nn\\&&\times\,\frac1{s!}\,\sum_{\pi\in S_s}~\prod_{i=1}^s\,
\delta_{l_i\,,\,l'_{\pi(i)}}~\delta^{(2)}\left(x_i-x_{\pi(i)}'\right)
\ ,
\label{Trexcitedfinal}\eea
with the level number $N$ defined by (\ref{levelnumber}) and the index
$\nu_g$ defined in section~\ref{Tension}. Thus the
states (\ref{Psisexcited}) are orthogonal, and Dirac delta-function
normalizable with norm given by $(pq)^{g/2}\,(4\pi/k)^{(N-\nu_g)/2}$. Note that
(\ref{Trexcitedfinal}) depends explicitly on the actual insertion
points on $\Sigma$, in contrast to what occurs in the vacuum sector
where the gauge theory is topological. The formula
(\ref{Trexcitedfinal}) can be loosely thought of as describing the
modification of the dimension of the vacuum Hilbert space from the
excited membrane states. In the next subsection we show how to make
this statement more precise, and use it to provide another derivation
of the D-brane tension formula from the topological membrane approach.

\subsection{The Regularized Dimension}

In~\cite{TD} it was shown that there is a universal formula for the
tension of a D-brane in terms of the suitably regulated dimension of
the space of physical open string states of the associated boundary
conformal field theory. In this subsection we will show that, quite
remarkably, there is an analog of this dimension formula in the
topological membrane approach. The key feature is that once we
consider the entire Hilbert space ${\cal
  H}_{g,s,k}(\lambda_1,\dots,\lambda_s)=\bigoplus_{\{l_i\}}{\cal
  H}_{g,s,k}(\lambda_1,\dots,\lambda_s;l_1,\dots,l_s)$ of all states
of the form (\ref{Psisexcited}) with definite $U(1)$ charges, the
Hamiltonian no longer vanishes and has energy spectrum given by
(\ref{calE0VD},\ref{LandaucalE}). This implies that the statistical
mechanical partition function $\Tr_{{\cal
    H}_{g,s,k}(\lambda_1,\dots,\lambda_s)}(\e^{-\beta H_D})$ is no
longer independent of the inverse temperature $\beta$ which is the
circumference of the thermal circle $\sphere^1$. However, in the
high-temperature limit $\beta\to0$ whereby the circle shrinks to a
point, the partition function is the trace of the identity operator
and yields the dimension of the physical state space
${\cal H}_{g,s,k}(\lambda_1,\dots,\lambda_s)$. Of course this
dimension is formally infinite above the vacuum sector (containing
infinitely many Landau levels) as it includes
propagating states of the three-dimensional photon which has mass
(\ref{photonmass}). However, by carefully taking the limit we will see
that the temperature compactification provides a well-defined
regularization and yields a dimension formula which reproduces the
D-brane mass (\ref{effS}). This is completely analogous to the
way the dimension formula gives the brane tension in~\cite{TD}. In the
context of the topological membrane approach, this derivation is far
more desirable than our previous one, because it provides a more
intrinsic three-dimensional definition of the tension in terms of the
mass scales of the bulk theory, with no explicit reference to its
string theoretic origin.

Having to sum over all states (\ref{Psisexcited}) required to compute
the trace over ${\cal H}_{g,s,k}(\lambda_1,\dots,\lambda_s)$ would be an
arduous task. We can, however, simply the calculation
by the following observation. Since the electric field
operators obey
$[E_z(\vb{z}),E_z(\vb{z}')]=0=[E_\bz(\vb{z}),E_\bz(\vb{z}')]$,  we can
take a generic wavefunctional (\ref{Psisexcited}) and adiabatically
transport every one of its insertion points to a single point
$z\in\Sigma$. The only effect this operation will have is to induce phase
factors due to the linkings of the charged particle trajectories in
the bulk (see section~\ref{Surgical}). Although these phase factors
constrain the charge spectrum of the quantum gauge theory~\cite{TM_15}
(and hence the momentum lattice of string theory), for the calculation
that follows they play no role and will simply drop out of all
formulas with the appropriate normalizations. Thus without loss of
generality we may lift some of the infinite degeneracy at each level,
such that the contribution to the trace over physical states depends
on only a single \textit{representative} excited wavefunctional for
each Landau level $N$. In other words, we compute the trace only over
the single insertion states (\ref{Psisexcited}) which have $s=1$ and
$N=l$. We denote the Hilbert space in which these states live as
${\cal H}_{g,k}(\lambda)=\bigoplus_{N\geq0}{\cal
  H}_{g,1,k}(\lambda;N)$. In the string theory picture, this reduction
simply means that the mass is measured by the coupling of a brane
boundary state with the graviton vertex operator. Its relation to
Ishibashi states and the Cardy formula is obtained through the
analysis of section~\ref{FundWave}.

The regulated dimension of the Hilbert space is defined as
\be
{\rm reg}\,\dim{\cal H}_{g,k}(\lambda)=\lim_{\beta\to0}\,\Tr^{~}_{{\cal
    H}_{g,k}(\lambda)}\left(\e^{-\beta H_D}\right) \ .
\label{dimcalHbetadef}\ee
The first thing we need to do is define precisely the trace over $s=1$
states (\ref{Psisexcited}) contributing to the statistical mechanical
partition function in (\ref{dimcalHbetadef}). We sum over all
Landau levels $N\geq0$ and positions $x\in\Sigma$ of external particles. As
there are $N$ identical ways to insert the particles via the electric
field creation operator in (\ref{Psisexcited}), we need to divide by
the total number $N!$ of permutations in each Landau
level. Furthermore, we normalize the trace by dividing it by $(pq)^g$,
so that (\ref{dimcalHbetadef}) coincides with the number of linearly
independent topological wavefunctions when restricted to the vacuum
sector of the topologically massive gauge theory. Thus the formal
definition of (\ref{dimcalHbetadef}) is given by
\be
{\rm reg}\,\dim{\cal H}_{g,k}(\lambda)=\lim_{\beta\to0}\,\sum_{N=0}^\infty
\frac1{N!}\,\int\limits_\Sigma d^2x~\frac{\left\langle\left.\Psi_1^1
\left\{\matrix{x\cr Q\cr N\cr}\right\}\right|\e^{-\beta H_D}
\left|\Psi_0^1\left\{
\matrix{x\cr Q\cr N\cr}\right\}\right.\right\rangle}{(pq)^g} \ .
\label{dimcalHformaldef}\ee

Substituting the energy eigenvalues (\ref{LandaucalE}) and the
wavefunctional normalizations (\ref{Trexcitedfinal}) into
(\ref{dimcalHformaldef}), and summing over all Landau levels thereby
yields
\be
{\rm reg}\,\dim{\cal H}_{g,k}(\lambda)=\left(\frac k{4\pi}\right)^{\nu_g}\,
\lim_{\beta\to0}\,{\cal A}_\Sigma~\delta^{(2)}(0)~
\exp\left\{-\beta{\cal E}_0+\frac{4\pi}k~\e^{-\beta\mu/
    \langle D^4\rangle}\right\} \ ,
\label{dimcalHsubstsum}\ee
where ${\cal A}_\Sigma$ is the area of the Riemann surface $\Sigma$
and the ground state energy ${\cal E}_0$ is given by
(\ref{calE0VD}). As expected from bulk charge conservation, the
expression (\ref{dimcalHsubstsum}) is independent of $Q$. However, it
is singular. Due to the Dirac delta-function normalizability of the
excited physical states, the contribution from each Landau level diverges as
$\delta^{(2)}(0)$. To get a finite result, we remove these quantum
field theoretic infinities by an appropriate renormalization of the
gravitational sector of the bulk theory at finite temperature. For
this, we regulate the delta-function at the origin as ${\cal
  A}_\Sigma~\delta^{(2)}(0)=\Lambda_{\rm uv}/\mu$, where $\Lambda_{\rm
  uv}\to\infty$ is a fundamental ultraviolet cutoff on the string worldsheet
$\Sigma$. In the given non-trivial dilaton background, it follows from
(\ref{dimcalHsubstsum}) that the three-dimensional Planck mass
$\kappa$ should then scale logarithmically with the cutoff and linearly
with the temperature as
\be
\kappa~\Bigl\langle V(D,\omega)\Bigr\rangle=\frac1\beta\,\ln\left(\frac{
    \Lambda_{\rm uv}}\mu\right)
\label{kappascale}\ee
in the limits $\beta\to0$ and $\Lambda_{\rm uv}\to\infty$. Then our
derivation of the dimension formula is non-singular provided that the mass
scale of the gravitational sector is logarithmically close to the
ultraviolet cutoff (and the topological photon mass) and linearly
close to the (infinite) temperature.

However, this is not quite what we want, as one easily checks that
even after this renormalization, the formula (\ref{dimcalHsubstsum})
does not reproduce the correct target space radius dependence of the D-brane
tension that we found in section~\ref{Tension}, except in the
topological limit $\mu\to\infty$ or $\langle D^2\rangle=0$. The reason
for this is that from the point of view of the bulk three-dimensional
dynamics, we should be decompactifying the temperature circle rather
than shrinking it to a point, in order to map the excited state
contributions onto those appropriate for the topological membrane
dynamics. Thus we instead compactify the Euclidean time direction on
the {\it dual} circle of radius $\tilde\beta=1/\mu^2\beta$, and take
the equivalent limit $\tilde\beta\to\infty$. The expression
(\ref{dimcalHbetadef}) is thereby computed as
\be
{\rm reg}\,\dim{\cal H}_{g,k}(\lambda)=\lim_{\tilde\beta\to\infty}\,
\Tr^{~}_{{\cal H}_{g,k}(\lambda)}\left(\e^{-\tilde\beta H_D}\right) \ .
\label{dimcalHbetadefdual}\ee
Everything proceeds in precisely the same way as above but with the temperature
replaced by its dual. In particular, by taking the limits
$\tilde\beta\to\infty$ and $\Lambda_{\rm uv}\to\infty$ with
$\ln(\Lambda_{\rm uv})/\tilde\beta$ held fixed, the topological
graviton mass now undergoes a {\it finite} renormalization. With this,
the above calculations show that only the lowest Landau level $N=0$
contributes in the limit $\tilde\beta\to\infty$, and we arrive
at the final expression
\be
{\rm reg}\,\dim{\cal H}_{g,k}(\lambda)~=~\left\langle\left.\Psi_1^1
\left\{\matrix{x\cr Q\cr 0\cr}\right\}\right|\Psi_0^1\left\{
\matrix{x\cr Q\cr 0\cr}\right\}\right\rangle_{\rm ren}~=~\left(\frac k
{4\pi}\right)^{\nu_g}
\label{regdimfinal}\ee
for the regulated dimension of the physical Hilbert space. Note that
this formula could also have been obtained by taking the instead the
limit $\beta\to1/\mu$ in (\ref{dimcalHsubstsum}), appropriate to the
membrane geometry $\Sigma\times[0,1]$, in the topological sector where
$\langle D^2\rangle=0$.

Comparing with (\ref{effS}), we thereby arrive at the dimension
formula
\be
{\cal M}^2=\left(\frac\pi{64}\right)^{\nu_g/2}\,\alpha'\,
(g_s)^{-\chi(\Sigma)}~\sqrt{{\rm reg}\,\dim{\cal H}_{g,k}(\lambda)} \ .
\label{tendimformula}\ee
The appearence of the square root here is not surprising when we apply
the orbifold correspondence rule (\ref{orbsqrtclosed}) to the
closed string amplitude which appears in (\ref{regdimfinal}), i.e. it
produces the appropriate regularized dimension for the {\it open}
string sector. This applies to the $\sf PT$ orbifold, corresponding to
a D-brane wrapping the target space circle $\sphere^1$ of radius $R$
given by the formula (\ref{CScouplingid}), in which case the physical
state normalization formula (\ref{Trexcitedfinal}) may be applied directly.
In the case of the $\sf PCT$ orbifold, corresponding to a Neumann
brane along the target space circle, one needs to modify the
contribution of harmonic modes to the normalization functional
integral (\ref{excitedshifts}) along the lines described in
section~\ref{Tension}. It is straightforward to then show in a
completely analogous way that the regulated dimension of the space of
physical membrane states is given by ${\rm reg}\,\dim{\cal
  H}_{g,k}(\lambda)=(4\pi/k)^{\nu_g}$, and hence that once again the
formula (\ref{tendimformula}) holds. Thus (\ref{tendimformula}) is a
universal result for the physical states of the topological membrane.

\subsection{Worldsheet Duality}

The truncation to the lowest Landau level $N=0$ used to arrive at
(\ref{regdimfinal}) is the analog of the fact that the dominant
contribution to the regularized dimension of the open string Hilbert
space in the associated boundary conformal field theory comes from the
identity representation in the closed string picture~\cite{TD}. This is
completely consistent with the membrane interpretation, as these are
the only contributions relevant to the orbifold of the amplitude
(\ref{amplcomplset}). In fact, the passing to the dual temperature
compactification $\beta\to\tilde\beta$ is the analog of the worldsheet
modular transformation that is used to compute the boundary conformal
field theoretic dimension of states and which induces the usual
worldsheet duality between the open and closed string sectors. The
Cardy condition in this context simply states the equality
of the closed string cylinder amplitude between two boundary states
and the one-loop annulus amplitude with corresponding prescribed open
string boundary conditions. This is essentially what was computed in
section~\ref{Cardy}. It follows from the calculation of the previous
subsection that this worldsheet duality is simply a duality between
Wilson and Polyakov loop correlators in the three-dimensional gauge
theory. In fact, as we now explain, the topological membrane approach
gives an interesting new perspective on this duality.

Consider the one-loop open string vacuum diagram with boundary
conditions labelled by $\alpha$ and $\beta$. In the membrane picture,
it corresponds to an annulus $\Sigma^{\rm o}=\cyl^2$ which sits at
three-dimensional time $t=1/2$ (see fig.~\ref{fig.orbi}). Its
Schottky double is the torus $\Sigma=\torus^2$ which sits at the times
$t=0$ and $t=1$. In the open string channel, as the string propagates
around the cycle of the annulus, each point on it is described by {\it
  two} pre-images on the torus. In the closed string channel, it
thereby indeed corresponds to closed string propagation, but now
between the two corresponding Ishibashi states. With left and right
moving worldsheets glued onto each other in the orbifold picture,
i.e. $\Sigma_0\equiv\Sigma_1$, the propagation in worldsheet time
$\tau\in[0,1]$ is interlaced with the membrane propagation time
$t\in[0,1]$ in such a way that the equivalence between the open and
closed string channels is clear. These processes are all depicted in
fig.~\ref{Duality}.

\fig{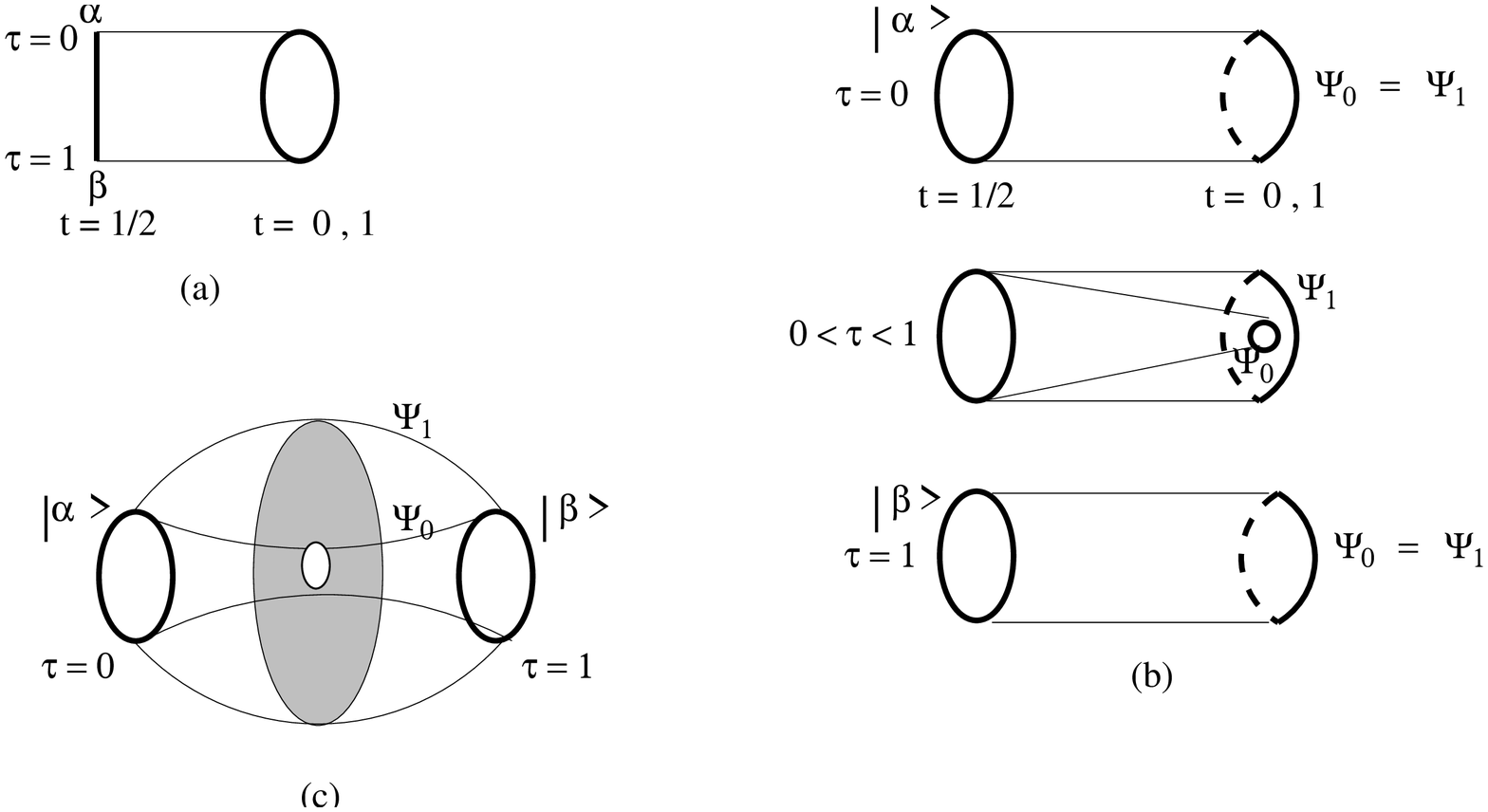}{Membrane representation of worldsheet duality. {\rm (a)}
  An open string has two closed string pre-images. {\rm (b)} The interlacing
  between worldsheet and membrane time evolutions. Propagation of the
  string between its initial and final Ishibashi states
  $|\alpha\rangle\!\rangle$ and $|\beta\rangle\!\rangle$ requires a
  superposition of the initial and final membrane wavefunctionals
  $\Psi_0$ and $\Psi_1$. Here the horizontal direction is propagation
  along the membrane time $t$. {\rm (c)} The propagation of the closed
  string between the Ishibashi states naturally induces a worldsheet
  annulus diagram. Here the horizontal direction is propagation along
  the worldsheet time $\tau$.}{Duality}

\subsection*{Acknowledgments}

The work of P.C.F. was supported by Grant SFRH/BPD/5638/2001 from
FCT-MCT~(Portugal). The work of I.I.K. was supported by PPARC Grant
PPA/G/0/1998/00567 and EU Grant FMRXCT960090. The work of R.J.S. was
supported by a PPARC Advanced Fellowship.

\end{document}